\documentclass[a4paper,11pt]{article}
\usepackage{jcappub} 
\usepackage{lineno} 
\usepackage{appendix}
\usepackage{orcidlink}
\usepackage{hyperref}
\usepackage{mathtools}
\usepackage{bm}

\usepackage{comment}

\usepackage{cleveref}

\newcommand{\beq}{\begin{equation}}
\newcommand{\eeq}{\end{equation}}

\newcommand{\beqn}{\begin{eqnarray}}
\newcommand{\eeqn}{\end{eqnarray}}

\usepackage{footnote}

\begin{document}

\title{
Gravitational Wave Echoes of the First Order Phase Transition in a Kination-Induced Big Bang 
}
\date{}

\author[a,b]{Richard~Casey,}
 \author[b,c]{Katherine~Freese}
 \author[b,d]{and Evangelos~I.~Sfakianakis}
 \affiliation[a]{Department of Physics and Astronomy, Colgate University, Hamilton, NY 13346, USA}
\affiliation[b]{Texas Center for Cosmology and Astroparticle Physics,
Weinberg Institute for Theoretical Physics, Department of Physics,
The University of Texas at Austin, Austin, TX 78712, USA}
 \affiliation[c]{The Oskar Klein Centre, Department of Astronomy, Stockholm University, 106 91 Stockholm, Sweden}
\affiliation[d]{Department of Physics, Harvard University, Cambridge, MA, 02131, USA}

\emailAdd{rcasey@colgate.edu}
\emailAdd{ktfreese@utexas.edu}
\emailAdd{evangelos.sfakianakis@austin.utexas.edu}

\abstract
{
Gravitational waves (GWs) produced during first-order phase transitions (FOPTs) in the early universe provide a powerful probe of nonstandard cosmological histories. We study GW production from a FOPT ending a kination-dominated epoch in the \emph{Kination-Induced Big Bang} scenario, in which a period of kination domination terminates through a phase transition that reheats the universe into radiation domination. 
A rolling scalar field drives the kination epoch. In the specific model we consider, its derivative coupling to a second scalar (tunneling field) dynamically traps the latter in a false vacuum, with the phase transition triggered as the kination field slows due to Hubble friction.
We compute the resulting stochastic GW background from bubble nucleation and collisions, presenting analytic estimates and numerical results for the peak amplitude and frequency. In all cases we find  an upper bound $\Omega_{\rm GW} h^2 \lesssim 2\times10^{-7}$ from the bubble percolation condition. In the case where the false vacuum energy dominates at the transition (yet the kination field drives the FOPT), we find $\Omega_{\rm GW} h^2 \gtrsim 10^{-12}$.  We further find that the Hubble scale during the phase transition across a broad set of model parameters is bounded by ${\cal O}(10^{-13})  M^2/M_{\rm Pl} \lesssim H_\ast \lesssim {\cal O}(0.1)  M^2/M_{\rm Pl}$, where $M$ is the mass-scale controlling the strength of the interaction between the kination and tunneling fields. The predicted signal spans frequencies from nHz to MHz, allowing the model to explain the signal reported by Pulsar Timing Array experiments and to be constrained or probed by interferometers such as LISA, Advanced LIGO, Cosmic Explorer, and BBO.
Interestingly, a FOPT can occur even if the bare tunneling potential has a single minimum, as metastability is generated dynamically by the  coupling between the tunneling  and the kination field.
}

\maketitle

\section{Introduction}

The 2015 detection of gravitational waves (GWs) by the LIGO/Virgo collaboration \cite{LIGOScientific:2016aoc} marked the beginning of a new era in observational astronomy.
Additionally, recent observations by Pulsar Timing Arrays (PTAs), including the NANOGrav collaboration \cite{NANOGrav:2023gor}, 
The European Pulsar Timing Array (EPTA) and Indian Pulsar Timing Array (InPTA)~\cite{EPTA:2023fyk, EPTA:2023xxk}, the Parkes Pulsar Timing Array (PPTA)~\cite{Reardon:2023gzh} and the Chinese Pulsar Timing Array (CPTA)~\cite{Xu:2023wog} 
 have collectively reported compelling evidence of a stochastic gravitational wave background (SGWB) signal.
While the ``standard" explanation of the pulsar timing array signals is black hole mergers,\footnote{It is a matter of debate whether merging SMBHs may suffer from completing their mergers due to the ``final parsec problem"~\cite{Milosavljevic:2002ht, Tiruvaskar:2025lkq}.}
alternative cosmological explanations are also possible, including a first order phase transition (FOPT) in the early universe.
In fact, Freese and Winkler~\cite{Winkler:2024olr}, by means of a frequentist hypothesis test, found that the observed gravitational wave (GW) spectrum prefers a first-order phase transition at $2 - 3 \sigma$ significance compared to black hole mergers (depending on the underlying black hole model). This mild preference is linked to the 
larger amplitude and the spectral shape of the predicted GW signal.

Overall, gravitational waves (GWs) have  emerged as a powerful observational window into the
dynamics of the early universe, providing unique insights into phenomena otherwise inaccessible to conventional electromagnetic astronomy, opening up extraordinary opportunities for breakthroughs in both particle physics and cosmology. 
Future missions such as LISA \cite{LISA:2017pwj, Caprini:2015zlo}, BBO~\cite{Crowder:2005nr}, DECIGO \cite{Kawamura:2020pcg}, Cosmic Explorer~\cite{Reitze:2019iox} and the Einstein Telescope \cite{Punturo:2010zz}, promise to illuminate high-energy processes occurring shortly after the Big Bang, including phase transitions, cosmic inflation, and exotic cosmological epochs. 
For
for GW predictions in LIGO from FOPTs see~\cite{Lopez:2013mqa} and subsequent work, e.g.~\cite{Caprini:2015zlo, Hindmarsh2021Phase, Athron:2023xlk, Caldwell:2022qsj, Binetruy:2012ze, Freese:2022qrl, Winkler:2024olr}.

In some models of inflation, a FOPT provides the transition from the inflationary phase, where the universe expands quasi exponentially and the energy density remains almost constant, into the hot Big Bang, defined as the moment where the energy stored in the fields driving inflation is transferred (in part) to the radiation and elementary particles that make up the universe. 
One difficulty of ending inflation in a FOPT is ensuring that the phase transition occurs quickly enough so that the bubbles of new phase do in fact collide and thermalize (reheat) the universe, avoiding the so-called ``empty universe" problem \cite{guth1983could}. One proposed solution to this problem is double field inflation, where a rolling scalar field (coupled to the tunneling field) serves to make the bubble nucleation rate of the tunneling field time dependent \cite{adams1991double,linde1990eternal}.
In this model, the tunneling rate is initially slow so that
inflation can last long enough to solve the horizon and flatness problems; subsequently the tunneling rate becomes fast so that the phase transition completes throughout the entire universe, leading to  
successful reheating. The two fields are coupled, e.g. via a polynomial coupling; the rolling field triggers the FOPT, by changing the properties of the tunneling potential.  This is an appealing solution, since extra scalar fields in the early universe arise naturally in high energy theories, like string theory and supergravity, which are usually associated with the energy scales relevant for inflation, even though a fully explicit embedding of inflation in string theory remains elusive~\cite{Baumann:2014nda, Linde:1990flp}.

Well-motivated nonstandard early-universe histories beyond the simplest inflationary scenarios also exist. In this paper we study the alternative known as ``kination,"~\cite{Spokoiny1993Deflationary, Joyce1997Cosmological},  in which 
the energy density of the universe, and hence its cosmological evolution, is dominated by the kinetic energy of a rolling scalar field.
In particular we focus on a ``Kination-Induced Big Bang" (proposed by one of us \cite{Freese:2022qrl}), in which a period of kination domination ends with a FOPT that reheats into the ordinary radiation-dominated early history of our universe.  This model
builds on the framework of double field inflation, in that the behavior of one scalar field controls the rate of a FOPT in a second tunneling field and eventually triggers the phase transition.
In this paper, it is the  kinetic energy of the rolling field that  triggers the FOPT of the tunneling field, as described in detail in the following section. 

A Kination Induced Big Bang can be especially interesting as a way to reheat quintessential inflation \cite{peebles1999quintessential},
a cosmological model in which a single scalar field is  responsible for both primordial inflation and the current accelerated expansion of  the universe.
While a major motivation of a Kination-Induced Big Bang is to provide a simple model that reheats quintessential inflation, a Kination-Induced Big Bang is a general mechanism that can (in principle) reheat any cosmological model that runs through a period of kinetic energy domination  \cite{Freese:2022qrl}. Thus, our analysis of the gravitational wave signatures produced from  a Kination-Induced Big Bang is applicable in a wide array of models.

The purpose of this paper is to investigate the gravitational wave (GW) signatures arising from a FOPT occurring during or shortly after a kination epoch, where the kination-driving scalar field is coupled  to and triggers a second scalar field responsible for tunneling from a false vacuum to a true vacuum. As  bubbles of the true vacuum expand and collide, spherical symmetry is broken,   time-varying quadrupole moments appear in the stress-energy tensor, and GWs are produced.
In Section \ref{sec:model}, we present a model for the Kination-Induced Big Bang and equations for the peak amplitude and frequency of GW production.
In Section~\ref{sec:analytic} we derive approximate analytic results, which apply to a specific limit of the parameter space. In Section~\ref{sec:parameterscan} we  perform an exhaustive parameter scan. In Section \ref{sec:experiments}, we turn to the major current and future GW experiments and assess their ability to probe the parameter space of our model.
We demonstrate that, for different values of the underlying parameters, our model can cover the whole frequency-amplitude observable space, with frequencies ranging from nano-Hz to MHz (and beyond). The amplitude is bounded from above by $\Omega_{GW}h^2\lesssim {2\times}10^{-7}$ and can get arbitrarily small values, albeit with $\Omega_{GW}h^2\sim  10^{-12}$ being a special value, related to our analytic approximation.
Finally, we summarize our results and outline avenues for future work in Section~\ref{sec:conclusions}.

\section{Model}\label{sec:model}

We consider a model for a Kination-Induced Big Bang comprised of two real scalar fields $\phi,\chi$, where
the field $\phi$ is responsible for the kination dominated epoch and is derivatively coupled to a tunneling field $\chi$ which has a double well potential,
via the Lagrangian \cite{Freese:2022qrl}
\begin{equation}
    {\cal L} =\left ( {1\over 2} - {\chi^2\over M^2}\right )
\partial_\mu\phi \partial^\mu\phi+ {1\over 2} \partial_\mu\chi \partial^\mu\chi -V(\phi) - {1\over 2}m_\chi^2\chi^2 +\mu\chi^3 -\lambda^2 \chi^4 - V_0 \, .
\label{eq:lagrangian}
\end{equation}
valid below\footnote{Since we are mostly interested in the Hot Big Bang phase transition after
kination, it is suﬃcient to describe this period in the eﬀective theory given here (there is no need
to consider an explicit ultraviolet completion).} the scale $M$. 
We take the $(+,-,-,-)$ convention for the spacetime metric signature.
The derivative interaction between the kination field $\phi$ and the tunneling field $\chi$ is suppressed by a mass scale $M$, which should be interpreted as the cutoff of the effective field theory. In this respect, $M^{-2}$ plays a role analogous to the Fermi constant\footnote{The Fermi four-fermion operator ${\cal L}_{\rm{Fermi}}\sim G_F J^{\mu\dagger} J_\mu$, where $J_\mu$ is the charged weak current~\cite{Donoghue:2022wrw} and $G_F\sim m_W^{-2}$ is the Fermi constant, is the effective description of fermion interactions induced by 
$W$-boson exchange at energies well below 
the $W$-boson mass, $m_W$.} in the four-fermion description of weak interactions, parametrizing a higher-dimensional operator obtained after integrating out heavy degrees of freedom and controlling the strength of the interaction. Equivalently, the same interaction can be understood as arising from a nontrivial field-space metric, with $M$ setting the curvature scale of the scalar manifold, as discussed further in Appendix~\ref{appendix:fieldspace}.
Consistency of the theory in terms of having a positive kinetic energy requires 
\begin{equation}
\label{eq:pos KE}
{\chi^2\over M^2} <{1 \over 2} 
\end{equation}
 to avoid ghost-like terms\footnote{Appendix~\ref{appendix:fieldspace} discusses a complete Lagrangian, without the limitations of a kinetic term becoming negative, but in order to avoid introducing more structure than is needed for the dynamics of a kinetically induced first order phase transition, we  work with Eq.~\eqref{eq:lagrangian} in the main body of this work.}.

 The Universe starts out in a kination phase, in which the kinetic energy of the $\phi$ field dominates the energy density of the Universe.
A scalar field rolling in an FRW spacetime is described by the Klein Gordon equation
$
\ddot\phi + 3H\dot\phi +V_{,\phi}=0
$.
During kination we can neglect $V(\phi$)
since the potential energy of the kination field must be subdominant, $\dot\phi^2 /2\gg V(\phi)$, (otherwise it would not be kination). This leads to a stiff equation of state $w \equiv p_\phi /\rho_\phi \approx 1$. Neglecting the potential in the Klein Gordon equation leads to the simple (approximate) solution for the rolling field $\dot \phi \propto a^{-3}$, where $a(t)$ is the scale factor of the Universe, meaning that the kination field velocity is reduced over time (see  Appendix~\ref{appendix:kination} for a review of kination).  As we will now show, this evolution of $\dot \phi$ will drive the  field $\chi$ to undergo a FOPT.

The  potential of $\chi$, shown in Fig.~\eqref{fig:cartoonpotential}, has two minima separated by a potential barrier: a metastable minimum (false vacuum) at $\chi_{\rm {min},F} = 0$ and a stable minimum (true vacuum) at 
\begin{equation}
\chi_{\rm {min},T} ={ 3 \mu   + \sqrt{9\mu^2- 16 \lambda^2 m_\chi^2} \over 8 \lambda ^2} 
\end{equation}
 (see Appendix~\ref{appendix:potential} for details on the potential).
Plugging  $\chi_{\rm {min},T}$ into Eq. \eqref{eq:pos KE} restricts our parameters to (see Appendix~\ref{sec:EFTlimit}) 
\beq
\label{eq:boundonparamters}
{\mu\over M}  < \lambda^2  \, .
\eeq
We fix $V_0$ such that the potential energy of $\chi$
vanishes in the true minimum. For this choice, $V_0$ is equal to the false vacuum energy density.

The effective mass of the tunneling field is given by the sum of the two quadratic terms for $\chi^2$ in the Lagrangian,
\beq
m^2_{\chi,{\rm {eff}}}=m_\chi^2 + 2{\dot\phi^2\over M^2} \, .
\eeq 
For $m_{\chi,{\rm {eff}}}>3\mu/4\lambda$ only the minimum at the origin exists for $V(\chi)$; for $\mu/\sqrt{2}\lambda<m_{\chi,{\rm {eff}}}<3\mu/4\lambda$ a  second minimum exists but with a higher potential energy than the one at the origin; and  for $m_{\chi,{\rm {eff}}}<\mu/\sqrt{2}\lambda$  the second minimum (the true vacuum at $V(\chi_{\rm {min},T}) =0$)) becomes energetically favorable.

\begin{figure}
    \centering
    \includegraphics[width=.5\textwidth]{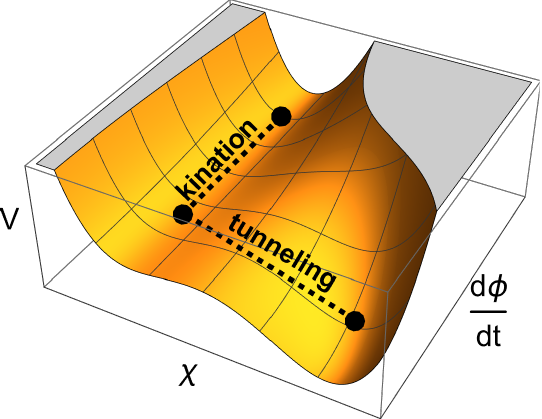}
\caption{The potential in a kinetically induced first order phase transition as a function of the tunneling field amplitude and the kination field velocity $V=V(\chi,\dot\phi)$.
The kination field $\phi$  initially has a large velocity $\dot\phi$.
The tunneling field $\chi$ is then trapped in a metastable minimum through its coupling to the kination field ${\cal L}_{\rm {int}} = -\chi^2(\partial\phi)^2/M^2$. While the
kination field slows  down through Hubble friction as $\dot\phi\propto a^{-3}$ a second deeper minimum occurs along the $\chi$-direction. Once the barrier between the two minima becomes sufficiently small $\chi$ tunnels into the true minimum and kination ends through a first order phase transition.}
\label{fig:potential3Dplot}
\end{figure}

The kinetic term $\dot \phi^2$ is initially large and then decreases with time, driving $m_{\chi,{\rm {eff}}}$ to smaller values as time goes on.
Specifically, at early times, during kination,  $\dot\phi^2$ can be as large as $M^4$, and 
the field $\chi$ is stabilized at the minimum $\chi=0$,
which is either energetically
favorable or separated by a large potential barrier (both the height and width of the barrier grow with $m_{\chi,{\rm {eff}}}$ as shown in Appendix~\ref{appendix:potential}). 
As the universe expands, the Hubble friction progressively reduces the velocity $\dot\phi^2$, which scales as $\dot\phi^2\propto a^{-6}$ during kination.
At some critical value $\dot\phi_c^2$ the potential barrier is small enough, allowing the $\chi$ field to start tunneling through it with a high enough rate and subsequently roll to the true vacuum \cite{Freese:2022qrl}; see Figure~\ref{fig:potential3Dplot}.  We define $t_*$ to be the time at which this phase transition takes place.  The FOPT leads to the formation of true vacuum bubbles which collide and reheat the universe in a Hot Big Bang. This mechanism is similar to that of double field inflation~\cite{Adams:1990ds}, where a rolling field triggers tunneling of a second field to the true vacuum and thus reheating of the universe. However the current model is different in that it is the kinetic energy of the rolling field during a kination epoch (rather than the value of the vacuum energy) that triggers the phase transition.

The critical time at which the phase transition takes place $t_*$ is defined in terms of the tunneling rate as we now discuss. 
Assuming that there is no (significant amount of) plasma present during the kination phase\footnote{This is  true for example if kination follows an inflationary phase. If there is a plasma present, two effects must be considered. Firstly, the ratio of the false vacuum energy to the total energy density of the Universe must include the radiation energy of the plasma, through the parameter $\alpha$, which is defined in Eq.~\eqref{eq:alpha}. Furthermore, the thermal transition rate between the false and true vacua is given by \cite{linde1981fate, linde1983decay} $\Gamma_T  = T^4 (S_3/2\pi T)^{3/2}e^{-S_3/T}$, where $S_3$ is the three dimensional Euclidean action interpolating between the two vacua.
The transition is dominated by the fastest of the two rates: the thermal one defined here or the vacuum one, defined in Eq.~\eqref{eq:Gamma}.} the transition rate per volume between the false and true vacua (quantum tunneling through the potential barrier at zero temperature) is given by
\beq
\Gamma = m_{\chi,{\rm {eff}}}^4 \left ({S_4\over 2\pi}\right)^2 e^{-S_4}
\, ,
\label{eq:Gamma}
\eeq
where $S_4$ is the four-dimensional Euclidean action of the bounce solution interpolating between the two vacua \cite{coleman1977fate, Callan:1977pt}. An approximate analytic  expression of $S_4$ for the potential of Eq.~\eqref{eq:lagrangian} is given in Eq.~\eqref{eq:S4sum}.

Since quantum mechanical tunneling is inherently a probabilistic process, every point in space has a probability of remaining in the false vacuum at time $t$, $P_{fv}(t) = e^{-I(t)}$, where $I(t)$ is the (expectation value of the) number of bubble nucleation sites in the past light cone of any chosen point~\cite{guth1980phase,guth1981cosmological}
\beq
\label{eq:Itintegral}
I(t) = {4\pi\over 3} \int_{0}^t dt'\,  \Gamma(t') a^3(t') r_{\rm {com}}^3 (t,t')
\eeq
 We can define the time of the phase transition $t_*$ as the time where each point has on average one bubble of true vaccuum nucleated within its past light cone; $I(t_*)=1$ \cite{Ellis:2018mja,Freese:2022qrl}.
 The comoving radius of the past light cone $r_{\rm {com}}$ is \cite{Freese:2022qrl}
 \beq
r_{\rm {com}}(t,t') = \int_{t'}^t {d\tilde t\over a\left (\tilde t\right )}
 \eeq
where $a\left (\tilde t\,\right )$ is the scale factor of the universe driven (in our case) by the vacuum energy of the false vacuum and the kinetic energy of the rolling field $\phi$.

An important parameter for a first order phase transition is  $\beta^{-1}$, which describes the duration of the phase transition and is  defined through \cite{Freese:2022qrl}
\beq
\beta \equiv \left . - {\dot P_{fv} \over   P_{fv} } \right |_{t=t_*} = \left . \dot I \right |_{t=t_*} \simeq \left .  {\dot\Gamma\over \Gamma }\right |_{t=t_*} \simeq - \left .\dot S_4 \right |_{t=t_*}
\label{eq:betadef}
\eeq
The last (approximate) equality arises by considering that the dominant contribution to $\dot\Gamma$ arises from the term $e^{-S_4}$ in Eq.~\eqref{eq:Gamma}. The second to last (approximate) equality arises because the time dependence of I(t) in Eq. \eqref{eq:Itintegral} is determined primarily by the exponential time-
dependence of $\Gamma$ rather than by the power-law time-dependence of the light cone volume.

In order for the first order phase transition to successfully complete and reheat the universe, the bubbles of true vacuum must percolate. This requires that the part of the Universe that is ``stuck" in the false vacuum   decreases with time \cite{turner1992bubble}. The volume of the false vacuum is $V_{fv} \propto a^3(t) P(t)$ and since $a(t)$ grows, we require a stronger condition than simply $P(t)$ decreasing with time (or $\beta<0$) \cite{ellis2019maximal}. In fact we require $d_t V_{fv} =   d_t \left (a^3(t) P(t)\right ) <0$ which  leads to 
\beq
\beta >3H_*
\eeq
\cite{Guth:1982pn,turner1992bubble} where $H_* \equiv H(t_*)$ is the Hubble parameter at the time of the phase transition. As we will see later, this requirement puts an upper bound on the magnitude of the stochastic GW signal produced during the FOPT.

 The energy density of the universe at the time $t_*$ of the phase-transition can be estimated as
\begin{equation}
\label{eq:rhostar}
\rho_{\text{tot}}(t_*) = \frac{\dot{\phi}_c^2}{2} + V_0\,.
\end{equation}
Depending on  parameter choices, the phase transition may occur during kination or shortly after kination. In the second case, the universe undergoes a second period of vacuum domination driven by $V_0$. Any second vacuum-domination can, however, only last very briefly. Otherwise $\dot{\phi}$ would be completely redshifted away making it implausible that the evolution of $\dot{\phi}$ triggers the phase transition.

\subsection{GW production: Peak Ampliude and Frequency}

As shown in Eq \eqref{eq:rhostar}, the energy density of the Universe shortly before the phase transition is distributed between the false vacuum $\rho_{\rm vac} = V_0$ and the kinetic energy of the rolling field (we take any radiation component before the phase transition to be subdominant). We define \cite{kamionkowski1994gravitational, Freese:2022qrl} the ratio of these two quantities at the time of the phase transition $t_*$ as 
\beq
\label{eq:alpha}
\alpha \equiv {\rho_{vac}\over \rho_{kin}(t_*) }
={V_0\over {1\over 2}\dot\phi_c^2}
 \, .
\eeq
As mentioned above, depending on  parameter choices, the phase transition may occur during kination ($\alpha < 1$) or shortly after kination. In the second case, the universe undergoes a second brief period of vacuum domination driven by $V_0$ (corresponding to $\alpha > 1$). One can show that the maximum value of $\alpha$ in the latter case scales as $\alpha\propto \lambda^{-2}$, which will be demonstrated numerically in Section~\ref{sec:experiments}. We consider both regimes ($\alpha\lessgtr 1$) in this paper.
During the phase transition,  bubbles of true vacuum form, expand and collide.  Then the vacuum energy and kinetic energy of the rolling field will be converted to thermal energy of the ensuing radiation. We denote the temperature of the radiation bath immediately after the phase transition as $T_*$ and assuming that the transition time is short compared to the Hubble time, we can equate the total energy density before and after the transition \cite{Freese:2022qrl}, 
\beq
\rho_r(T_*) \simeq \rho_{tot} \simeq \rho_{vac}+\rho_{kin}(t_*)
= \rho_{vac} {\alpha+1\over\alpha}
\eeq
We denote the temperature
of the radiation bath right after the transition by $T_*$.
Assuming immediate thermalization of the radiation after the FOPT, the temperature $T_*$ is given by \cite{Freese:2022qrl}
\beq
\label{eq:Temperature}
T_*^4 \simeq 
{90 \over \pi^2 g_{\rm{eff}}}H_*^2M_{\rm {Pl}}^2
=
{30 \over \pi^2 g_{\rm{eff}}
(T_*)} (\rho_{vac} + \rho_{kin}(t_*) )
= {\alpha+1\over \alpha}{30 \rho_{vac}\over \pi^2 g_{\rm{eff}}(T_*)} \, ,
\eeq
where $g_{\rm {eff}}$ is the effective number of radiation degrees of freedom. Throughout this work, we choose a single value of $g_{\rm {eff}}(T_*)=10$ for concreteness, leading to $T_* \simeq 0.75 V_0^{1/4} (\alpha+1)^{1/4}\alpha^{-1/4}$.

As  bubbles of the true vacuum expand and collide, spherical symmetry is broken  and thus time-varying quadrupole moments appear in the stress-energy tensor. The GW signal
has a  peak at a frequency related to the inverse time scale of the transition; see Eq.~\eqref{eq:frequency}.\footnote{In the presence of a radiation plasma during the transition, other sources of GWs include sound waves and turbulence in the plasma, which we ignore in our case.}
From the time of GW production until today, the frequencies are redshifted and the amplitude is diluted due to the expansion of the Universe.
The contribution of GW to the energy density of the universe today, evaluated at the peak of the GW spectrum, is  \cite{kosowsky1992gravitational,kosowsky1993gravitational}
\beq
\label{eq:omegaGW}
\Omega^{\rm{peak}}_{\rm {GW}}h^2 ={ 7.6 \times 10^{-5}\over g_{\rm {eff}}^{1/3}(T_*)}
\tilde \Omega \left ({H_*\over \beta}\right )^2 \left ({\kappa \, \alpha\over \alpha+1}\right )^2
\eeq
The parameter $\kappa$ is the energy fraction carried by the bubble walls at collision, which we set to unity for our model. The parameter $\tilde \Omega$ sets the overall normalization
of the spectrum, and we take $\tilde \Omega=0.05$,
which is between the result of lattice simulations and the envelope (thin wall) approximation\footnote{The uncertainty is about $\pm 50 \%$, which is not significant for our purposes, given that $\Omega_{\rm {GW}}$ can vary over several orders of magnitude. For a discussion of the different methods for computing the shape of the GW spectrum and their limitations see Ref.~\cite{Freese:2022qrl} and references therein.} \cite{Freese:2022qrl}.
Inserting the values of $\kappa$, $\tilde \Omega$ and setting $g_{\rm {eff}}=10$ for concreteness leads to
\beq
\label{eq:omegaGWsimplewithalpha}
\Omega^{\rm{peak}}_{\rm {GW}}h^2 ={ 1.8 \times 10^{-6}}
 \left ({H_*\over \beta}\right )^2 \left ({ \, \alpha\over \alpha+1}\right )^2 \, .
\eeq
 In this work we  focus primarily on the  amplitude and  frequency at the peak of the spectrum, except when using the full spectrum to compare to NANOGrav data in Section~\ref{sec:PTA}. We see that larger values of $\beta/H_*$ suppress the GW amplitude.  As described in the previous section, we require $\beta/H_*>3$ (from the percolation condition).
In the case of $\alpha\gg 1$   (equivalently, $\alpha/(1+\alpha) \approx 1$), 
we therefore get the upper limit on the GW amplitude from kination ending in a FOPT,
\beq
\label{eq:omegabound}
\Omega_{GW}^{peak}h^2 \lesssim 2\times 10^{-7}
\eeq
For $\alpha<1$, the upper bound of the GW amplitude is suppressed by $\alpha^2/(1+\alpha)^2\approx \alpha^2$.
Thus Eqn \eqref{eq:omegabound} provides an upper bound on the value of $\Omega_{\rm {GW}}h^2$ today that is applicable to all cases
for kination ending in a FOPT.

The parameter $\beta/H_*$ also determines the peak frequency of the GW spectrum at the time of production $f_{\rm {peak}}^* \simeq 0.2\beta$, which redshifted until today becomes \cite{huber2008gravitational,Freese:2022qrl} 
\beq
\label{eq:frequency}
f_{\rm {peak}}\simeq 1.54 \times 10^{-8} \left (g_{\rm {eff}}(T_*)\right )^{1/6} \left ({\beta\over H_*}\right ) \left (  {T_*\over {\rm {GeV}}}\right ) \, {\rm {Hz}}
\, .
\eeq
By  using Eq.~\eqref{eq:Temperature} and $g_{\rm {eff}}(T_*) =10$ the peak frequency of GW from kination ending in a FOPT becomes
\beq
f_{\rm {peak}}\simeq 2.3 \times 10^{-8} \left ({\beta\over H_*}\right ) \left (  {V_0^{1/4}\over {\rm {GeV}}}\right ) \left (
{\alpha+1\over \alpha}
\right )^{1/4}\, {\rm {Hz}} 
\, .
\eeq
We note that the dependence of $\Omega_{GW}$ on $\alpha$ is much stronger than the dependence of $f_{\rm {peak}}$ on $\alpha$.
The two main quantities that control the GW amplitude and frequency are  $\beta/H_*$ and $\alpha$.

\section{Analytic results in certain limits}
\label{sec:analytic}

In this section, we obtain analytical results for the  dynamics of the phase transition in some limits of the parameter space,
in order to gain intuition and provide simple approximate expressions for the amplitude and frequency of the produced GW's today. 
In Section~\ref{sec:phicandbeta} 
we compute the velocity of the kination field at the time of the transition $\dot \phi_c$, which will allow us to calculate $\beta/H_*$ (where $\beta^{-1}$ is the duration of the phase transition),
an essential quantity in determining the amplitude and peak frequency of the GWs; see Eqs.~\eqref{eq:omegaGW} and \eqref{eq:frequency}. 
In Section~\ref{sec:GWamplandfreq} we use these  values to derive simple analytic expressions for the frequency and amplitude of the produced GWs.
Finally, Section~\ref{sec:accuracy} shows the accuracy of our approximations.
Later, in Section~\ref{sec:parameterscan}, numerical work will extend these results beyond the parameter limits assumed in the current section.

\subsection{Dynamics of the Phase Transition}
\label{sec:phicandbeta}

For zero temperature tunneling in a general quartic-cubic-quadratic potential, including our $V(\chi)$, 
the four-dimensional Euclidean action for the bounce $S_4$, which controls the decay rate $\Gamma$ through Eq.~\eqref{eq:Gamma},  has been shown to be well approximated  by the fitting function \cite{Adams:1993zs} 
\beq
\label{eq:S4sum}
S_4={\pi^2\mu^6 \over 24\lambda^2 
\left (
\mu^2 - 2\lambda^2 m_{\chi,{\rm{eff}}}^2
\right )^3
}
\sum_{i=1}^3 A_i \left ({
\lambda m_{\chi,{\rm{eff}}}\over \mu}
\right )^{2i} \, .
\eeq
The term $\sum_{i=1}^3 A_i \left ({
x}
\right )^{2i}$ contains the constants $A_i = \{55.328, -173.104, 132.896  \}$ and peaks at $x\simeq 0.46$ with a peak value of around $5.2$ and vanishes (for the first time) at $x\simeq 0.75$. 
The denominator of $S_4$ vanishes at $\mu = \sqrt{2} \lambda m_{\chi,{\rm {eff}}}$, when the two minima of the potential become degenerate and thus small deviations from this condition would lead to thin-wall bubbles (which is generically not the case for our model). 
The value of the argument varies from $0<x\ll 1$ (thick wall regime) to $x=1/\sqrt{2}\simeq 0.7$ (degenerate vacua limit).

In order to compute the dynamics of the FOPT, we now make assumptions about the parameters which will allow us to make significant analytical progress.
Firstly, we consider the limit of small coefficient for the quartic term $\lambda \ll 1 $ in the potential for the $\chi$ field.
The width of the barrier is $\Delta\chi \simeq{m^2_{\chi,{\rm {eff}}}/ 2\mu} + {\cal O}(\lambda^2)$ and its height is $\Delta V \simeq {m^6_{\chi,{\rm {eff}}}} / {54 \mu ^2} +{\cal O}\left (\lambda^2\right )$. We see that to lowest order in $\lambda\ll 1$ the potential barrier is independent of $\lambda$. While taking a very small value of $\lambda$ can be considered fine-tuned, simply requiring perturbativity\footnote{The perturbativity constraint for the quartic tree  level term to dominate over the one loop correction in the parametrization of Eq.~\eqref{eq:lagrangian} is $\lambda<\sqrt{\pi/6}\simeq 0.7$, which is the value quoted in the text.} for  the quartic term in the Lagrangian of Eq.~\eqref{eq:lagrangian} leads to $\lambda \lesssim 0.7$. A coupling that is one order of magnitude below the perturbativity bound is enough to put our analysis in the $\lambda\ll 1 $ regime. We will  see that this limit allows for significant analytic progress, which will help us  build intuition about the dynamics of the phase transition, while at the same time providing us with ways to accelerate the (exact) general numerical calculations (including beyond this limit)  which  produce the GW results shown in Fig.~\ref{fig:GWscan1}.

We can now simplify the Euclidean bounce action $S_4$.
For small values of $\left (\lambda m_{\chi,{\rm{eff}}}/ \mu\right )$ we are justified to only keep the first term in the sum of Eq.~\eqref{eq:S4sum}. In the same limit, the denominator of the first term also simplifies to $\mu^2 - 2\lambda^2 m_{\chi,{\rm{eff}}}^2\simeq \mu^2$. This leads to the much simpler form of the action
\beq
\label{eq:S4smalllambda}
S_4 \simeq {\pi^2\over 24} A_1 {m_{\chi,{\rm{eff}}}^2\over \mu^2}
= {\pi^2 \over 24  }A_1 \left ({m_\chi^2\over \mu^2} + 2 {\dot\phi^2\over M^2\mu^2}
 \right )  \, ,
\eeq
where  the constant $A_1=55.328$.
In obtaining this equation, we have used $\lambda m_{\chi,{\rm {eff}}}/\mu\ll 1$, which is indeed the case in our limit $\lambda \ll 1$, together with two conditions
 $m_\chi/\mu \lesssim {\cal O}(1)$ and  $\dot\phi/ M\mu = {\cal O}(1)$.  We will show below (see middle panel of Fig. 3 and Eq.~\eqref{eq:phidotsimplest}) that both of these
conditions are satisfied at the time of the phase transition.\footnote{In fact $m_\chi/\mu$ cannot be much larger than unity, because this typically leads to $\beta/H_*<3$ and thus the bubbles of true vacuum do not percolate. }

For the remainder of this section,  we employ the limit $m_\chi/\mu\ll 1$.  In this limit, as we will show below in Eq.~\eqref{eq:phidotsimplest} and visually demonstrate in the left panel of Figure~\ref{fig:abmvsratio},  $m_{\chi,\rm{eff}}$ is dominated by the kinetic energy term $m_{\chi,\rm{eff}}^2 \sim \dot\phi^2/M^2$.  Then 
the action $S_4$ in Eq.\eqref{eq:S4smalllambda} 
is further simplified as
\beq
S_4 \simeq 
 {\pi^2 \over 12  }A_1  {\dot\phi^2\over M^2\mu^2} \, .
\label{eq:S4simplest}
\eeq
Working with this most simplified version, where $S_4\propto \dot\phi^2$, allows us to analytically compute $\dot\phi_c$, the velocity of the  rolling scalar field (which is driving kination) at the time of the transition.

The  parameter  $\beta$, which describes the inverse duration (rate) of the phase transition, becomes
\beq
\label{eq:betaeqn}
\beta  \simeq - \left .\dot S_4  \right|_{t=t_*} \simeq 3H_*\dot\phi \left . {\partial S_4\over \partial \dot \phi} \right|_{\dot\phi = \dot\phi_c}
\simeq 6 H_*  {\pi^2 \over 12  }A_1  {\dot\phi_c^2\over M^2\mu^2}
\eeq
Combining Eqs.~\eqref{eq:S4simplest} and \eqref{eq:betaeqn}, we find 
\begin{equation}
\label{eq:simple}
{\beta\over H_*} = 6   {\pi^2 \over 12  }A_1  {\dot\phi_c^2\over M^2\mu^2}= 6S_4 \, .
\end{equation} 

Plugging this simple relation into  
Eq. \eqref{eq:omegaGW}, we see that the GW amplitude from the FOPT is determined by 
 \beq
 \Omega_{GW}\propto (\beta/H_*)^{-2}\propto S_4^{-2} \, .
\eeq 
To obtain the value of $\Omega_{GW}$, we thus need to compute $S_4$,  
 or equivalently $\dot\phi_c^2$ through Eq.~\eqref{eq:S4simplest}.
For the remainder of this section we will work towards obtaining the simple expression for $\dot\phi_c^2$ given in Eq.~\eqref{eq:phidotsimplest}, 
the speed of the kination field at the time of the PT, 
as well
as $\beta/H_*$ (the inverse duration of the PT) given in Eq.~\eqref{eq:betavaluesimplest}, which controls the frequency and amplitude of
the GWs produced by bubble collisions during the FOPT.

Let us return to the time of the transition,   defined as $I(t_*) =1$. 
Going back to the integral of Eq.~\eqref{eq:Itintegral}
we note it is strongly dominated by times around $t_*$. We can thus approximate  $a(t') r_{\text{com}}(t, t') \simeq  (t - t') $ in the integral. Furthermore, expanding the bounce action in the exponent of Eq.~\eqref{eq:Gamma} around \( t = t_* \) and using Eq.~\eqref{eq:betadef}, we approximate
$
\Gamma(t') \simeq \Gamma(t_*) e^{\beta(t' - t_*)}
$. The integral $I(t)$ at the transition time becomes
\beq
I(t_*) \simeq {4\pi\over 3}\Gamma(t_*) 
\int_0^{t_*} dt' 
e^{\beta(t' - t_*)} (t_*-t')^3
\, .
\eeq
The exponential function is monotonically growing and thus the integral is dominated by the upper boundary value $t_*$. By changing variables $t'-t_*=-s/\beta$ and taking the limit $\beta \gg |t'-t_*|^{-1}$ the integral becomes
\beq
I(t_*) \sim {4\pi\over 3}\Gamma(t_*) {1\over\beta^4} 
\int_0^\infty ds \, e^{-s} s^3 = 
{8\pi \Gamma(t_*)\over\beta^4}
\eeq
where we used the known integral $\int_0^\infty ds \, e^{-s} s^3 = 6$. Since $I(t_*)=1$ at the time of the transition, we obtain
\beq
\Gamma(t_*) = {\beta^4\over 8\pi}
\label{eq:gammaeqbeta4} \, .
\eeq
Using Eq.~\eqref{eq:simple}, we have
\beq
\Gamma(t_*)  \simeq
{1\over 8\pi} 6^4 H^4 S_4^4 \, .
\label{eq:gammaeqbeta4approx}
\eeq
We use the same approximations as before; from Eq.~\eqref{eq:Gamma} and taking the limit $m_\chi \ll \mu$,
\beq
\Gamma(t_*) \simeq m_{\chi,{\rm{eff}}}^4 {S_4^2\over 4\pi^2} e^{-S_4}
\simeq \left (2{\dot\phi_c^2\over M^2} \right)^2 {S_4^2\over 4\pi^2} e^{-S_4}=
 \left ( {24\mu^2\over \pi^2 A_1} S_4\right)^2{S_4^2\over 4\pi^2} e^{-S_4} 
 = \mu^4 {144\over \pi^6A_1^2}S_4^4 e^{-S_4} \, .
 \label{eq:gammaS4approx}
\eeq
The second (approximate) equality follows because we used the assumption that $\dot\phi_c / M\mu ={\cal O}(1)$, meaning that the velocity term dominates over $m_\chi^2$ in the $\chi$ effective mass in the $m_\chi\ll \mu$ limit.  The third equality follows from Eq.~\eqref{eq:S4simplest} which allowed us to swap $\dot\phi_c^2$ for $S_4$.

\begin{figure}
\centering
\includegraphics[width=.7\textwidth]{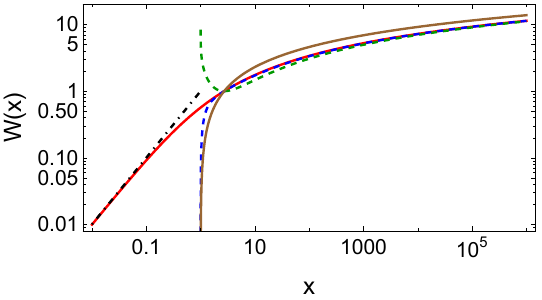}
\caption{The exact Lambert $W$ function is shown in red. The small-argument approximation $W(x)\simeq x$ is shown in black dot-dashed. The full large-argument approximation 
$W(x)\simeq \ln (x) \left(1-\frac{\ln (\ln (x))}{1+ \ln (x)}\right)$
 is shown in blue-dashed. Successively more drastic approximations are shown in green-dashed $W(x)\simeq \ln (x) \left(1-\frac{\ln (\ln (x))}{\ln (x)}\right)$ and then  the brown-solid $W(x)\simeq \ln(x)$.
Our parameters lie in the $x\gg 1$ regime and we will thus use the successively more drastic approximations given by the green-dashed and brown lines in Eqs.~\eqref{eq:phidotcAB} and \eqref{eq:phidotfinalsimplification} respectively.
}
\label{fig:lambertW}
\end{figure}

We now equate the two above expressions of $\Gamma(t_*)$ given in Eqs.~\eqref{eq:gammaS4approx} and \eqref{eq:gammaeqbeta4approx}. 
Intruigingly, working in the limit of $m_\chi\ll \mu$,  the $S_4^4$ term cancels, leaving only the $e^{-S_4}$ term.  This leads to a very simple expression for the action
\beq
S_4 = \ln\left(
{8\over 9\pi^5A_1^2}
\right ) - 4\ln(H/\mu)
\label{eq:S4}
\eeq
which can be equated with $S_4\propto \dot\phi_c^2$ in Eq.~\eqref{eq:S4simplest}.
The Hubble scale can also be written as a function of $\dot\phi$ and potential parameters
\beq
H^2 ={ {1\over 2}\dot\phi^2 + V_0\over 3 M_{\rm {Pl}}^2}
\simeq {1\over 3 M_{\rm {Pl}}^2}
\left (
{1\over 2}\dot\phi^2 + {27\over 256}{\mu^4\over \lambda^6}
\right )
\, .
\label{eq:hubble}
\eeq
Here, we we have used $V_0= 27\mu^4/256\lambda^6$, which arises by demanding that the potential vanishes at the true vacuum and taking the limit $\lambda\ll 1$. The derivation is presented in Appendix~\ref{appendix:potential} and specifically Eq.~\eqref{eq:V0simplest}.
We can now solve the combined set of Eqs.~\eqref{eq:S4}, \eqref{eq:S4simplest} and \eqref{eq:hubble} to derive the velocity $\dot\phi_c$ solely as a function of the potential parameters, leading to
\beq
\dot\phi_c^2 = 
 {24M^2\mu^2\over \pi^2 A_1}
W \left (
{M_{\rm{Pl}}^2\over M^2}
{e^{9 A_1\pi^2 \mu^2 \over 1024 M^2\lambda^6}
\over 
3\sqrt{2\pi} }
\right )
-{27\mu^4\over 128 \lambda^6} \, ,
\eeq
where $W(x)$ is the Lambert $W$ function (or product logarithm).
We can re-write this, using the combination $\dot\phi^2/ M^2\mu^2$, since that is what appears in the calculation of $S_4$ and therefore $\beta/H$
\beq
{\dot\phi_c^2\over M^2\mu^2} = 
 {24\over \pi^2 A_1}
W \left (
{M_{\rm{Pl}}^2\over M^2} {1\over 3 \sqrt{2\pi}}
e^{9 A_1\pi^2 \mu^2 \over 1024 M^2\lambda^6}
\right )
-{27\over 128 \lambda^6}{\mu^2\over M^2}
\, .
\eeq
 Let us make a small detour to describe the behavior of the product logarithm. As shown in Fig.~\ref{fig:lambertW}, $W(x)\simeq x$ for $x<1$ and 
$W(x)\simeq \ln(x) \left (
1-\frac{\ln \left (\ln (x)\right )}{\ln (x)+1}\right )$ for $x>1$. Depending on the size of $x\gg 1$, further approximations can be made, with different regimes of validity and accuracy, as shown in the figure.

The argument in the product logarithm in our case is generally large, since it is proportional to $M_{\rm {Pl}}^2/M^2\gg 1$, thus we can approximate the solution as
\beq
\label{eq:phidotcAB}
{\dot\phi_c^2\over M^2\mu^2} \simeq
 {24\over \pi^2 A_1}
\ln\left (A e^{B}\right )\left (
1-\frac{\ln \left (\ln (A e^{B})\right )}{\ln (A e^{B})+1}\right )
-{27\over 128 \lambda^6}{\mu^2\over M^2} \, ,
\eeq
where
\beq
A \equiv {1\over 3 \sqrt{2\pi}} {M^2_{\rm {Pl}}\over M^2}\, ,~ B\equiv
{9 A_1\pi^2 \mu^2 \over 1024 M^2\lambda^6} \, .
\eeq
The dominant term is 
\beq
{24\over \pi^2 A_1 }\ln(Ae^B)={24\over \pi^2 A_1 } \ln (A) + {24\over \pi^2 A_1 }B
=
{24\over \pi^2 A_1 } \ln\left (  {1\over 3 \sqrt{2\pi}} {M^2_{\rm {Pl}}\over M^2}\right ) + 
{27\over 128  } { \mu^2 \over  M^2\lambda^6} \, .
\eeq
We see that the last term ${27\over 128  } { \mu^2 \over  M^2\lambda^6}$ cancels exactly with the corresponding term 
$-{27\over 128  } { \mu^2 \over  M^2\lambda^6}$ in Eq.~\eqref{eq:phidotcAB}. Taking the most drastic approximation $W(x)\simeq \ln(x)$ for $x\gg 1$, in order to get a simple expression (we will examine the accuracy in Section~\ref{sec:accuracy}), we arrive at the very simple expression
\beq
\label{eq:phidotfinalsimplification}
{\dot\phi_c^2\over M^2\mu^2} \simeq
{24\over \pi^2 A_1} \ln\left (  {1\over 3 \sqrt{2\pi}} {M^2_{\rm {Pl}}\over M^2}\right ) \, .
\eeq
As a reminder, $A_1=55.328$. Inserting all numerical values, the above equation becomes
\beq
\label{eq:phidotsimplest}
{\dot\phi_c^2\over M^2\mu^2}\approx 0.088 \ln\left ({M_{\rm{Pl}} \over M }\right ) \, .
\eeq
Even for $M=0.1 M_{\rm {Pl}}$ we get ${\dot\phi_c^2/ M^2\mu^2} \approx 0.1 $, and this quantity grows for smaller values of $M$, becoming unity for $M\approx 10^{-10}M_{\rm {Pl}} \simeq 10^9 \, \rm{GeV} $. Therefore, the assumption that for $m_\chi/\mu\ll1$, the velocity term dominates in $m_{\chi, \rm{eff}}$ is fully justified. As a side-note, the percolation condition $\beta/H_*>3$ puts a very weak condition on the mass scale $M\lesssim 0.9 M_{\rm {Pl}}$, practically allowing us to choose $M$ arbitrarily high. 

Using the approximate expression for $\dot\phi^2_c$ given in Eq.~\eqref{eq:phidotsimplest}, the inverse duration $\beta$ of the PT, given by Eq.~\eqref{eq:betaeqn} becomes
\beq
\label{eq:betavaluesimplest}
{\beta\over H_*} \simeq 24 \ln\left ({M_{\rm{Pl}} \over M }\right ) \, .
\eeq
Given the weak (logarithmic) dependence on $M$ and taking $M\gtrsim 1$ GeV  (see Section~\ref{sec:experiments}), we get an estimate of $\beta/H_*\lesssim 1000$. As a reminder, this equation has been derived assuming $m_\chi/\mu\ll 1$.  As we will see in Section~\ref{sec:parameterscan}, when we take $m_\chi/\mu\gtrsim 2$ the value of $\beta/H$ decreases, making Eq.~\eqref{eq:betavaluesimplest} an upper limit. 
Quantitatively, as we increase $m_\chi/\mu$, the inverse duration approaches $\beta/H\to3$ from above.
Thus ${\beta/ H_*}$ is bounded from above and below within three orders of magnitude.
The duration of the transition $\Delta t\sim \beta^{-1}$ therefore lies in the interval
\begin{equation}
0.001 H_*^{-1}\lesssim \Delta t\lesssim 0.33 H_*^{-1} \, ,
\end{equation}
where $H_*^{-1}$ is  the Hubble time at the time of the transition.

In this subsection we have provided analytical results for the  quantities that describe the dynamics of the phase transition in the limits of $\lambda\ll 1$ and $m_\chi/\mu\ll 1$. Specifically we have computed $\dot\phi_c$, which is the speed of the kination field at the time of the PT, as well as $\beta/H_*$ (the inverse duration of the PT), which  controls the frequency and amplitude of the GWs produced by bubble collisions during the FOPT. The final simplest approximate expressions are given in Eq.~\eqref{eq:phidotsimplest} for $\dot\phi_c$ and Eq.~\eqref{eq:betavaluesimplest} for $\beta/H_*$. 

\subsection{Analytic Estimates for GW amplitude and frequency}
\label{sec:GWamplandfreq}

We are now ready to analytically compute the peak frequency and amplitude of the GW spectrum produced during the phase transition, in the limits of $\lambda\ll 1$ and $m_\chi/\mu\ll1$.
GWs produced during the FOPT are constrained  by the available vacuum energy that goes into accelerating the bubble walls (as opposed to the kinetic energy of the kination field that is not ``inherited" by the bubble walls).

In the limits of $\lambda\ll 1$ and $m_\chi/\mu\ll1$, let us examine the distribution of the energy budget between the vacuum and kinetic energies through the parameter $\alpha$ defined in Eq.~\eqref{eq:alpha}
\beq
\alpha \equiv {\rho_{\rm vac}\over \rho_{\rm kin}(t_*)}= {V_0\over {1\over 2}\dot\phi_c^2}
 = {2 V_0 / M^2\mu^2 \over\dot\phi_c^2/M^2\mu^2 }
 \simeq {
0.21 \lambda^{-6} (\mu^2/M^2)
\over 0.088 \ln(M_{\rm{Pl}}/M)
 }
\simeq 2.4 {1\over \lambda^6} {\mu^2\over M^2}{1\over \ln(M_{\rm{Pl}}/M)}
\, .
\label{eq:alphacalc}
\eeq
Here we used $V_0= 27\mu^4/256\lambda^6$, which is true for $\lambda\ll 1$, setting the potential at the true vacuum to zero (see Eq.~\eqref{eq:V0simplest}), and took $\dot\phi_c$ from Eq.~\eqref{eq:phidotsimplest}.
Using the consistency  condition for the positivity of the $\phi$ kinetic term in Eq.~\eqref{eq:lagrangian}, $\mu\lesssim \lambda^2 M$, we can provide an upper limit 
\beq
\alpha \lesssim 
{1\over \lambda^2} {1\over \ln(M_{\rm{Pl}}/M)}
\, ,
\eeq
where we dropped the ${\cal O}(1)$ prefactor.

The amplitude of the GWs depends on the value of $\alpha$, which can be split into two branches,  or equivalently, using the last equality in Eq.~\eqref{eq:alphacalc}, two different relations among the parameters in the potential\footnote{The  consistency condition $\mu\lesssim \lambda^2 M$ for the positivity of the $\phi$ kinetic term has also been used as the upper limit in the expression for the first branch}:
\begin{equation}
\label{eq:alphabranches}
\begin{split}
 \alpha \gtrsim 1 ~,&~ {\rm corresponding} \,\, {\rm to} \,\, 3\lambda^3\lesssim {\mu\over M}\lesssim \lambda^2
\\
\alpha 
\lesssim 1 ~,&~ {\rm corresponding} \,\, {\rm to} \,\,  3\lambda^3 \gtrsim {\mu\over M}
\end{split}
\, ,
\end{equation}
where we took $2.4/\ln(M_{\rm{Pl}}/M)={\cal O}(0.1)$ for the parameter range of interest. 
 We note that the two scales $\mu$ and $M$ have completely different origin, and thus it is not possible to predict the relation between them and consequently the value of $\alpha$; in principle a large hierarchy between them is not unnatural. 
For the marginal case $\alpha=1$ the condition becomes $ \mu/M\approx 3\lambda^3$.
Hence, given parameters in the potential, one can determine right away which branch is relevant. 
We see that, for a given size of the cubic term $\mu$ and quartic coupling $\lambda$, the kinetic coupling scale $M$ can have a significant effect on $\alpha$ and through that on $\Omega_{GW}$\footnote{$M$ can be viewed as an ad hoc kinetic coupling  parameter or as the curvature of a positively curved field-space, as discussed in Appendix~\ref{appendix:fieldspace}. The latter possibility creates an interesting connection between the field-space geometry and the amplitude of GW's, where large values of $M$ (weakly curved manifolds) tend to lead to smaller values of $\Omega_{GW}$ and vice versa.}.
In the first branch, where $\alpha \gtrsim 1$ at the time of the phase transition, one can take $\alpha/(1+\alpha)\simeq 1$ in Eqs.~\eqref{eq:omegaGW} and \eqref{eq:frequency}.
On the other hand, in the second branch, values of $\alpha \lesssim 1$ suppress  the resulting GW amplitude as ${\cal O} (\alpha^2)$.

Let us first focus on  the case $\alpha\gg 1$ at the time of the phase transition, so that $\alpha/(1+\alpha)\simeq 1$.
From Eq.~\eqref{eq:alphacalc} one can see that the $\lambda$ dependence generically leads to $\alpha\gg 1$ implying $\lambda\ll 1$, the limit taken in Section~\ref{sec:phicandbeta}\footnote{
The phase transition is triggered when the kination-induced mass (due to $\dot\phi$) and physical mass become comparable. For O(1) couplings, this happens roughly when the kinetic energy becomes comparable to the vacuum energy i.e. $\alpha \sim 1$. However, for even slightly different couplings, this value may shift. For quartic coupling $\lambda \ll 1$, the mass scale is  smaller than the vacuum energy scale. In such a case the phase transition happens slightly after kination, following  a short period of vacuum domination, corresponding to $\alpha \gg 1$ at the time of the phase transition.}.
The limit of $\alpha\gg 1$ corresponds to the case where the phase transition takes place  slightly after kination, during  a short period of vacuum domination.
We emphasize that  $\dot \phi$ nevertheless plays the dominant role in triggering the phase transition.  It controls the tunneling rate  of the  $\chi$ field (see Eq.~\eqref{eq:simple}) due to its presence in $m_{\chi,\rm eff}$ and  in
the shape of the potential 
as shown in Appendix~\ref{appendix:potential}.
Whereas, in inflation, the field at first rolls slowly and then speeds up as reheating takes place to ordinary matter and radiation, in the case of kination, the field at
first rolls very quickly and then slows down (see Appendix~\ref{appendix:kination}).
Since during kination the kinetic energy of the rolling field dominates over its potential, the Klein-Gordon equation becomes $\ddot\phi +3H\dot\phi\simeq 0$, leading to $\dot\phi\propto a(t)^{-3}$, thus the kination velocity decreases  as the universe expands.  As the kination field slows down, the barrier height and barrier width, which scale approximately with powers of $m_{\chi,{\rm {eff}}}$ are reduced, 
so that the tunneling probability increases.
Setting $g_{\rm {eff}}(T_*)=10$ and taking the limit $\alpha \gg 1$ in Eqs.~\eqref{eq:omegaGW} and \eqref{eq:frequency}, the GW at the peak of the spectrum is described by
\begin{eqnarray}
\label{eq:omegasimplest}
    \left . \Omega^{\rm{peak}}_{\rm {GW}}h^2 
    \right |_{\alpha\gg 1} &\simeq &1.8\times 10^{-6} \left ({H_*\over \beta}  \right )^2 \, ,
    \\
   \left . f_{\rm{peak}}    \right |_{\alpha\gg 1} &\simeq & 1.7\times 10^{-8} \left ({\beta\over H_*} \right ) \left (
{\rho_{\rm {vac}}^{1/4} \over {\rm {GeV}}}
    \right) \, {\rm {Hz}}
    \, ,
    \label{eq:freqsimplest}
\end{eqnarray}
where we used the estimate of the temperature right after the transition $T_*\simeq 0.75 \rho_{\rm {vac}}^{1/4}$ in the expression for the peak GW frequency given in Eq.~\eqref{eq:frequency}.

We can express the peak frequency as a function of the potential parameters.
Plugging Eq.~\eqref{eq:betavaluesimplest} into Eq.~\eqref{eq:freqsimplest}, the peak frequency in the limits $\lambda\ll 1$ and $m_\chi/\mu\ll 1$ becomes
\beq
\left . f_{\rm{peak}}\right |_{\alpha\gg 1}\approx 4 \times 10^{-7} \ln \left (
{M_{\rm {Pl}}\over M}
\right )
\left (
\rho_{\rm {vac}}^{1/4} \over {\rm {GeV}}
\right ) \, {\rm {Hz}}
= {2.3\times 10^{-7}\over \lambda^{3/2}} 
\ln \left (
{M_{\rm {Pl}}\over M}
\right )
{\mu\over {\rm {GeV}}} \, {\rm {Hz}} \, ,
\label{eq:freqsimplestvalue}
\eeq
where we used $\rho_{\rm vac} =V_0  \simeq {27 \mu ^4}/{256 \lambda ^6} $ from  Eq.~\eqref{eq:V0simplest}.

The corresponding GW amplitude is similarly easy to estimate using the same approximations. In particular we use Eq.~\eqref{eq:omegasimplest} for the case of $\alpha>1$ and multiply it by $\alpha^2$ for the case $\alpha<1$, leading, in the limits $\lambda\ll 1$ and $ m_\chi/\mu\ll 1$, to
\beq
\Omega^{\rm{peak}}_{\rm {GW}}h^2\approx  \begin{cases}
3.1\times 10^{-9}\left ( {1\over \ln(M_{\rm{Pl}} /M)} \right )^2 &,~{\mu\over M}> 3\lambda^3 \,\,\,\,\, (\alpha > 1)\\
17.9\times 10^{-9}
{1\over \lambda^{12}} {\mu^4\over M^4}\left ( {1\over \ln(M_{\rm{Pl}} /M)} \right )^4 &, ~{\mu\over M}< 3\lambda^3 \,\,\,\,\, (\alpha < 1)
\end{cases}
\label{eq:Omegabranches}
\eeq
where we used the expression for $\alpha$ given in Eq.~\eqref{eq:alphacalc}\footnote{Let us note that Eq.~\eqref{eq:alphacalc} shows a ``tug of war" between $\lambda^6$ and $\mu^2/M^2$. If the former is small, it pushes $\alpha$ to large value, whereas if the latter is small it pushes $\alpha$ to small values. Hence $\alpha < 1$ can result for $\lambda \ll 1$ if $\mu/M$ is small enough.
}.

Regarding the first branch ($\alpha > 1$):
Eq.~\eqref{eq:Omegabranches} shows that for $\alpha\gg 1$, the GW amplitude $\Omega_{GW}^{peak}$ does not depend on $\lambda$, $m_\chi$, or $\mu$,  and it has only a  logarithmic dependence on $M$. In reality there is a mild dependence on $\mu$ (even for $\alpha \gg 1$), which we neglected when  approximating  $\dot\phi_c$ through Eq.~\eqref{eq:phidotsimplest}.
That being said, Eq.~\eqref{eq:Omegabranches} gives a simple order of magnitude estimate for $\Omega_{\rm {GW}}$, which agrees with the full numerical result, as we will see in Section~\ref{sec:parameterscan}.  For $M$ in the range $1 - 10^{12}$ GeV, we find
\beq
\label{eq:typical}
\Omega^{\rm{peak}}_{\rm {GW}}h^2 \sim 10^{-12} {\rm GeV} \,\,\,\,\,\, ({\rm for} \,\,\,\, \alpha > 1, \, \lambda \ll 1, \, m_\chi/\mu\ll 1) \, .
\eeq
Eq.~\eqref{eq:typical} provides a rough ``floor" for the anticipated GW signal in the $\alpha > 1$ limit.  For $m_\chi/\mu \gtrsim 1$, the amplitude becomes larger.
Below in Section~\ref{sec:parameterscan}, when we run full numerical parameter scans for parameters beyond the limits imposed to obtain the above equation, we will find gravitational wave amplitudes larger than this number, all the way up to  the percolation bound of $\Omega^{\rm{peak}}_{\rm {GW}}h^2 \lesssim 10^{-7}$ given in Eq.~\eqref{eq:omegabound}.
In the case of $\alpha>1$,
as the quartic term in the potential gets larger, there is a shrinking of the parameter space that produces a substantial GW amplitude, due to the shrinking of the range $3\lambda^3<\mu/M<\lambda^2$. 
The lower bound on the ratio in this equation is required for vacuum energy to dominate over kinetic, and the upper bound is
required by positivity of the kinetic term in the Lagrangian.  
For large values of $\lambda$, i.e., close to
the perturbativity bound of $\lambda\simeq 0.7$, there is no value of the ratio $\mu/M$ that ensures positivity of the $\phi$ kinetic term and causes no suppression, since $3\lambda^3>\lambda^2$ for $\lambda\gtrsim 0.3$ (see Eq.~\eqref{eq:alphabranches}). Overall, increasing $\lambda$ and looking for $\alpha>1$ leads to a  very fine tuned regime, unless $\mu$ and $M$ arise in a correlated way from some UV theory.

Regarding the second branch ($\alpha < 1$):
The suppression of the peak GW amplitude $\Omega_{GW} \propto \alpha^2$ in the case of $\mu/M<3\lambda^3$ is evident from the lower branch of Eq.~\eqref{eq:Omegabranches}. 
Since $\alpha \propto \lambda^{-6} (\mu^2/M^2)$, for a given value of $\lambda$, as the ratio $\mu/M$ decreases, so does $\alpha$ and consequently $\Omega_{GW}$.
 In order to get a significant suppression, either $\lambda$ must be not very small, or $\mu/M$ must be much smaller than $3\lambda^3$. 
For $\alpha \ll 1$  the GW signal can be far below that of Eq. \eqref{eq:typical}.

For both branches,  if we examine the peak frequency in Eq.~\eqref{eq:freqsimplestvalue}, we see that the most important parameter dependence is on the scale $\mu$, since $M$ enters only logarithmically and $\lambda$ is taken to vary over only a few orders of magnitude.  
Of course $\mu$ is not a completely independent parameter, since the limit of positivity of the $\phi$ kinetic term  in Eq.~\eqref{eq:lagrangian} demands that $\mu < \lambda^2 M$. Let us introduce a dimensionless parameter that measures the distance of $\mu$ from the case where $\chi \approx \chi_{min,T}$ leads to a vanishingly small kinetic term for $\phi$: 
\beq
\label{eq:definer}
r \equiv { \mu \over \lambda^2 M } \, ,
\eeq
 where $0<r\le 1$, with $r\simeq 1$ being the value that saturates the  bound in Eq. \eqref{eq:boundonparamters} obtained by requiring 
$\chi_{min,T}^2 \leq M^2/2$ to always keep the kinetic term for $\phi$ positive.
Swapping $\mu$ for $r$ in Eq.~\eqref{eq:freqsimplestvalue}, the peak GW frequency (for $\alpha\gg 1$) is written as
\beq
\left . f_{\rm{peak}}\right |_{\alpha\gg 1}\simeq 2.3\times 10^{-7}\lambda^{1/2} 
\ln \left ( {M_{\rm {Pl}}\over M}\right ) \left ({ r M \over {\rm {GeV}} } \right ) \, {\rm {Hz}} \, .
\label{eq:freqwithr}
\eeq
In Section~\ref{sec:parameterscan}, we will  present in detail the parameter dependence of the GW frequency and amplitude and the degeneracies that arise.

\subsection{Accuracy of the approximations}
\label{sec:accuracy}

\begin{figure}
\centering
\includegraphics[width=.45\textwidth]{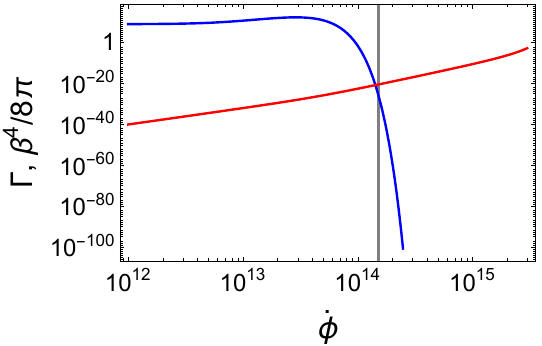}
\includegraphics[width=.45\textwidth]{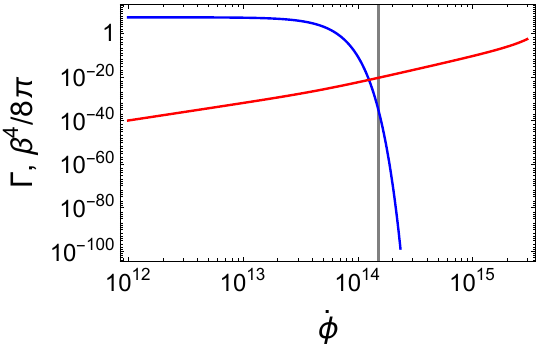}
\\
\includegraphics[width=.45\textwidth]{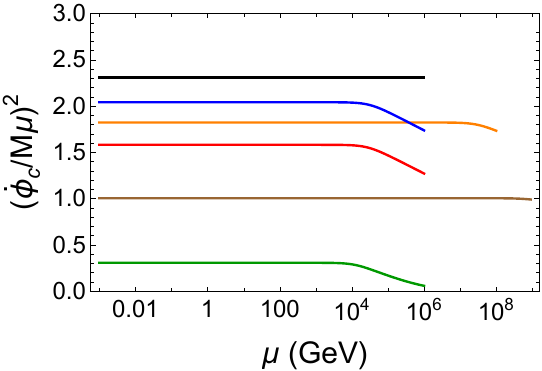}
\includegraphics[width=.45\textwidth]{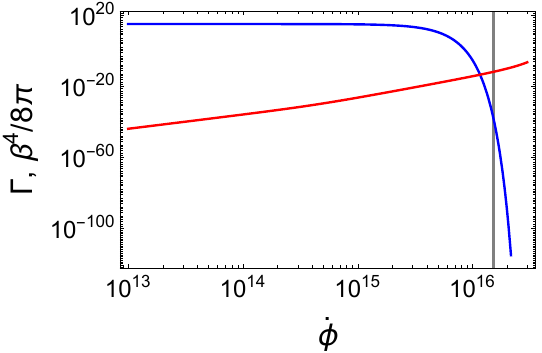}
\caption{Three panels, clockwise from the top left: The determination of $\dot\phi_c$ through the intersection of the numerically derived curves for $\Gamma$ (blue) and $\beta^4/8\pi$ (red) for
$(M,\mu, m_\chi, \lambda) = (10^{10}{\rm {MeV}}, 0.01 \lambda^2 M, 0.1\mu, 0.01)$, 
$(10^{10}{\rm {MeV}}, 0.01 \lambda^2 M, 2\mu, 0.01)$ and 
$(10^{10}{\rm {MeV}}, 0.01 \lambda^2 M, 0.1\mu, 0.1)$. The vertical gray lines show our analytic approximation for $\dot{\phi_c}$. One can see that our analytic approximation can be used
as a good starting point for
 finding the exact value of $\dot\phi_c$, which is necessary for numerically computing the peak frequency and amplitude of GW spectra (see Section~\ref{sec:parameterscan}). 
 \\
The bottom left panel shows the ratio $\dot\phi_c^2 / M^2\mu^2$ as a function of $\mu$. The   approximate value of Eq.~\eqref{eq:phidotsimplest}, which is derived  using our analytic approximations for  $\lambda, m_\chi/\mu\ll 1$, is shown in black. The numerical value in the same regime with $\lambda=0.01$ and $m_\chi=0.1\mu$ is shown in blue. The other colors refer to cases not compatible with the limits of our approximate solution in the black line. The red curve has $\lambda=0.01$ but $m_\chi=\mu$ and the green curve has $\lambda=0.01$ and $m_\chi=2\mu$. The orange and brown curves have $\lambda=0.1, 0.3$ respectively, while keeping $m_\chi=0.1\mu$. In all cases, $\dot\phi_c^2 / M^2\mu^2 = O(1)$. 
}
\label{fig:phidot}
\end{figure}

In this section, we compare the approximate analytic results for some of the basic parameters obtained in various limiting regimes in  Section~\eqref{sec:phicandbeta} with exact numerical computations of the same quantities.
To begin, we turn our attention to the numerical evaluation of the critical velocity $\dot\phi_c$, from which all other quantities derive, and compare our analytic approximation to the exact result. The first three panels of Figure~\ref{fig:phidot}, clockwise from left,  show the difference between the exact solution and approximate expression for $\dot\phi_c$.  Numerically  obtained curves showing  $\Gamma$ (blue) and $\beta^4/8\pi$ (red) intersect at $\dot\phi_c$ (see Eq.~\eqref{eq:gammaeqbeta4})
for a variety of parameter choices in the Lagrangian as seen in the caption. The vertical gray lines show our analytic approximation for $\dot{\phi_c}$. 
 On the practical front, the approximate value for $\dot\phi_c$ is close enough to the exact value, that using it as the starting point of a simple numerical root search algorithm makes it converge quickly and reliably to the true solution. It is therefore very useful, even in the cases where it cannot be used on its own to accurately compute the frequency and amplitude of GWs,  because it simplifies and accelerates the only computationally non-trivial part, which is finding $\dot\phi_c$. 
 Further, in the lower left panel, we see that $\dot\phi_c^2/M^2\mu^2={\cal O}(1)$ for all parameter choices, becoming slightly smaller for $m_\chi/\mu\gtrsim 1$; hence our assumption of this relation in Section~\ref{sec:phicandbeta} above is justified.

\begin{figure}
\centering
\includegraphics[width=.315\textwidth]{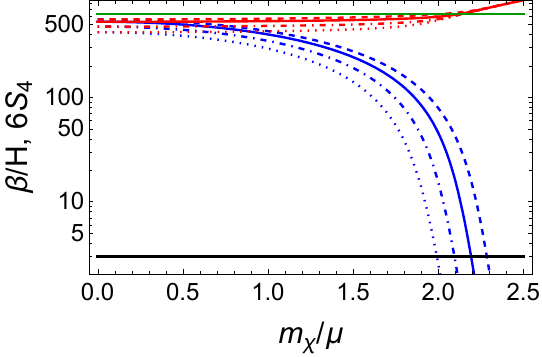}
\includegraphics[width=.35\textwidth]{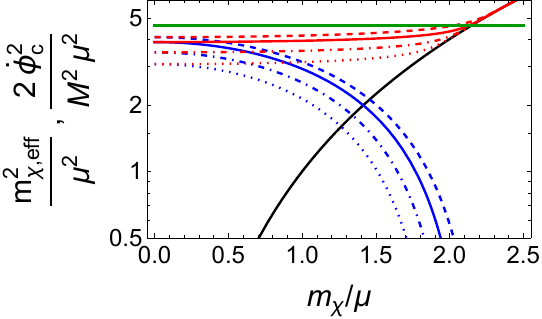}
\includegraphics[width=.315
\textwidth]{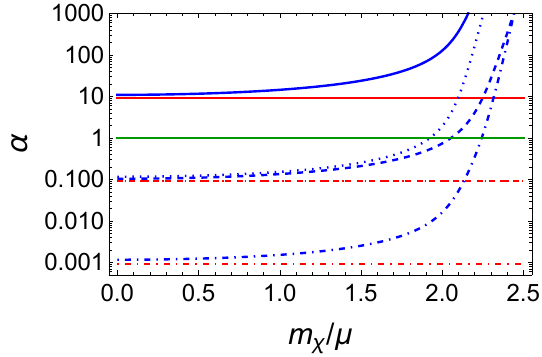}
\caption{
Various quantities of interest as a function of the ratio $m_\chi/\mu$. For all panels $M=10^{10}$ MeV, solid curves correspond to $\mu/M=0.1\lambda^2$, $\lambda=0.01$.
Dashed curves correspond to $\mu/M=0.01\lambda^2$, $\lambda=0.01$,
dotted curves correspond to $\mu/M=0.1\lambda^2$, $\lambda=0.1$ and dot-dashed curves correspond to $\mu/M=0.01\lambda^2$, $\lambda=0.1$. 
\\
{\it Left:} The  inverse duration of the phase transition  $\beta/H$ (blue) and $6S_4$ (red). In the limit  $m_\chi / \mu\ll 1$ we expect $\beta/H = 6S_4$, which is verified by the overlapping blue and red curves in that limit. The horizontal green curve corresponds to the analytic prediction for $\beta/H$ in the limits $\lambda \ll 1$ and $m_\chi/\mu\ll 1$ (the regime of the analytic calculations). The horizontal black line shows the lower bound  $\beta/H=3$ required for successful bubble percolation.
\\
{\it Middle:} The effective mass (red) and the rescaled critical velocity (blue). The black curve shows $m_\chi^2/\mu^2$ and the horizontal green line corresponds to the approximate expression of $2 \dot\phi_c^2/M^2\mu^2$ per Eq.~\eqref{eq:phidotsimplest}.
In the limit of $m_\chi / \mu\ll 1$, $m_{\chi,\rm{eff}}^2 \simeq 2\dot\phi^2_c/M^2$, as can be seen by the fact that the black curve is subdominant and the red and blue curves overlap in that limit.
\\
{\it Right:} The energy ratio  $\alpha$ (blue), along with the analytical approximation of Eq.~\eqref{eq:alphacalc} (red). The horizontal green line shows $\alpha=1$. 
}
\label{fig:abmvsratio}
\end{figure}

We now move to the analytical estimate and numerical computation of the two main parameters that control (through $\dot\phi_c^2$) the GW amplitude and frequency: the transition speed $\beta/H_*$ and the suppression factor $\alpha$, which are shown in Figure~\ref{fig:abmvsratio}. Some trends are easily visible from these plots and will be useful for understanding the results from the full numerical parameter scan  for the GW frequency and amplitude, which are presented in Section~\ref{sec:parameterscan}.  

The first observation from the left panel of Figure~\ref{fig:abmvsratio} is the excellent agreement of $\beta/H_*$ (shown in blue)  to our simple analytic estimate, which equates it to $6S_4$ (shown in red), based on Eqs.~\eqref{eq:phidotsimplest} and \eqref{eq:betavaluesimplest} for $m_\chi/\mu\ll 1$.
In the same limit, we expected  $m_{\chi,{\rm eff}}^2\simeq 2\dot\phi_c^2/M^2$, as verified by the middle panel of Figure~\ref{fig:abmvsratio} for $m_\chi/\mu\ll 1$.
In practical terms, $m_\chi/\mu<1$ is enough for our approximations to provide a good fit to the full numerical evaluation. For $m_\chi/\mu>1$ we see a steep decrease in the critical velocity $\dot\phi_c$ and correspondingly the rate of the transition $\beta/H_*$. From this monotonic behavior we can predict that the frequency will decrease for increasing $m_\chi/\mu$, with all other parameters being constant, and correspondingly the peak amplitude $\Omega_{GW}^{peak}$ will sharply increase in the same regime. In particular, for increasing $m_\chi/\mu$, the value  of $\dot\phi_c$ and $\beta/H_*$ becomes increasingly sensitive on the the exact value of the ratio. 
Of course, the decrease of $\beta/H_*$ with increasing $m_\chi/\mu$ cannot continue ad infinitum, since the condition for the bubbles to percolate  translate to $\beta/H_*>3$. This typically restricts the mass ratio to $m_\chi/\mu\lesssim 3$ and the corresponding GW peak amplitude to $\Omega_{GW}^{peak}h^2\lesssim 10^{-7}$.
For $m_\chi/\mu<1$ we expect all quantities to be largely insensitive to the exact value of the ratio.

In the third panel of Figure~\ref{fig:abmvsratio}, we examine the energy ratio  $\alpha$, again as a function of $m_\chi/\mu$. We see  that $\alpha$ is almost flat for $m_\chi/\mu<1$ and rises sharply for $m_\chi/\mu>1$, eventually becoming larger than unity. It is important to note that since $\alpha$ grows and $\beta/H_*$ falls with increasing $m_\chi/\mu$, it is possible that $\alpha$ does not exceed unity before $\beta/H_*=3$ and thus there are cases where, for certain choices of $\lambda, \mu, M$, the allowed maximum value of $m_\chi/\mu$, derived through $\beta/H_*=3$ leads to $\alpha<1$. As stated previously, in all cases   $\Omega_{GW}^{peak} h^2\lesssim10^{-7}$.

Concluding this section, we remark that the analytic approximations for all quantities are in excellent agreement with their numerical counterparts for $\lambda\ll 1$ and $m_\chi/\mu\ll 1$. That being said, the numerical results start deviating strongly from the analytic approximations for $m_\chi/\mu\gtrsim 1$. This is especially evident in the case of $\beta/H_*$, which drops rapidly with increasing $m_\chi/\mu$ and reaches the percolation threshold of $\beta/H_*=3$ for $m_\chi/\mu\sim 3$. However, the analytic approximation for $\dot\phi_c$ shows a remarkable robustness in terms of remaining a good approximation to its true numerical value for all values of $m_\chi/\mu$. Even though the exact value can differ from the analytic approximation by an ${\cal O}(1)$ factor, we can still use the simple expression of Eq.~\eqref{eq:phidotsimplest} for $\dot{\phi}$ as the starting point for finding the exact value when performing a parameter scan in the next section. 

\section{Parameter scan}
\label{sec:parameterscan}

\begin{figure}
\centering
\includegraphics[width=.75\textwidth]{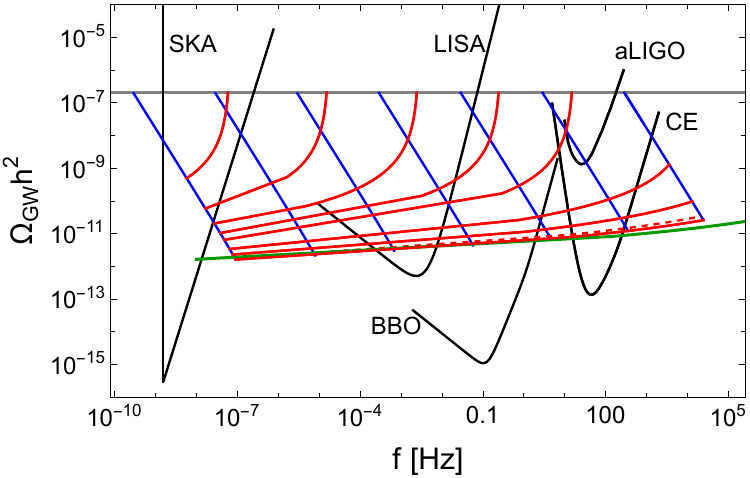}
\\
\includegraphics[width=.395\textwidth]{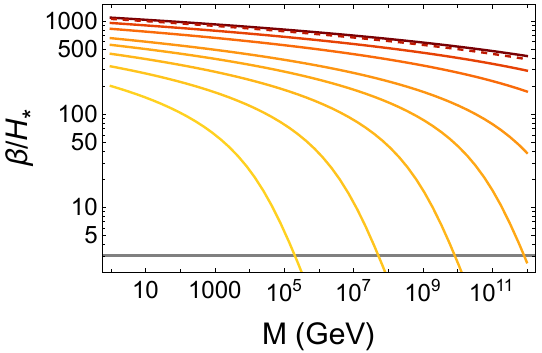}
\includegraphics[width=.08\textwidth]{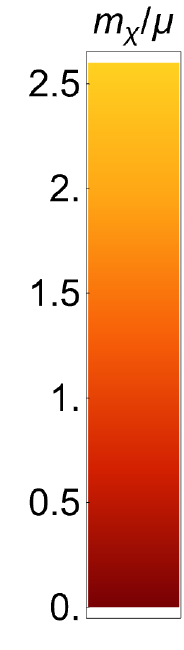}
\includegraphics[width=.395\textwidth]{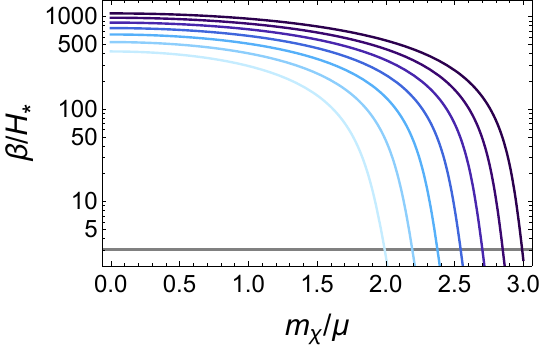}
\includegraphics[width=.1\textwidth]{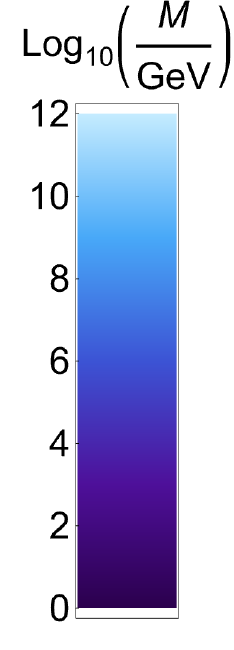}
\caption{GW production for the case $\alpha>1$ (where the FOPT triggered by the kination field takes place during a brief period of vacuum domination subsequent to the kination domination):
{\it Upper panel:} The peak amplitude and frequency of the stochastic GW spectrum in our model of a kinetically-induced FOPT, along with the sensitivity curves for various experiments shown as black curves.
The upper horizontal gray line corresponds to the absolute maximum value of $\Omega_{GW}^{peak}h^2 \simeq 2\times 10^{-7}$, which is reached at the threshold of bubble percolation $\beta/H_*=3$.
For all colored curves shown we fix $\lambda=0.01$ and $r=\mu/(\lambda^2 M)=0.1$, leading to $\alpha>1$ and thus no suppression of GW production.
The green curve corresponds to our analytic estimate of Eq.~\eqref{eq:omegafapprox}, which is a function solely of $M$, obtained in the limit $m_\chi / \mu \ll 1$ for fixed $\lambda, r$. The red curves correspond (bottom to top) to $m_\chi/\mu =0,0.5,1,1.4,1.8,2,2.2,2.4,2.6$ with the curve $m_\chi/\mu=0.5$ being dashed and all others being solid. For each red curve, $M$ grows from left to right in the range $1\le M/{\rm {GeV}}\le 10^{12}$ (although some reach the percolation threshold, and correspondingly the horizontal gray line, before reaching  $M= 10^{12}\,{\rm {GeV}}$).
The  blue curves correspond (left to right) to a fixed value of $M/{\rm {GeV}}= 1,10^2,10^4,10^6,10^8,10^{10},10^{12}$ with $m_\chi/\mu$ growing from $0$ at the bottom until each curve reaches the percolation threshold.
\\
{\it Lower panels:} The quantity $\beta/H_*$ (where $\beta^{-1}$ is the duration of the phase transition) as a function of $M$ (left) and $m_\chi/\mu$ (right) for the same parameters as in the upper panel. The horizontal gray line shows the percolation threshold at $\beta/H_* = 3$.
In the lower left panel, different colors  corresponds to different values of $m_\chi/\mu$ ranging from $m_\chi/\mu=0$ (top, dark red) to  $m_\chi/\mu=2.6$ (bottom, yellow). The dashed curve corresponds to $m_\chi/\mu=0.5$ and we see that it is almost indistinguishable from $m_\chi/\mu=0$. 
In the lower right panel, different shades of blue correspond to different values of $M\in [1,10^{12}]\, {\rm {GeV}}$, the same range used in the upper panel.
In the lower right panel, we see that $m_\chi/\mu$ could only exceed $\sim 3$ for values of the EFT cutoff $M < 1$GeV, too low to accommodate BBN after the phase transition.
}
\label{fig:GWscan1}
\end{figure}

\begin{figure}
\centering
\includegraphics[width=.75\textwidth]{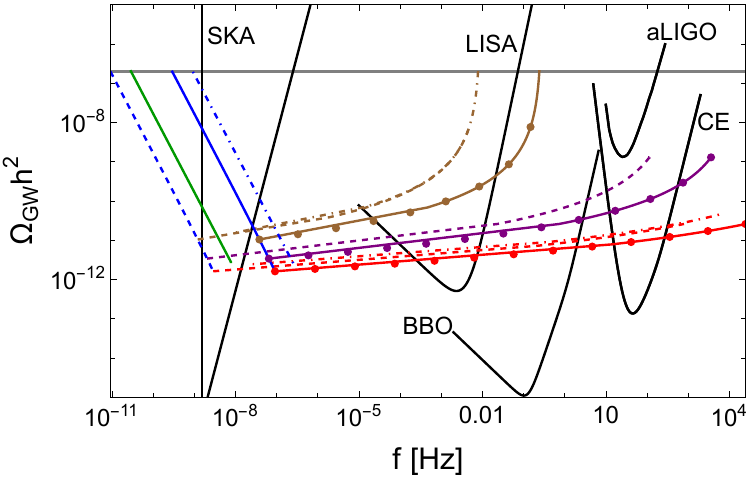}
\caption{Comparing the effect of $\lambda$ and $r=\mu/(\lambda^2M)$ on GW production for the case $\alpha>1$. We follow the scanning strategy of Figure~\ref{fig:GWscan1}, but use fewer sets of parameters for visual clarity. The red, purple and brown curves correspond to $m_\chi/\mu =0,1.4,2$ respectively. The solid curves have $\lambda=0.01$ and $r=0.1$ (the case shown in Figure~\ref{fig:GWscan1}), the dashed curves  have $\lambda=0.001$ and $r=0.01$  and finally the dot-dashed curves have $\lambda=0.001$ and $r=0.1$ (the dashed and dot-dashed curves are virtually identical). For all curves $M$ varies from $1$ GeV (at the left) to $10^{12}$ GeV (at the right). One can see that for lower $\lambda$ the curves shift to lower frequency. We also see that for $\lambda\ll 1$, changing $r$ does not change the results. Some curves reach the bubble percolation threshold for $M<10^{12}\, {\rm {GeV}}$, similarly to Figure~\ref{fig:GWscan1}, which corresponds to the horizontal gray line of $\Omega_{GW}h^2 \simeq 2 \times 10^{-7}$.  The dots along the red, purple and brown solid curves correspond to $M=1,10,100,.., 10^{12} \,{\rm {GeV}} $.
The blue solid, dashed and dot-dashed curves correspond to the same values of $\lambda$ and $r$ as their red, brown and purple counterparts. However, for the blue curves we choose $M=1$ GeV and vary $m_\chi/\mu$. The solid green line corresponds to the same parameters as the blue solid with the difference of $\lambda=10^{-4}$.
One can see that lower $\lambda$ shifts the curves to lower frequency. }
\label{fig:GWscandifflambda}
\end{figure}

\begin{figure}
\centering
\includegraphics[width=.75\textwidth]{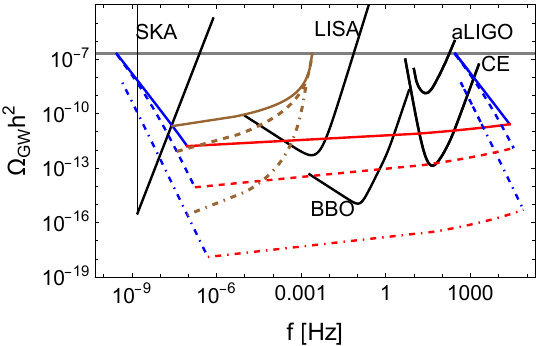}
\caption{ GW production for $\alpha<1$. 
All curves are produced with $\lambda=0.01$. The red (brown) curves correspond to $m_\chi/\mu=0$ ($m_\chi/\mu=2.2$) with $M$ varying from $M=1$~GeV (left) until  $M=10^{12}$~GeV (right); for the brown curves the percolation threshold is reached for $M<10^{12}$~GeV.
The blue curves correspond to a fixed value of $M/{\rm{GeV}} = 1,10^{12}$
with $m_\chi/\mu$ growing from 0 at the bottom until each curve reaches the percolation threshold.
Each curve, solid, dashed and dot-dashed, corresponds to a different value of $r=0.1,0.01, 0.001$ respectively. For comparative purposes with Figures~\ref{fig:GWscan1} and~\ref{fig:GWscandifflambda}, the value $r=0.1$ (solid curves) correspond to $\alpha >1$.  All all other values of $r$ put the system in the $\alpha<1$ regime, either for the whole curve or for part of it. 
}
\label{fig:GWscanlowalpha}
\end{figure}

In this section we present a full numerical parameter scan,  covering the entire frequency range of current and future GW experiments.
For reasons of visual clarity we overlay the predicted peak amplitude and frequency of the GW signal with the sensitivity curves\footnote{There are different prescriptions for the sensitivity curves of GW experiments~\cite{Moore:2014lga, Schmitz:2020syl, Thrane:2013oya, Smith:2019wny} that rely on assumptions, such as the required SNR and the integrated time, that can lead to minor differences between them. These are not important, as they only lead to minor changes in the observable parameter range.} of a few selected experiments, in particular  Square Kilometre Array
(SKA)~\cite{DewdneySKA, Moore:2014lga}, LISA~\cite{LISA:2017pwj, Smith:2019wny}, Big Bang Observer (BBO)~\cite{Phinney2004BBO}, Cosmic Explorer (CE)~\cite{LIGOScientific:2016wof, Evans:2021gyd} and Advanced LIGO (aLIGO)~\cite{LIGOScientific:2016jlg} in Figures~\ref{fig:GWscan1},~\ref{fig:GWscandifflambda} and~\ref{fig:GWscanlowalpha}.
Furthermore, we provide a systematic and intuitive way for understanding how changing each potential parameter leads to a change of the GW peak characteristics, thereby ``moving" the predictions in the frequency-amplitude plane.

Our Lagrangian in Eq.~\eqref{eq:lagrangian} has {\bf four parameters:  three mass-scales, $M$ (the limit of the EFT), $\mu$ (the cubic term), and $m_\chi$ (the bare mass of the tunneling field), and a dimensionless quartic coupling $\lambda$}. Our goal is to scan over this entire four-dimensional parameter space when computing the GW signal resulting from the Kination-Induced FOPT, and
to make comparisons with current and upcoming GW detectors.  

We initially study the regime where $\alpha>1$, which means that the vacuum energy dominates over the kinetic energy of the rolling kination field at the time of the FOPT and thus the produced GWs can reach their maximal amplitude (as a reminder, for $\alpha <1$, the GW amplitude is suppressed by $\alpha^2$). As a  starting point for understanding the full parameter scan, we remind the reader of the equations analytically derived above (see Eqs.~\eqref{eq:Omegabranches} and \eqref{eq:freqwithr}) that describe the GW production in the combined limits $\lambda\ll 1$ and $m_\chi/\mu\ll 1$:
\begin{align}
\begin{split}
\left . \Omega_{GW}^{\rm{peak}}\right |_{\alpha>1} &\simeq 3.1\times 10^{-9}\left ( {1\over \ln(M_{\rm{Pl}} /M)} \right )^2 \, ,
\\
\left . f_{\rm{peak}}\right |_{\alpha>1}&\simeq 2.3\times 10^{-7}\lambda^{1/2} 
\ln \left ( {M_{\rm {Pl}}\over M}\right ) \left ({ r M \over {\rm {GeV}} } \right ) \, {\rm {Hz}} \, .
\end{split}
\label{eq:omegafapprox}
\end{align}
Here the dimensionless quantity $r$ defined in Eq.~\eqref{eq:definer} 
 must be in the range $3\lambda < r < 1$.
 The upper bound $r< 1 $  ensures positivity of the $\phi$ kinetic term and the lower bound arises in the case $\alpha > 1$
as discussed in Eq.~\eqref{eq:alphabranches} 
  (which implies $\mu/M > 3\lambda^3$).

In our numerical studies of the parameter space with $\alpha > 1$,
for concreteness we fix $\lambda=0.01$ (which falls in the $\lambda\ll 1$ regime) and $r=0.1$. 
Since perturbativity requires $\lambda < 0.7$ and the results don't change much as long as $\lambda \ll 1$, this specific choice of $\lambda$ is not restrictive.
With these choices, we have $\mu/M = 10 \lambda^2$ and through Eq.~\eqref{eq:alphabranches} the energy ratio condition $\alpha>1$ is satisfied. By fixing $r$ and $\lambda$ we have fixed the ratio $\mu/M$, and we are thus left with only two free parameters in the problem: $m_\chi$ (more conveniently $m_\chi/\mu$) and either $\mu$ or $M$. 

Figure~\ref{fig:GWscan1} shows the results of our parameter scan in the remaining two variables $m_\chi/\mu$ and $M$.
The upper panel is the amplitude-frequency plane for the peak of the GW signal, in addition to showing the sensitivity curves for various GW experiments as marked.
The green curve corresponds to our analytic estimate in Eq.~\eqref{eq:omegafapprox}, which is a one-parameter line (depending only on $M$) obtained in the limit $m_\chi/\mu \ll 1$. 
The horizontal gray line corresponds to the maximum allowed value of $\Omega_{GW}^{peak}h^2 \simeq 2\times 10^{-7}$ set by the threshold value for successful bubble percolation, $\beta/H_*=3$.
We perform a scan by choosing discrete values for $m_\chi/\mu$ and scanning over $M$, obtaining the family of red curves. The red curves correspond (bottom to top) to $m_\chi/\mu =0,0.5,1,1.4,1.8,2,2.2,2.4,2.6$  (the curve for $m_\chi/\mu=0.5$ is dashed red and all others are solid red). In each of the red curves, $M$ grows from left to right in the range 1 - $10^{12}$GeV.  
We see that for $m_\chi/\mu\lesssim 0.5$ the analytic approximation of Eq.~\eqref{eq:omegafapprox} is in excellent agreement with the numerical data, even capturing the frequency dependence of the peak GW amplitude over many orders of magnitude. As we choose larger values of $m_\chi/\mu$ we see two effects. Firstly, the overall GW amplitude increases. More importantly, the frequency dependence of the GW amplitude becomes much sharper, to the point where increasing the frequency (through increasing $M$) leads to  sharply hitting the percolation bound. 
In fact, as we will show in the discussion of the lower right panel, the value of $m_\chi/\mu$ cannot exceed the values up to $\sim  3$ plotted here without violating the percolation bound.
In the upper panel, we also show the results of a second parameter scan, where we assign discrete values to $M$ and scan over $m_\chi/\mu$, obtaining the family of blue curves.
  Each of the blue curves corresponds (left to right) to a fixed value of $M/{\rm {GeV}}= 1,10^2,10^4,10^6,10^8,10^{10},10^{12}$
with $m_\chi/\mu$ growing from $0$ at the bottom until each curve reaches the percolation threshold.

The lower panels plot  $\beta/H_*$ as a function of $M$ (left) and of $m_\chi/\mu$ (right) for the same parameters as in the upper panel. The horizontal gray line shows the minimum allowed
value of  $\beta/H_* = 3$ for successful percolation at the end of the phase transition.
In the lower left panel, the top (bottom)  line corresponds to $m_\chi/\mu=0$ ($m_\chi/\mu=2.6$).
In the lower right panel, the top (bottom) blue line corresponds to $M=1\, {\rm {GeV}}$ ($M=10^{12}\, {\rm {GeV}}$).

From the lower right panel, one can see that there is an upper bound of
\beq
{m_\chi \over \mu} \lesssim 3 \, ,
\eeq
the maximum value allowed (without violating the percolation bound) for the case where the EFT cutoff is at $M= 1$~GeV.  To get to higher values of $m_\chi/\mu$ would require  lower values of $M$; however $M$ much lower than $\sim$GeV would imply that the Universe at the end of the FOPT would reheat to a temperature too low to accommodate standard Big Bang Nucleosynthesis (BBN).  Thus  $m_\chi/\mu \sim 3 $ arises as the maximum value  in the upper panel and lower left panel.

Overall, Figure~\ref{fig:GWscan1} 
shows that, in the case of $\alpha > 1$ the GW amplitude lies in the range 
$ 10^{-12}\lesssim \Omega_{GW}h^2 \lesssim 10^{-7}$. The lower value ($10^{-12}$) is obtained for  the parameter space in which $m_\chi/\mu < 1$. 
On the other hand, to get $\Omega_{GW}\gg 10^{-12}$ requires $m_\chi/\mu\sim 2$ (close to the upper bound from BBN); thus in developing some UV theory for this case, one would have to  account for this extra ``symmetry" between the two mass-scales.\footnote{{\label{fn:paramlikelihood}
It is worth taking a look into the  likelihood of different parameter values (regions of parameter space).
Specifically, we can ask, what values of $m_\chi/\mu$ are reasonable, or even likely?
 Requiring a true vacuum at $\chi\ne 0$ forces the constraint $m_\chi/\mu <1/(\sqrt{2}\lambda)$. With $\lambda\ll 1$ this still leaves the possibility for a large hierarchy between them. 
We note that, if we set the mass term to zero,
radiative corrections due to the cubic term (diagrams with one loop and two three-point vertices) will generate a mass term with $\delta m^2 \sim \mu^2/16\pi^2$, unless protected by some symmetry. 
In order to determine the likelihood that the mass ratio $m_\chi/\mu$ lies within certain values, we would have to define a prior for $m_\chi$ and $\mu$. If we took the probability density function of their magnitude to be uniformly distributed, then $m_\chi/\mu<1$ will cover an ${\cal O}(1)$ fraction (equal or larger than $\sim 50\%$) of the allowed parameter space. On the other hand, if we take these quantities to be log-uniformly distributed, the outcome of $m_\chi/\mu<1$, becomes much more likely (even exponentially so). Since we do not attempt to provide a full UV realization of our model, we will explore the entire parameter space, in order to describe the possible observational signatures, without attempting to quantify how likely they are in a given realization of our model.
}}

In order to support the general applicability of  Figure~\ref{fig:GWscan1} as providing a robust  picture for this model, we will show how the results change if we change $\lambda$ and $r=\mu/(\lambda^2M)$, while making sure that we remain in the $\alpha>1$ regime. Figure~\ref{fig:GWscandifflambda} 
shows sets of curves on the GW frequency-amplitude plane (red, purple and brown), each having a fixed value of $\lambda$, $r$ and $m_\chi/\mu$ and varying $M$. We start at $M=1$ GeV on the left and as $M$ increases we transverse each curve towards the right. The maximal value we considered is $M=10^{12}$ GeV, although some curves reach the bubble percolation threshold (and correspondingly the horizontal gray curve $\Omega_{GW} \simeq 2 \times 10^{-7}$) before that.
We choose three values of $m_\chi/\mu = 0,1.4, 2$ shown respectively in red, purple and brown. The solid curves correspond to $\lambda=0.01$ and $r=0.1$,  The dot-dashed curves correspond to $\lambda=0.001$ and $r=0.1$ and finally the dashed curves correspond to $\lambda=0.001$ and $r=0.01$. All values shown fall in the $\alpha>1$ regime. We see that changing $\lambda$ has a larger effect for larger values of the mass ratio $m_\chi/\mu$, but the qualitative behavior is unchanged. Lower values of $\lambda$ shift the frequency to the left, as expected from Eq.~\eqref{eq:omegafapprox}.\footnote{{Let us briefly note, that  in the opposite regime of $\alpha<1$, the peak frequency 
also shifts to a lower value for smaller values of $\lambda$ for constant $M$ and $r$. 
In particular, the peak frequency given in Eq.~\eqref{eq:freqwithr} is  multiplied by approximately $\alpha^{-1/4}$ for $\alpha<1$. As we will see later, in Eq.~\eqref{eq:alphavsr}, $\alpha\propto r^2/\lambda^2$, meaning that for $\alpha<1$ the peak frequency becomes $f_{\rm {peak}}|_{\alpha<1}\sim 
10^{-7} (\lambda/\sqrt{r}) \ln(M_{\rm {Pl}}/M) (M/{\rm {GeV}})\,{\rm{Hz}}$. Thus, with $M$ and $r$ being constant, reducing $\lambda$ reduces the peak frequency regardless of $\alpha \lessgtr1$.
}} Furthermore, we see that for $\lambda\ll 1$, changing $r$ does not change the results; the dashed and dot-dashed curves are visually indistinguishable.  
As a further exploration of the (in)dependence of our results on the exact value of $\lambda\ll 1$, we direct the readers' attention to the blue and green curves of Figure~\ref{fig:GWscandifflambda}.
Along each of these curves, we vary the ratio $m_\chi/\mu$, while keeping $M$, $\mu$ and $\lambda$ fixed.
We see that the frequency-amplitude curve for varying  $m_\chi/\mu$ are shifted along the frequency axis if we change $\lambda$ or $r$, in a way exactly following Eq.~\eqref{eq:omegafapprox}. We must note that Eq.~\eqref{eq:omegafapprox} was derived only for $m_\chi/\mu\ll 1$ and thus the fact that the whole curve is shifted according to it could not have been anticipated. Furthermore, the GW amplitude $\Omega_{GW}$ grows faster  as a function of $m_\chi/\mu$ along the green curve (smaller $\lambda$) compared to the blue one, similarly to  the brown and purple curves.

We are now ready to proceed to the final part of our parameter scan and study the regime of $\alpha<1$ and its similarities and differences to the regime $\alpha>1$ shown in Figures~\ref{fig:GWscan1} and \ref{fig:GWscandifflambda}.
Since $\Omega_{GW}^{peak} \propto \alpha^2$, as $\alpha$ decreases, so does the GW amplitude.
Our results are shown in Figure~\ref{fig:GWscanlowalpha}. The solid blue, solid red and solid brown curves are taken from Figure~\ref{fig:GWscan1} (the brown curve was depicted in red in Figure~\ref{fig:GWscan1}). 
For the solid blue curves we fix $r=0.1$,  $\lambda=0.001$ and $M=1, 10^{12}$ GeV (left and right curves respectively), while letting $m_\chi/\mu$ vary. 
The blue dashed and dot-dashed curves correspond to decreasing  the value of $r=\mu/(\lambda^2M)$, thereby pushing $\mu$ and $M$ further apart, leading to $\alpha<1$ and thus a suppression of the GW amplitude. As noted previously, Since $\alpha \propto \lambda^{-6} (\mu^2/M^2)$, for a given value of $\lambda$, as the ratio $\mu/M$ decreases, so does $\alpha$ and consequently $\Omega_{GW}$.
As $m_\chi / \mu$ increases, each blue dashed curve eventually reaches $\Omega_{GW}^{peak}h^2\simeq {2\times}10^{-7}$, even though at its starting point (for the same value of $M$) it is about $100$ times smaller than its solid blue counterpart. 
As we have seen in Figure~\ref{fig:abmvsratio}, as one increases $m_\chi/\mu$ the  parameter $\alpha$ increases sharply. If it reaches $\alpha>1$ before $\beta/H_*=3$, the maximum value of $\Omega_{GW}^{peak}H^2\simeq {2\times}10^{-7}$ can be reached, as is the case for the blue dashed curve (where $r=0.01$). For the blue dot-dashed curves, where $r=0.001$, the initial suppression is so large, that $\beta/H_*=3$ before $\alpha>1$, therefore the maximum $\Omega_{GW}$ is lower than the upper limit of ${\cal O}(10^{-7})$.

 The  behavior exhibited by the blue dashed and dot-dashed curves is also seen for the  corresponding red and brown curves  curves. For the solid ones, we fix $r=0.1$, $\lambda=0.01$ and $m_\chi/\mu=0,2.2$ (red and brown respectively), while letting $M$ vary. The dashed and dot-dashed curves correspond to changing $r=0.01,0.001$ respectively, again leading to $\alpha<1$.
The red dashed and dot-dashed curves are always suppressed with respect to their solid counterpart in the frequency range of interest, because for small values of $m_\chi/\mu$, changing $M$ does not lead to significant change in the dynamics.
 However, the situation is qualitatively and quantitatively different for the brown curves, which correspond to $m_\chi/\mu=2.2>1$. There, the brown dashed curve merges with its solid brown counterpart for large values of $M$, where $\alpha>1$ for both of them. The same is not true for the brown dot-dashed curve, which reaches the percolation limit at $\beta/H_*=3$ while still $\alpha<1$ and thus exhibits an ${\cal O}(\alpha^2)$ suppression of the produced GW signal, leading to $\Omega_{GW}^{peak}h^2\lesssim 10^{-8}$.

In summary, our parameter scan shows that a kinetically induced first-order phase transition 
can populate the observable GW frequency–amplitude plane over many orders of magnitude. In the case $\alpha>1$,  
we find typical signals around $\Omega_{\rm GW}^{\rm peak} h^2 \sim 10^{-12}$ for $m_\chi/\mu \lesssim 1$ 
and enhancements up to $\Omega_{\rm GW}^{\rm peak} h^2 \sim 10^{-7}$ as the system approaches the 
percolation bound $\beta/H_* = 3$, which occurs for $m_\chi/\mu \gtrsim 2$. The dependence on 
$\lambda$ and $r$ remains mild so long as $\lambda \ll 1$ and $\alpha > 1$. In the 
opposite regime, $\alpha < 1$, the GW signal is suppressed  $\sim\alpha^2$ as expected from the reduced potential 
energy available to drive the bubbles before their collision. Taken together, 
these results delineate the regions that can be accessed by current and future GW observatories, 
and they highlight the generic ability of this class of models to produce detectable spectra 
across a wide range of experimental sensitivities.

\section{Testing the Model with Current and Future Gravitational Wave Detectors}
\label{sec:experiments}

In this section, we will focus on the major current and future GW  experiment types one by one and assess their ability to probe certain parts of the parameter space of our model. 
We  split our analysis into different parts. In Section~\ref{sec:LISAetal} we will show the parameter combinations of the Kination-Induced Big Bang model that are detectable by  GW Interferometers: LISA, advanced LIGO,   CE/ET and BBO/DECIGO\footnote{CE and ET have similar projected sensitivity curves. We will use the CE sensitivity curves as a proxy for both. Similarly BBO and DECIGO target the same frequency band with competing sensitivities, depending on the design details. We will only use BBO as a proxy for both.}.
 Then, in Section~\ref{sec:PTA} we turn to existing Pulsar Timing Array (PTA) experiments and show the parameter combinations that can explain the data. In Section~\ref{sec:experimentsall}, we provide an overview of the whole parameter space that is accessible by current and future experiments, using a combination of  cosmological and microphysical quantities.  Finally, in Section~\ref{sec:degeneracy} we demonstrate the parameter degeneracy arising from a  GW detection, given the large parameter space of our model.
The numerical scanning procedure is similar to that of Section~\ref{sec:parameterscan}, albeit with a much denser (finer) grid in parameter-space, in order to capture the region of parameter space of the model that falls into  the basin of observability of each different experiment. Furthermore, we do not present the results in the GW amplitude-frequency plane, but we instead extract and show the  combinations of model parameters that can be accessed with the sensitivity reach of each of the experiments.

\subsection{Detectability by Interferometers: LISA, Advanced LIGO, CE and BBO}
\label{sec:LISAetal}

For each of the GW wave detectors that we study, we ask the following question: What are the parameter combinations that lead to GW spectra with the peak being inside the ``basin" of the various experiments,\footnote{Let us note that a prediction of a GW signal with a peak above the sensitivity curve of a given experiment provides a good hint for the possibility of detection, but not a guarantee. For example, in the case of LISA, 
 different GW sources are expected to be  densely packed both in time as well as in frequency. Therefore, they will need to be modeled  simultaneously in a ``global fit", which poses challenges for SGWB detection and is a topic of intense research~\cite{Rosati:2024lcs,Hindmarsh:2024ttn, Caprini:2024hue, LISACosmologyWorkingGroup:2022jok, Smith:2019wny}.
} as shown for example in Figure~\ref{fig:GWscan1}.

In our parameter scan, we have scanned over three mass scales $M,m_\chi,\mu$ and the coupling constant $\lambda$ in the Lagrangian.
Given the size of the four-dimensional parameter space, we again need to develop a strategy for scanning, which elucidates the main dependency of the observables on the parameters. Having shown that the self-coupling $\lambda$  affects the results in a controlled way (mostly by shifting the frequency as shown in Figure~\ref{fig:GWscandifflambda}), we choose three values of $\lambda=10^{-3},10^{-2},10^{-1}$ and present the results separately for each of them. We should note that we do not push $\lambda$ to even lower values, since that could be considered a very small, perhaps fine-tuned, self-coupling (note that the quartic term in the potential of Eq.~\eqref{eq:lagrangian} is ``$\lambda^2\chi^4$") and also our results can be extrapolated in this regime by shifting the GW frequency. 

The three remaining mass-scales are treated in the following way. Instead of the tunneling field mass $m_\chi$, we use the ratio $m_\chi/\mu$ and choose a number of discrete values, similarly to the red curves of Figure~\ref{fig:GWscan1}. We are then left with $M$ and $\mu$. We scan over $M$ and instead of $\mu$, we scan over $r=\mu/(M\lambda^2)$, which is the ``distance" from the limit of validity of the EFT of Eq.~\eqref{eq:lagrangian}. 
Using the definition of $r$, we can eliminate $\mu$ from  Eq.~\eqref{eq:alphacalc}, which holds for  $m_\chi/\mu\ll 1$, leading to
\beq
\alpha = {2.4\over\ln(M/M_{\rm {Pl}})} {r^2\over \lambda^2} \,~ , ~ ( {\rm {for}} ~ m_\chi\ll \mu)
\, .
\label{eq:alphavsr}
\eeq
The parameter $r$ is now linked to the ratio of energy densities $\alpha$  through $\alpha \sim 0.1 r^2 /\lambda^2$ for the relevant range of $M$.

We will present our results as 
contours of parameters in the $M-r$ plane, inside which the parameters lead to (in principle) detectable GWs with the four detectors considered in this subsection. This can be seen as a ``translation" of the detector sensitivity curve from the $f_{GW}-\Omega_{GW}$ plane onto the $M-r$ plane for different values of $\lambda$ and $m_\chi/\mu$. 
Further, we
also present results in the $M-\alpha$ and $M-H_*$ planes.

Of particular interest in a cosmological setting is our presentation of results in the
$M-H_*$  plane, which is a combination of the model parameter M and the cosmological
parameter $H_*$, revealing concurrently the energy-scale of the kination sector (through $M$)
and the time of the FOPT (through $H_*$). Overall the combination of $H_*$ and $M$ allows us to quickly connect a cosmologically important quantity --the time when the radiation dominated history of the universe started-- with the  microphysics of our model through the coupling strength between the kination and tunneling fields (see Section~\ref{sec:experimentsall}). In particular, the Hubble scale at the time of the transition $H_*$ can be related to the mass-scale of the derivative coupling $M$ in a very simple way.
In the  case 
of $\alpha>1$, so that the Hubble scale is dominated by the false vacuum potential energy, $H^2_*\simeq V_0/3M_{\rm {Pl}}^2$.
Further restricting to the case $r=1$ and
using Eq.~\eqref{eq:V0simplest} for $V_0$, 
 we arrive at
$H_* = 0.5 \lambda M^2/M_{\rm {Pl}}$ (for $\alpha>1$ and $r=1$).
More generally, for smaller values of $\alpha$ or $r$, one instead finds the upper bound 
\beq
\label{eq:LISAHubble}
H_*  \lesssim 0.5  \lambda {M^2\over M_{\rm {Pl}}} 
\eeq
(n.b. for $r\le 1$, Eq.~\eqref{eq:definer} leads to $\mu \le \lambda^2 M$ in Eq.~\eqref{eq:V0simplest} for $V_0$). 
Since the phase transition ends the kination era and reheats the universe into a standard radiation--dominated state, the Hubble scale at the transition, \(H_*\),  fixes the initial temperature and therefore the redshift at which the usual hot Big Bang history begins. In radiation domination one has
$
H(z) = H_0 \sqrt{\Omega_{r,0}}\,(1+z)^2 
$,
so that \(1+z_* \propto H_*^{1/2}\). Thus, specifying \(H_*\) determines the epoch of the transition, the temperature through
 $T_* \sim \sqrt{H_* M_{\rm {Pl}}}$ 
 and the redshift of the universe at the onset of radiation domination
and the redshift through $1+z_* \sim 10^{20} \sqrt{H_*/{\rm {MeV}}}$. 
  This relation allows us to map the microscopic parameters of the model onto a definite cosmic time or redshift. We should also note that $H_*$  largely determines the peak frequency of the GWs through $f_{\rm {peak}}\sim 2.3\times 10^{-8} (\beta/H_*) (T_*/{\rm{GeV}}) \, {\rm {Hz}}\sim 34.5 (\beta/H_*) \sqrt{H_* / {\rm {GeV}}} \, {\rm {Hz}}$ (see Eq.~\eqref{eq:frequency}). Since across the entire parameter space we have found the speed of the transition to lie in the interval $3\le \beta/H_*\lesssim 1000$  we get 
\beq
100  \sqrt{H_* / {\rm {GeV}}} \, {\rm {Hz}} \lesssim f_{\rm {peak}}\lesssim 3.5 \times 10^4 \sqrt{H_* / {\rm {GeV}}} \, {\rm {Hz}} \, .
\eeq 

Overall, the Hubble scale at the time of the phase transition, $H_*$, is closely related to
 both the mass-scale $M$ (controlling the derivative interaction strength) and the GW peak frequency $f_{\rm {peak}}$. 
The upper bound in Eq.~\eqref{eq:LISAHubble} is set by $H_* \sim \lambda M^2 / M_{\rm Pl}$. The relation
$f_{\rm peak} \propto \sqrt{H_*}$ implies that each GW experiment (with different frequency range) will probe a characteristic range of cosmic
times through $H_*$.
Further, based on the relation between $H_*$ and $M$ shown in Eq.~\eqref{eq:LISAHubble}, each GW experiment will also constrain different ranges of the  strength of the kination--tunneling coupling encoded in $M$.

\subsubsection{LISA}

Let us start with LISA, which will be in principle able to detect GW signals with $\Omega_{GW}> 10^{-13}$ and a frequency ranging from $\sim 10^{-5}$ Hz to $\sim 0.1$ Hz. 
For example, our model can generate a GW signal peaked at $f\sim 10^{-3}$ Hz and $\Omega_{GW}^{\rm {peak}}\gtrsim 10^{-12}$ by choosing $\lambda=0.01$, $r=0.03$ and $M\sim 10^{8}\, {\rm MeV}$, putting it inside the LISA band for $0\le m_\chi/\mu\lesssim 2.4$.
Given the size of the four-dimensional parameter space, we again need to develop a strategy for scanning, which elucidates the main dependency of the observables on the parameters. 
We will see that for our model to produce a stochastic GW signal detectable by LISA requires $10^4 \, {\rm {MeV}}< M < 10^{14}\, {\rm {MeV}}$ for $0.001\le \lambda \le 0.1$.

\begin{figure}
\centering
\begin{minipage}[c]{1\textwidth}  

    \centering
    \includegraphics[width=.32\textwidth]{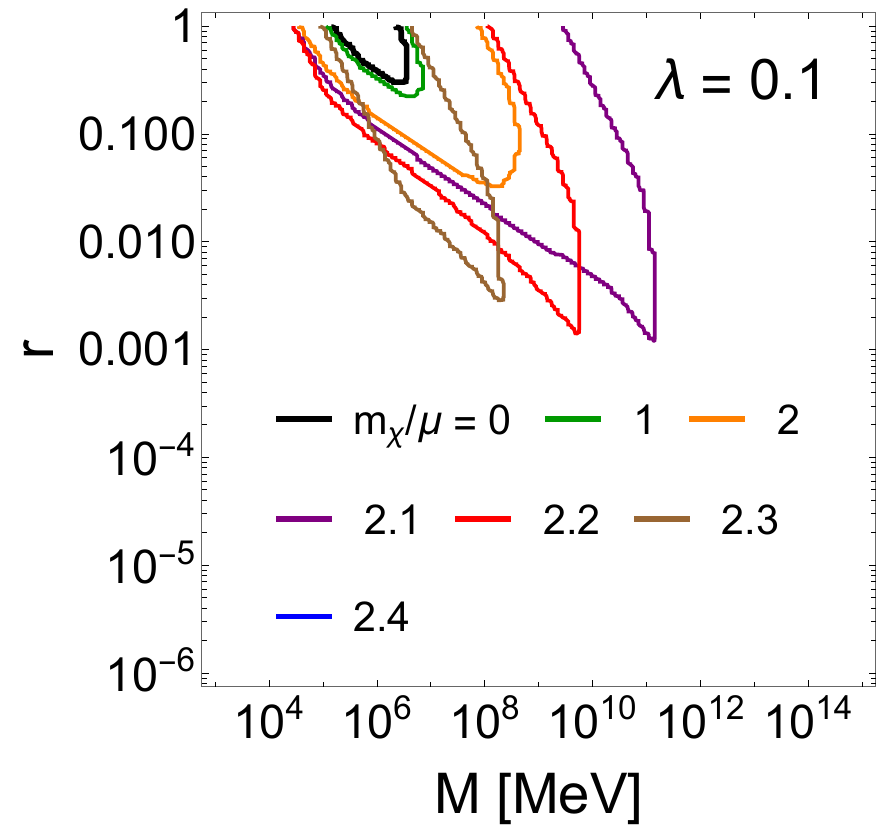}
    \includegraphics[width=.32\textwidth]{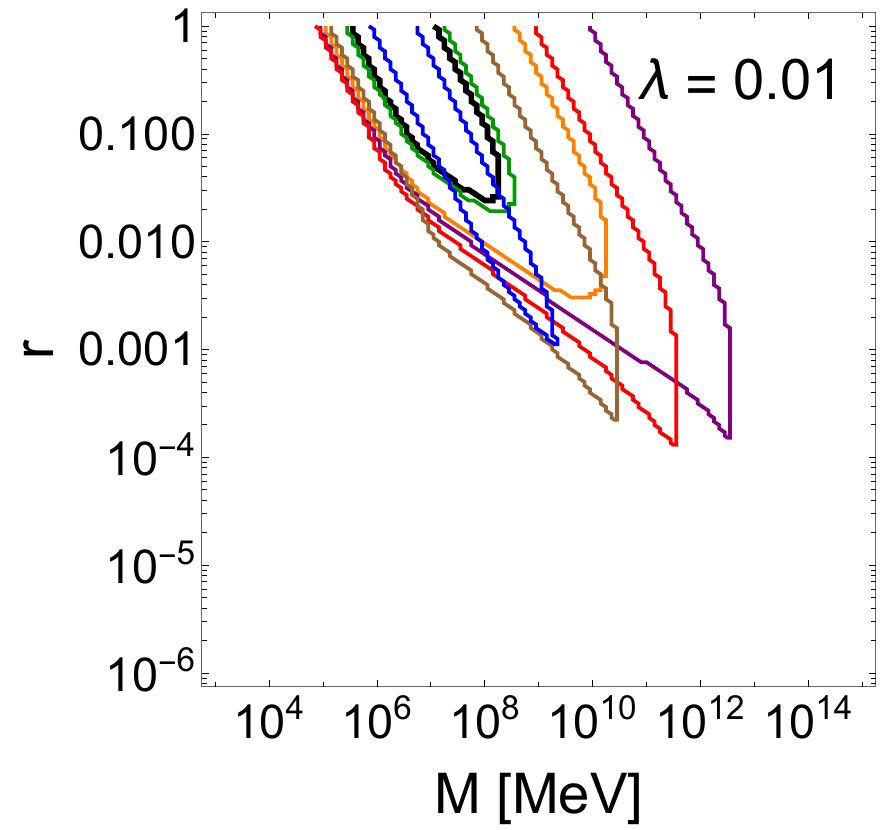} 
      \includegraphics[width=.32\textwidth]{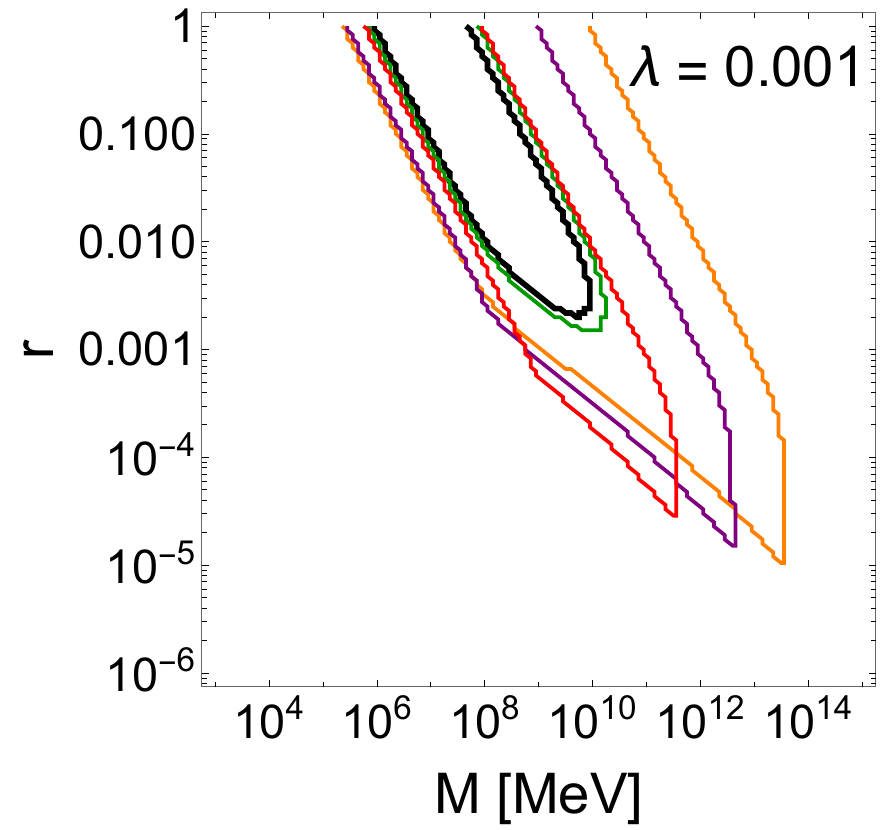}
\end{minipage}
\hspace{0.001\textwidth}

\caption{
LISA: The contours in the $M-r$ plane, inside which the parameters lead to GW spectra with the maximum inside the LISA band. The self-coupling is chosen as $\lambda=0.1,0.01,0.001$ (left to right). The different colors correspond to different values of $m_\chi/\mu=0,1,2,2.1, 2.2, 2.3, 2.4$ (blue, red, green, brown respectively).
Here $M$ is the limit of the effective field theory and $r=\mu/(M\lambda^2)$.
}
\label{fig:LISA}
\end{figure}

Figure~\ref{fig:LISA} shows the contours of parameters on the $M-r$ plane, inside which the parameters lead to (in principle) detectable GWs with LISA. This can be seen as a ``translation" of the LISA sensitivity curve from the $f_{GW}-\Omega_{GW}$ plane onto the $M-r$ plane for different values of $\lambda$ and $m_\chi/\mu$. Points that are inside (above) the LISA sensitivity curves (e.g. Figure~\ref{fig:GWscan1}) are inside (above) the corresponding contours of Figure~\ref{fig:LISA}.
Several physical properties of the system can be learned from this figure. 

First of all, we can explain why the allowed region gets larger as the value of $m_\chi/\mu$ increases, and in the process we will find the lowest $r$ value detectable by LISA.
In this paragraph, we begin  by examining  the case of $m_\chi/\mu=0$, which is shown as the black curve in the figure (and then in the next paragraph we turn to larger values of $m_\chi/\mu>0$). 
Our analytic result in Eqs.~\eqref{eq:omegafapprox} holds for $\alpha>1$, 
where we found  $\Omega_{GW}h^2\approx 5\times 10^{-12}$.
However, the GW amplitude at the lowest point of the LISA band  is an order of magnitude lower,\footnote{Let us note here that different assumptions for the LISA runtime ($T$) and different requirements for the Signal to Noise Ratio (SNR) lead to slightly different sensitivity curves, since $\Omega_{GW}\propto \sqrt{{\rm{SNR}}/T}$; see~\cite{Smith:2019wny}. Small changes in the sensitivity curve used will not significantly affect our results and all parameter scalings will remain valid.  The same holds for all other experiments that we will consider.}  $\Omega_{GW}h^2\approx 5\times 10^{-13}$. 
This means  that the lowest LISA GW value is obtained for $\alpha\approx 0.3$.
It is at this lowest point of the LISA band that the $m_\chi/\mu=0$ band ends and
the parameter space ``pinches off."
Using $\alpha\approx 0.3$ in the expression for $\alpha$ in Eq.~\eqref{eq:alphacalc} and the estimate $2.4/\ln(M_{\rm {Pl}}/M) \approx 0.1$, we arrive at
$0.33 \approx \lambda^{-6} \mu^2/M^2$. We can substitute $\mu$ for $r=\mu/(\lambda^2M)$ leading to
$r\approx 2\lambda$ at the point on the $M-r$ plane where the $m_\chi/\mu=0$ contour pinches off (ends). This is in very good agreement with the numerical contours presented in Figure~\ref{fig:LISA}.

Next we turn to larger values of $m_\chi/\mu > 0$:  one can see in Figure~\ref{fig:LISA} that the allowed parameter space in the $M-r$ plane opens up, becoming wider at the top and pinching at a lower value of $r$. To understand this behavior, let us use our intuition built in Section~\ref{sec:parameterscan} and point at Figures~\ref{fig:GWscan1} and \ref{fig:GWscanlowalpha}. 
We see in Figure~\ref{fig:GWscan1} that larger values of $m_\chi/\mu$, especially exceeding unity, will lead to larger $\Omega_{GW}^{peak}$ for $\alpha>1$. Since the LISA sensitivity curve ``opens up" for larger values of  $\Omega_{GW}$, encompassing more frequencies, this is immediately translated to a larger interval of $M$ that is detectable for $m_\chi/\mu \gtrsim 1$ and $r=1$ (note that $r=1$ corresponds to $\alpha>1$, making Figure~\ref{fig:GWscan1} applicable). This is seen in all panels of Figure~\ref{fig:LISA}. Furthermore, the contours of $m_\chi/\mu$ on the $M-r$ plane will pinch at a lower value of $r$, which we can visualize in the following way. Let us choose a fiducial point on the $\Omega_{GW}-f$ plane where we fix all parameters, including $m_\chi/\mu>1$ and $r$, such that $\alpha>1$. This corresponds schematically to starting somewhere on the solid brown curve of Figure~\ref{fig:GWscanlowalpha}. Now let us keep $M$, $\lambda$ and $m_\chi/\mu$ fixed and lower $r$. Since  $\alpha\propto r^2$, as we lower $r$, the point on the $\Omega_{GW}-f$ plane will move towards lower values of $\Omega_{GW}$, schematically it will move from the solid brown towards the dashed and dot-dashed brown curves.
At some point the value of $\alpha$ (equivalently $r$) will be so small that $\Omega_{GW}$ will be below the LISA sensitivity and thus we can visualize the point as  ``exiting" the LISA band. Since for $m_\chi/\mu>1$ the fiducial point on the $\Omega_{GW}-f$ plane starts at a  larger value of $\Omega_{GW}$ (the brown instead of the red solid curve on Figure~\ref{fig:GWscanlowalpha}), it can sustain more suppression ($\alpha\propto r^2$) before exiting the LISA sensitivity band.
This is  seen in Figure~\ref{fig:LISA}. We can estimate the lowest $\alpha$ (equivalently $r$) by going back to Eq.~\eqref{eq:omegaGWsimplewithalpha}, which scales approximately as
$\Omega_{GW}^{peak} h^2 \sim 1.8\times 10^{-6} (H_*/\beta)^2\alpha^2$ for $\alpha < 1$.
We know that the lowest point detectable by LISA is $\Omega_{GW} h^2 \sim 10^{-13}$. In order to compute the lowest detectable $\alpha$, we use the maximal value of $H_*/\beta$ which comes from the percolation bound $\beta/H_*=3$. This immediately gives $\alpha^2_{\rm {min}}\sim 10^{-6}$.  In other words, in order for the signal to be inside the LISA band, $\alpha \gtrsim 0.001$ is required. Unfortunately we cannot use Eq.~\eqref{eq:alphacalc} to relate $\alpha$ to $r$, as this uses the expression for $\dot\phi_c$ which only holds for $m_\chi/\mu\ll 1$. That being said, we numerically find that for $m_\chi/\mu \gtrsim 2$, $\alpha\simeq 10^{-3}$ translates to $r\simeq 0.01 \lambda$. This is indeed verified by all panels of Figure~\ref{fig:LISA} as the lowest possible $r$ value detectable by LISA.

 Let us now examine the observable frequencies for different values of $\lambda$. Let us focus again on the $m_\chi/\mu=0$ case, which is captured perfectly by our analytics and also choose $r=1$ for simplicity. From the frequency equation of Eq.~\eqref{eq:omegafapprox}, we see that (ignoring the change in $\ln(M_{\rm{Pl}}/M)$), a constant GW peak frequency requires the product $\lambda^{1/2} M$ to be constant. This is verified from Figure~\ref{fig:LISA}. For example, the black contour for $\lambda=0.1$ at $r=1$ covers $10^5 {\rm {Hz}} \lesssim f \lesssim 10^{6.5} {\rm {Hz}}$ and the corresponding range for $\lambda=0.001$ is  $10^6 {\rm {Hz}} \lesssim f \lesssim 10^{7.5} {\rm {Hz}}$, meaning that the frequency is shifted by one order of magnitude for every two orders of magnitude of change in $\lambda$, as expected, further cementing the accuracy of our analytical approach in its regime of validity and allowing us to estimate the observable parameter range for different values of $\lambda$ without requiring extra numerical computations.

A final feature seen from the plots of Figure~\ref{fig:LISA} is that all bands follow the same slope (to good accuracy) $r\propto M^{-1}$ on the $M-r$ plane. This has to do with the translation of the edge of the LISA band from the $\Omega-f$ to the $M-r$ plane for fixed values of $\lambda$ and $m_\chi/\mu$. This has a profound consequence: Since systems with smaller $\lambda$ allow for smaller values of $r$, the scaling $r\propto M^{-1}$ implies a larger maximal value of $M$ for smaller values of $\lambda$. In particular
\beq
{M(r=1)\over M({r_{min}})} \simeq r_{min} \simeq 2\lambda
\label{eq:Mmaxvslambda}
\eeq
for the bands with $m_\chi/\mu>1$. This holds (approximately) for the bands shown in Figure~\ref{fig:LISA}.

\begin{figure}
\centering  
\includegraphics[width=.32\textwidth]{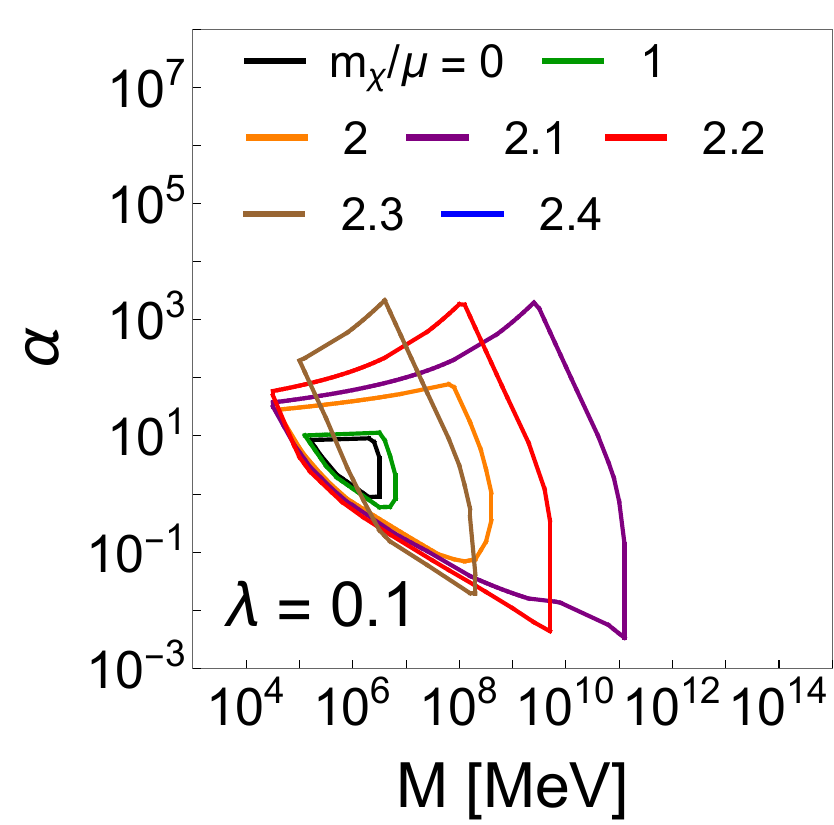}
    \includegraphics[width=.32\textwidth]{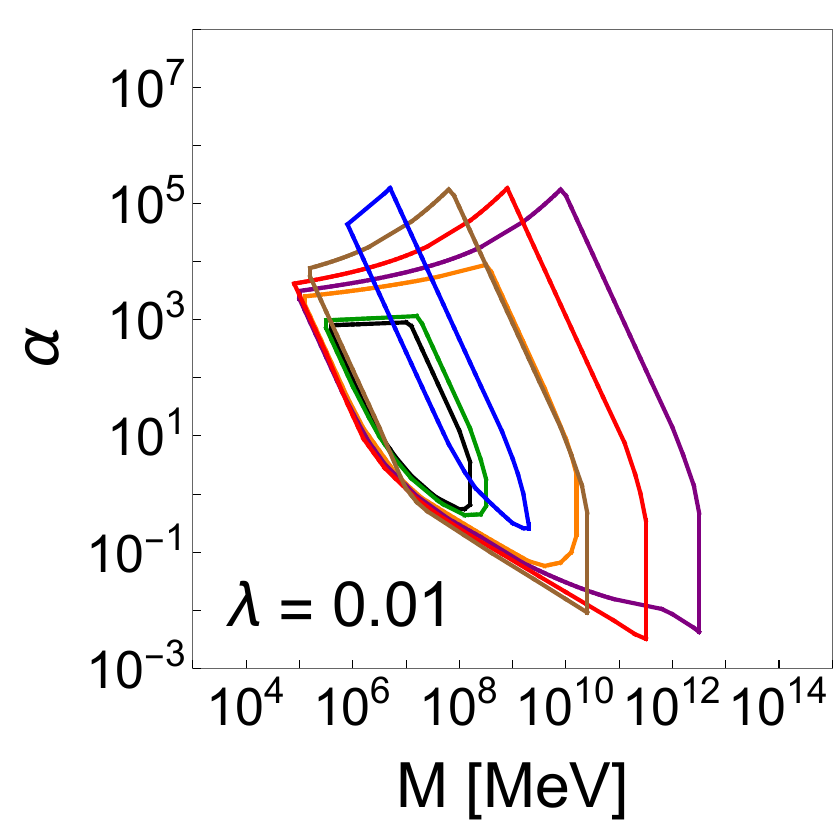} 
      \includegraphics[width=.32\textwidth]{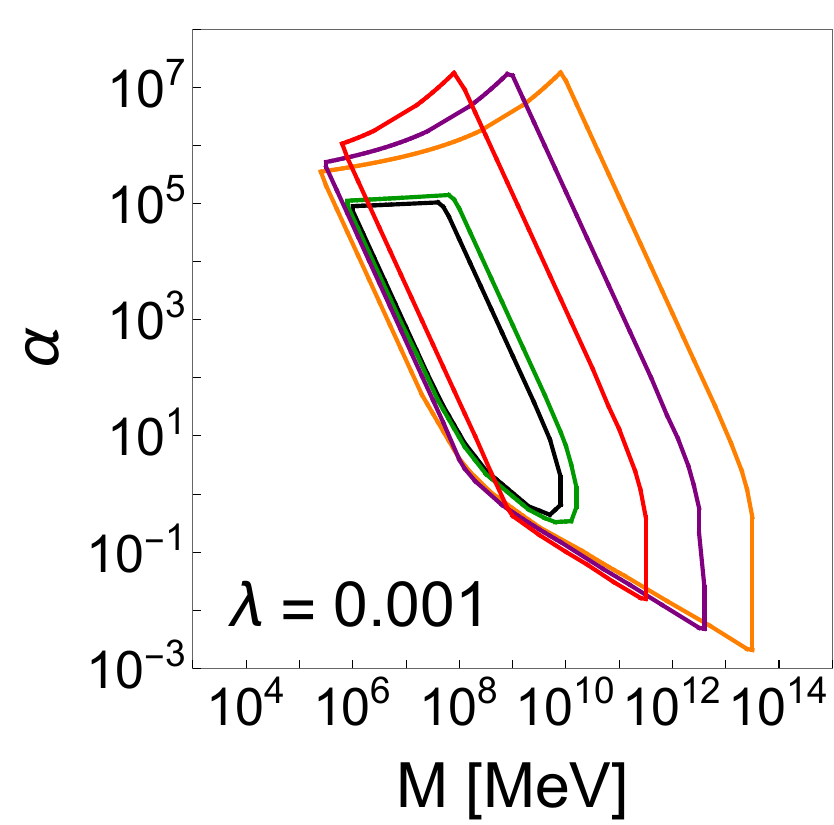}
      \\
      \includegraphics[width=.32\textwidth]{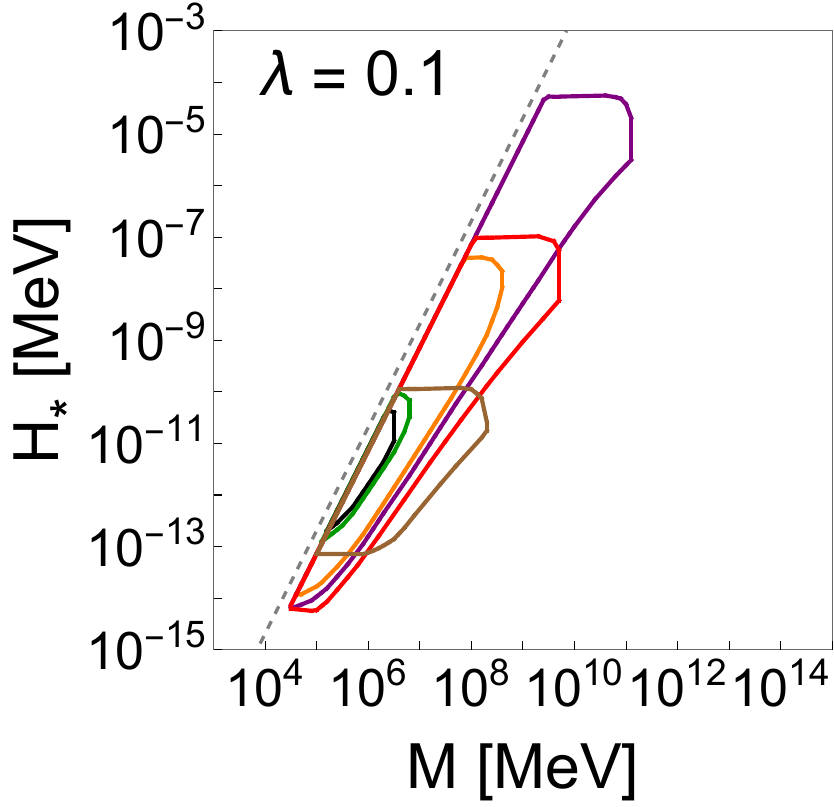}
    \includegraphics[width=.32\textwidth]{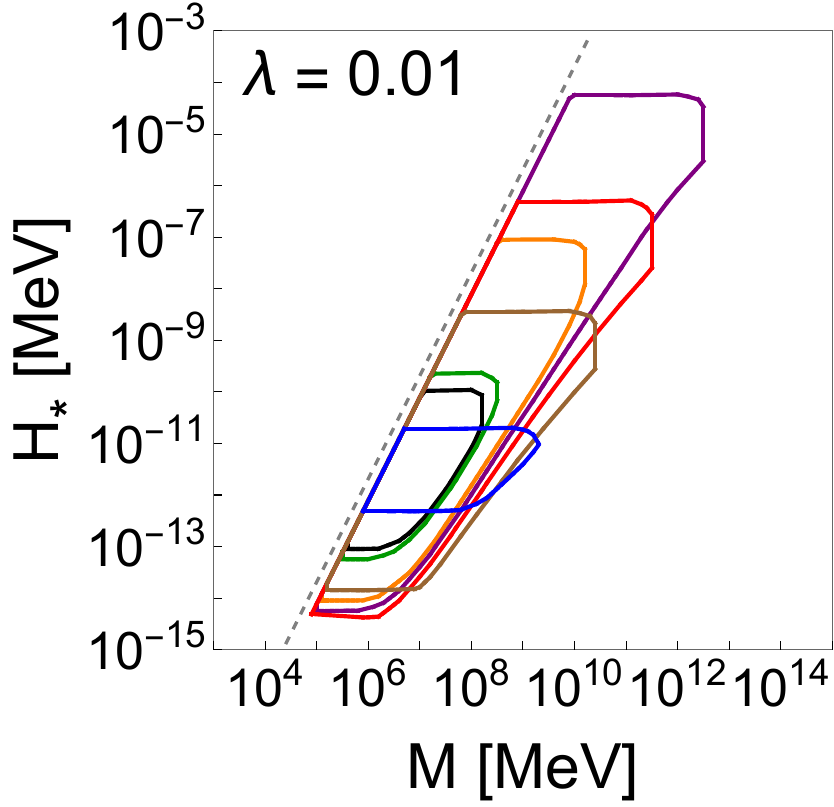} 
      \includegraphics[width=.32\textwidth]{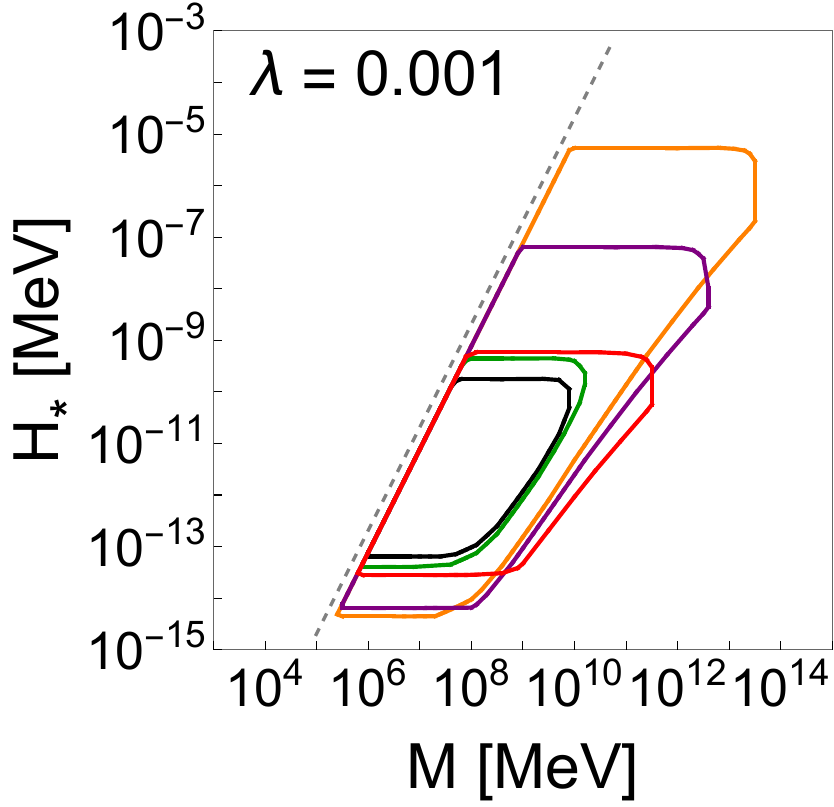}

\caption{LISA: The contours in the $M-\alpha$ (upper) and $M-H_*$ plane (lower), inside which the parameters lead to GW spectra with the maximum inside the LISA band. The self-coupling is chosen as $\lambda=0.1,0.01,0.001$ (left to right). The different colors correspond to different values of $m_\chi/\mu=0,1,2,2.1, 2.2, 2.3, 2.4$, following Figure~\ref{fig:LISA}.
The black dashed curve on the lower panels corresponds to the maximal Hubble scale $H_*=0.5 \lambda M^2/M_{\rm {Pl}}^2$, given in Eq.~\eqref{eq:LISAHubble}.
}
\label{fig:LISA2}
\end{figure}

Before concluding the analysis of the LISA band, it is worth examining the detectable parameter space in a different form. While the variable $r$ provides an easy computational tool and allows for direct contact with the potential of Eq.~\eqref{eq:lagrangian}, other parameters allow for faster connection to cosmology, specifically the Hubble scale at the time of the FOPT ($H_*$) and the ratio $\alpha$ of the false vacuum to kinetic energy of the kination field at the time of the phase transition. In Figure~\ref{fig:LISA2} we present the parameter space of the model accessible by the LISA detector in the $M-\alpha$ and $M-H_*$ planes.

Let us start by  examining how the  $M-r$ plane is mapped onto the $M-\alpha$ plane.
The quantity $\alpha$ is important, since it shows how the energy budget of the universe is distributed between the kination and potential energy sectors at the time of the transition. We start by
focusing  on the $\lambda=0.1$ case. For $\lambda\ll 1$ and $m_\chi/\mu < 1$ we can use Eq.~\eqref{eq:alphavsr}, immedately relating $r$ to $\alpha$. This provides an excellent fit to the $m_\chi/\mu=0,1$ contours of Figures~\ref{fig:LISA} and \ref{fig:LISA2}. We can immediately see that $r=1$ corresponds to $\alpha\simeq 10$ with a weak (logarithmic) dependence on $M$. The lowest point of these two contours was shown above to lie at $\alpha\simeq 0.3$. 
For the contours of $m_\chi/\mu>1$ we cannot use Eq.~\eqref{eq:alphacalc}. However, as we discussed above, we require $\alpha\gtrsim 0.001$ for the signal to be inside the LISA band regardless of the other parameters, which is verified by the lowest value of $\alpha$ at the ``tip" of the bands in the $M-\alpha$ panels of Figure~\ref{fig:LISA2}. 
A final interesting note, which shows the limitation of our analytical approach is the fact that for $m_\chi/\mu=0,1$, the upper boundary of the regions on the  $M-\alpha$ plane are (almost) horizontal, as they are the mapping of the $r=1$ line of the $M-r$ plane. However, the upper boundary of the bands for $m_\chi/\mu>1$ show a non-negligible dependence on the value of $M$, which is not predicted by our  analytical formulas. Other than that, the mapping of the $M-r$ onto the $M-\alpha$ plane is very intuitive.
For $\lambda=0.01,0.001$ the transformation from the $M-r$ onto the $M-\alpha$ plane shows the same features that we discussed for the $\lambda=0.1$ case. Most importantly, the lower limit of $\alpha\gtrsim 0.001$ is shown to be  independent of $\lambda$, as expected, making the $M-\alpha$ parametrization particularly useful when discussing the energy budget of the universe during the phase transition.

Figure~\ref{fig:LISA2} also shows  the parameter space of the model accessible by  the LISA detector in the $M-H_*$ plane
where $H_*$ is the Hubble scale at the time of the FOPT. While the Hubble scale can be computed as a combination of the various parameters in the potential, it is worth showing its 
behavior directly in this plot. 
We see in Figure~\ref{fig:LISA2} that Eq.~\eqref{eq:LISAHubble} provides an excellent estimate of the upper limit of $H_*$ across the entire range of $M$ for all values of $\lambda$ that we examined. The accessible parameter space by LISA on the (logarithmic) $M-H_*$ plane can be  described roughly as an obtuse trapezoid, which is shifted to higher values of $M$ by the scaling $M\propto \lambda^{-1/2}$ as discussed earlier.
Interestingly, we also find that the accessible region becomes wider in $M$ for smaller $\lambda$. This is due to the scaling described in Eq.~\eqref{eq:Mmaxvslambda}, where the range of $M$ that each band covers  grows inversely proportional to $\lambda$. Overall the allowed trapezoid on the $M-H_*$ plane is shifted horizontally by $1/\sqrt{\lambda}$ and varies in  width as $1/\lambda$, when one chooses a different value of $\lambda$.

Another point worth mentioning is our plotting choice for the values of $m_\chi/\mu$ shown  in Figures~\ref{fig:LISA} and~\ref{fig:LISA2}. For $m_\chi/\mu\lesssim 1$, we do not need to use many different values, since the results do not differ much.  However, for higher mass ratios $m_\chi/\mu\gtrsim 2$ that push $\beta/H_*$ close to the percolation ratio, the allowed parameter space (e.g. the allowed values of $M$)  opens up significantly and in practice is largely dominated by these values. Therefore, despite being a small part of the allowed parameter space (in the spirit of footnote~\ref{fn:paramlikelihood}), they are of high importance observationally and we therefore pay more attention at resolving them. 
We should also note that for $m_\chi/\mu$ larger than the values depicted in Figures~\ref{fig:LISA} and~\ref{fig:LISA2}, we did not find viable parameter combinations with observables within the LISA band.

In summary, we translated the LISA sensitivity curve from the $(f_{\rm GW},\Omega_{\rm GW})$ plane onto the $(M,r)$, $(M,\alpha)$ and $(M,H_*)$ planes, and identified the region of the four-dimensional parameter space for a kinetically induced FOPT that can be probed by LISA. We considered a discrete set of values of $0\le m_\chi/\mu \le 2.4$. We showed that the results are roughly the same for 
 $m_\chi /\mu= 0$ and $1$ (and therefore all values in between), which is not surprising given our earlier analytic estimates. 
For $m_\chi/\mu>2$ the GW amplitude gets progressively larger. 
 We also  considered a discrete set of values of 
 $\lambda = 0.001, 0.01, 0.1$ and for each one scanned over $M$ and $r$ for each discrete value of $m_\chi/\mu$. Scanning over $r$  is an effective way to scan over $\mu$, since $r=\mu/(M\lambda^2)$.   We found that detectability requires a minimal energy fraction $\alpha \gtrsim 10^{-3}$ (corresponding to $r \gtrsim 0.01\lambda$ for $m_\chi/\mu\gtrsim 2$). 
 We showed that 
 our model produces a stochastic GW signal detectable by LISA for $10^4 \, {\rm {MeV}}< M < 10^{14}\, {\rm {MeV}}$ if $0.001\le \lambda\le 0.1$.
 In addition, we found that the Hubble scale at the time of the FOPT is in the range $H_* \in (10^{-13}, 10^{-9})$ MeV for the case of $m_\chi/\mu \leq 1$ and as large as $10^{-5}$ MeV for $m_\chi/\mu \sim 2$.  Smaller values of $\lambda$ both shift the observable window to higher $M$ and broaden it, with the width in $M$ scaling approximately as $1/\lambda$. Together, these results provide a clear map between LISA’s sensitivity and the underlying  parameters (both microscopic such as $M$ and cosmological such as $H_*$) of this model of a  kinetically induced phase transition.

\subsubsection{Advanced LIGO}

We  now perform the same analysis for the  sensitivity of Advanced LIGO (aLIGO), shown e.g.~in Figure~\ref{fig:GWscan1}, with frequencies around $10-100$ Hz. We use the same scanning strategy as with LISA in order to map the parts of parameter space
that fall within the sensitivity curve of aLIGO, with the results are shown in Figure~\ref{fig:aLIGO}.
Given the non-detection of a stochastic GW background from data collected up to the aLIGO-VIRGO-KAGRA fourth Observing Run (O1-O4a runs)~\cite{LIGOScientific:2016jlg, LIGOScientific:2025bgj, LIGOScientific:2025kry}, the analysis of this Section will lead to an excluded, rather than a detectable region of parameter space.\footnote{Refs.~\cite{LIGOScientific:2016jlg, LIGOScientific:2025bgj} use power-law and flat spectra, inspired primarily by GWs produced from an non-distinguishable number of inspiring BHs using data up to the O3 and O4a runs respectively. The latest constraint from a flat spectrum is 
$\Omega_{GW} \le 2.8 \times 10^{-9}$ or
$\Omega_{GW}h^2 \le 1.25 \times 10^{-9}$.
Ref.~\cite{LIGOScientific:2025kry} uses spectra from FOPTS (among others), arriving at $\Omega_{GW} \le 2.3 \times 10^{-9}$, which is very similar to the result derived using a flat spectrum. }
In this section, we will present the parts of parameter space of our model that are ruled out through aLIGO.

In the frequency band of aLIGO (higher than for LISA), the values of $M$ that produce a detectable signal lie in the range
$10^{11}<M<10^{18}$ MeV for $0.001\le \lambda\le 0.1$. As expected, the values of $M$  are in general larger than the ones accessible by LISA.  There is a  key difference with the parameter space detectable by LISA.  Since the detectability band of aLIGO does not reach sensitivities of $\Omega_{GW}h^2 = {\cal O}(10^{-12})$, cases with $m_\chi/\mu\ll  1$ are unobservable with aLIGO. Building on our intuition from Section~\ref{sec:parameterscan}, we expect only parameter combinations within a small range of $m_\chi/\mu\sim 2$ to be detectable with aLIGO. We can perform the same exercise as with LISA to find the absolute lowest value of $r$ that is accessible by aLIGO, within the small range of $m_\chi/\mu$ (and corresponding values of $M$) that put the GW signal within the aLIGO frequency range. We see that the tip of the aLIGO sensitivity curve is at $\Omega_{GW}h^2 \sim 10^{-9}$. 
Using again the expression for peak GW amplitude $\Omega_{GW}^{peak}\sim 10^{-6}(H_*/\beta)^2 \alpha^2$ and taking the smallest value of $(\beta/H_*)_{\rm{min}}=3$, we can derive the smallest possible $\alpha$ that allows for observability in aLIGO, which is $\alpha_{\rm min}\simeq 0.1$.
By numerically evaluating $\dot\phi_c$ in the regime $m_\chi/\mu>1$, the constraint on $\alpha$ translates 
to $r\gtrsim 0.1 \lambda$. The calculated minimum observable values of $r$ and $\alpha$ agree agree with the contours of Figure~\ref{fig:aLIGO}. In particular, in all three  panels that show the accessible parameter space  in the $M-r$ plane, the lowest tip of the different bands are at $r_{\rm {min}}\simeq 0.1\lambda$, whereas on the $M-\alpha$ plane, the bands pinch at $\alpha\simeq 0.1$ for all values of $\lambda$ (with the approximation becoming better for $\lambda\le 0.01$. Furthermore, we see that, although the minimum allowed value of $\alpha$ in the aLIGO parameter space is larger than the one for LISA (as expected due to the difference in sensitivity), the maximum value scales as $\alpha_{\rm {max}} \sim 10/\lambda^2$ in both cases. The maximum value of $\alpha$ depends on  the behavior of $\dot\phi_c$ for $m_\chi/\mu>1$ (deviating from the expression given in Eq.~\eqref{eq:alphacalc}, which only holds for $m_\chi/\mu \ll 1$) and we thus expect this scaling of $\alpha_{\rm {max}}$ to hold universally for our model, as it is related to the behavior close to $\beta/H=3$. Finally, the accessible parameter space on the $M-H_*$ plane follows the expected pattern, lying to the right of the line defined through Eq.~\eqref{eq:LISAHubble} and shifted to larger values of $M$ and correspondingly $H_*$ compared to LISA.  The shape of the allowed region can be again  described as  a trapezoid shifted along $M$ by $1/\sqrt{\lambda}$ with a width approximately scaling as $1/\lambda$, further cementing the validity of these simple scaling arguments.

{Summing up, Advanced LIGO is able to probe values of the mass-scale $M$ in a relatively wide range, namely $10^{11}\, {\rm {MeV}}< M \lesssim 10^{17} \, {\rm {MeV}}$ and Hubble scales of $10^{-1} \, {\rm {MeV}} < H_* \le 10^3 \, {\rm {MeV}}$.
However, aLIGO is only sensitive to $m_\chi/\mu\sim 1.5 - 1.8$ (depending on the value of $\lambda$), a very narrow interval.
These two arguments: the wide range on $M$ and restricted range in $m_\chi/\mu$ may seem  contradictory  but are not, since most of the observable parameter space is dominated by the largest values of $m_\chi/\mu$. This can be understood as follows:   values of $m_\chi/\mu$ that put the FOPT dynamics close to the percolation bound $\beta/H_*=3$ allow for the other parameters of the system, like the energy ratio $\alpha$, which suppresses GW production, to vary more broadly while still producing GWs within a given frequency and amplitude range. 
{Given the non-detection of a stochastic background by aLIGO, we   view the contours of Figure~\ref{fig:aLIGO} as exclusion plots on the parameter space of our model.} Let us note again, that while a large range of $M$-values is excluded, this is true only for a small range of mass ratios $m_\chi/\mu$. Therefore, the non-detection of GWs from aLIGO does not significantly affect the viable parameter space of our model. This is visually demonstrated in the next section, specifically in Figure~\ref{fig:CE}.

\begin{figure}
\centering
    \includegraphics[width=.32\textwidth]{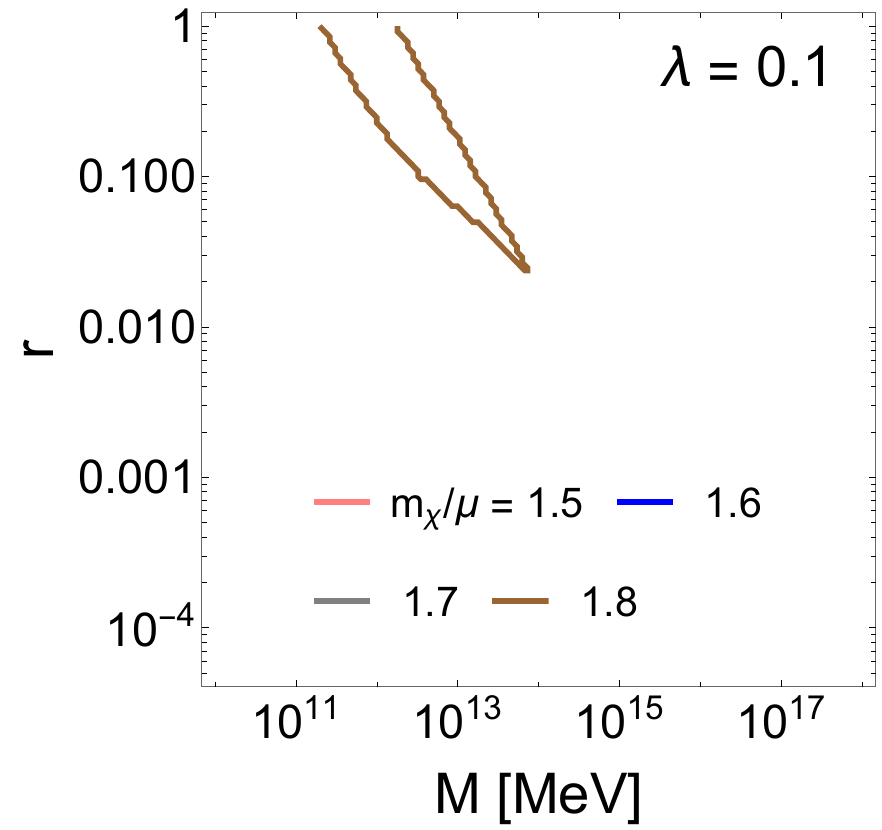}
\includegraphics[width=.32\textwidth]{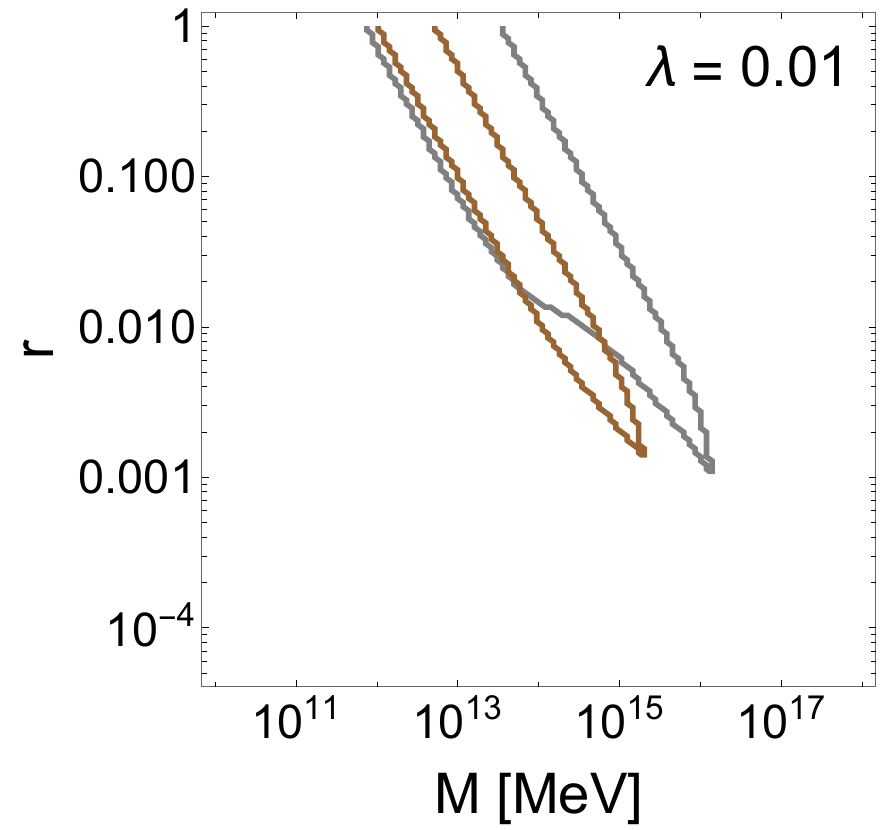} 
\includegraphics[width=.32\textwidth]{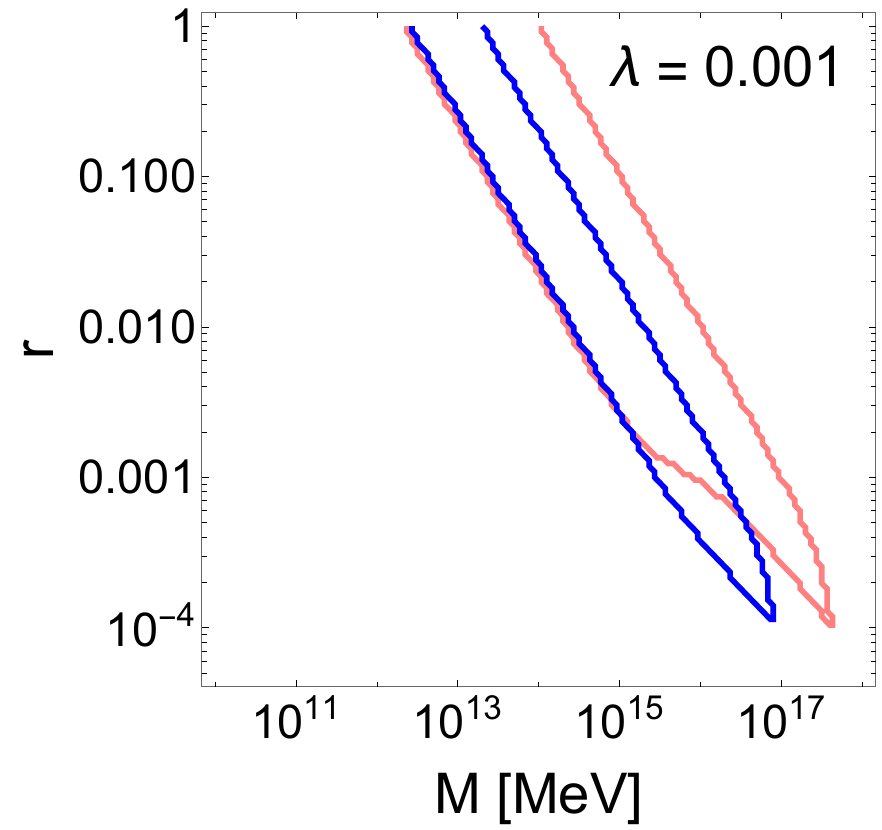}
    \\
\includegraphics[width=.32\textwidth]{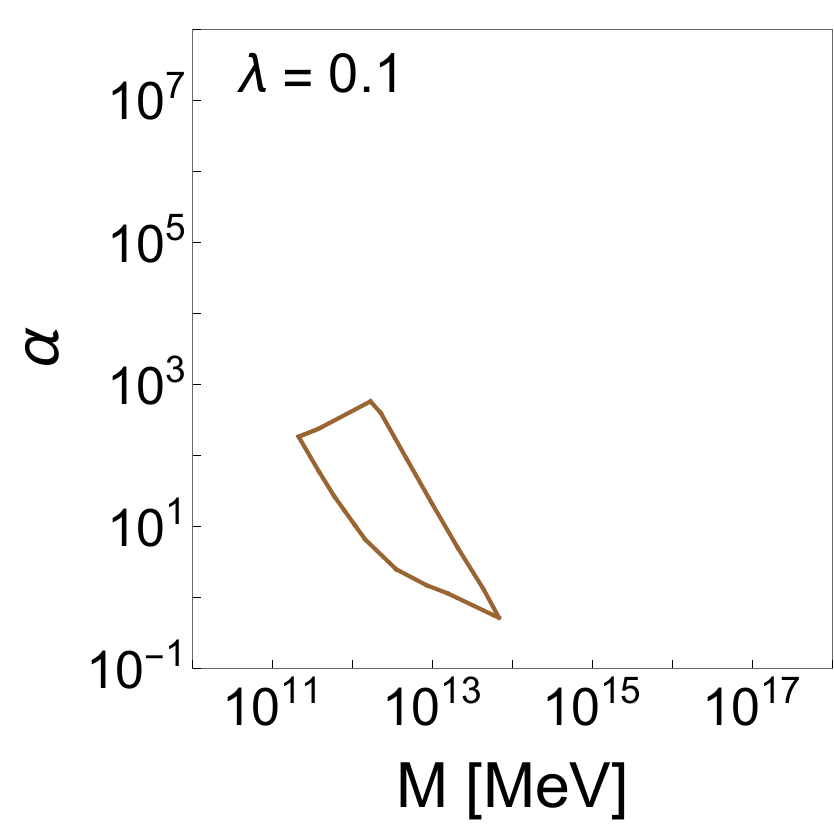}
\includegraphics[width=.32\textwidth]{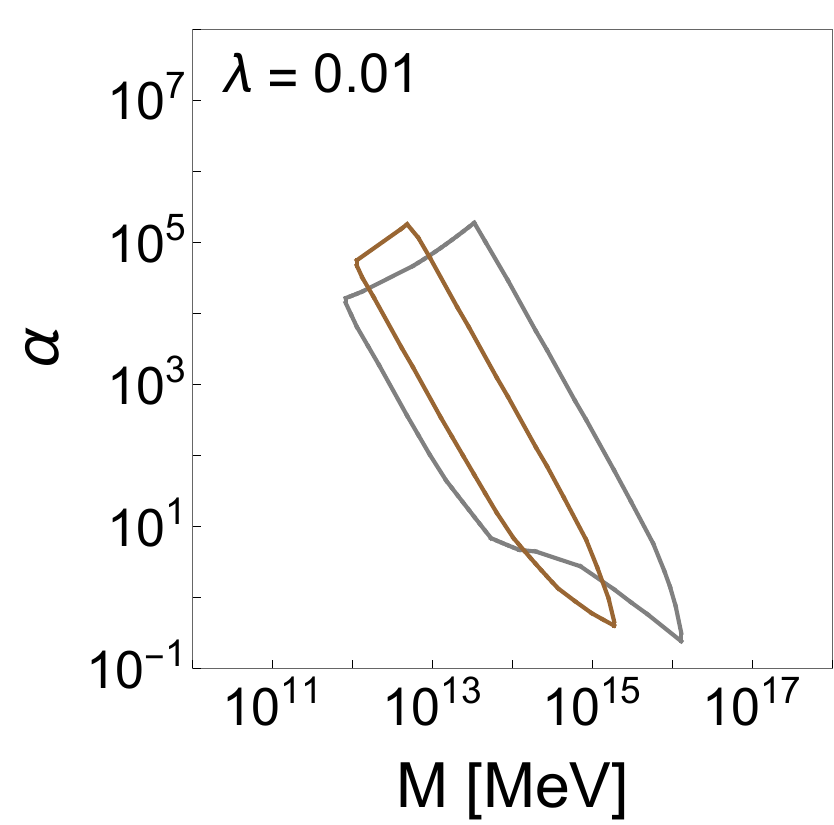} 
\includegraphics[width=.32\textwidth]{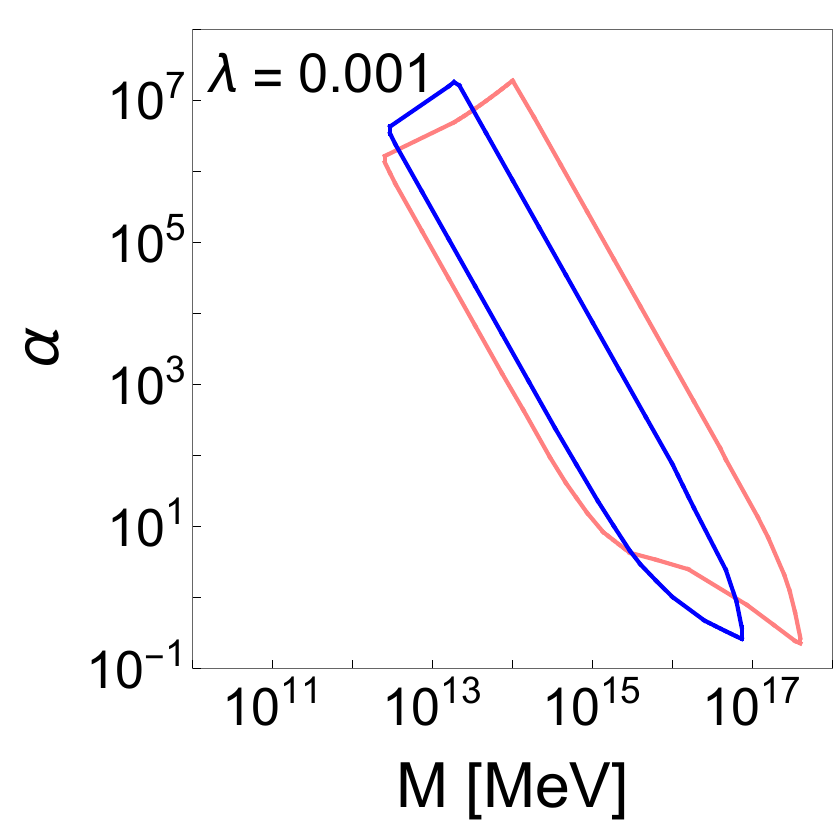}
    \\
\includegraphics[width=.32\textwidth]{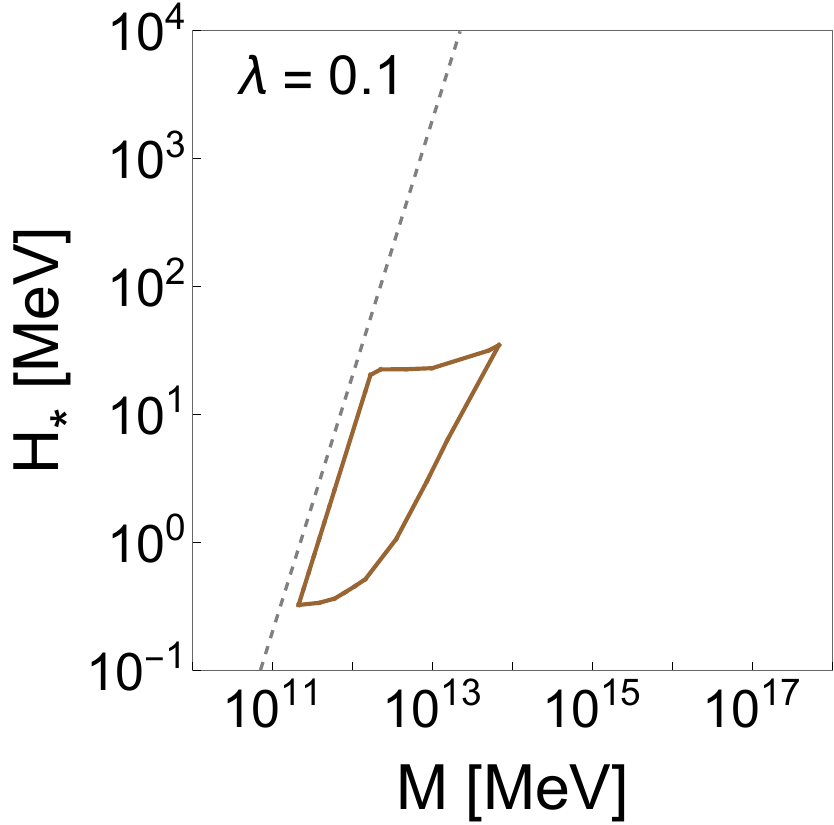} \includegraphics[width=.32\textwidth]{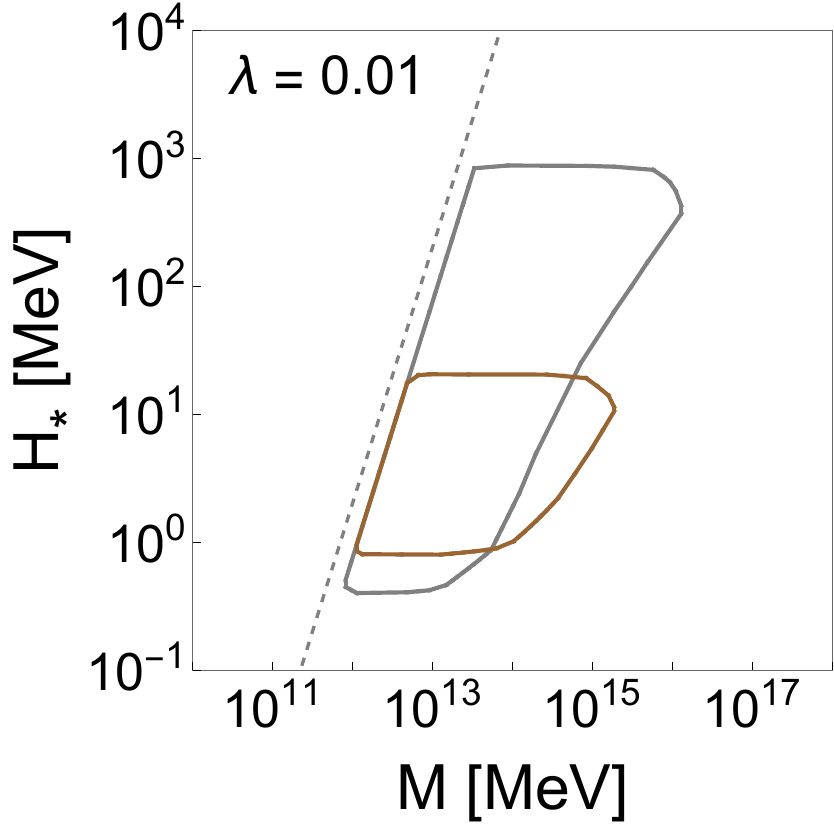}   \includegraphics[width=.32\textwidth]{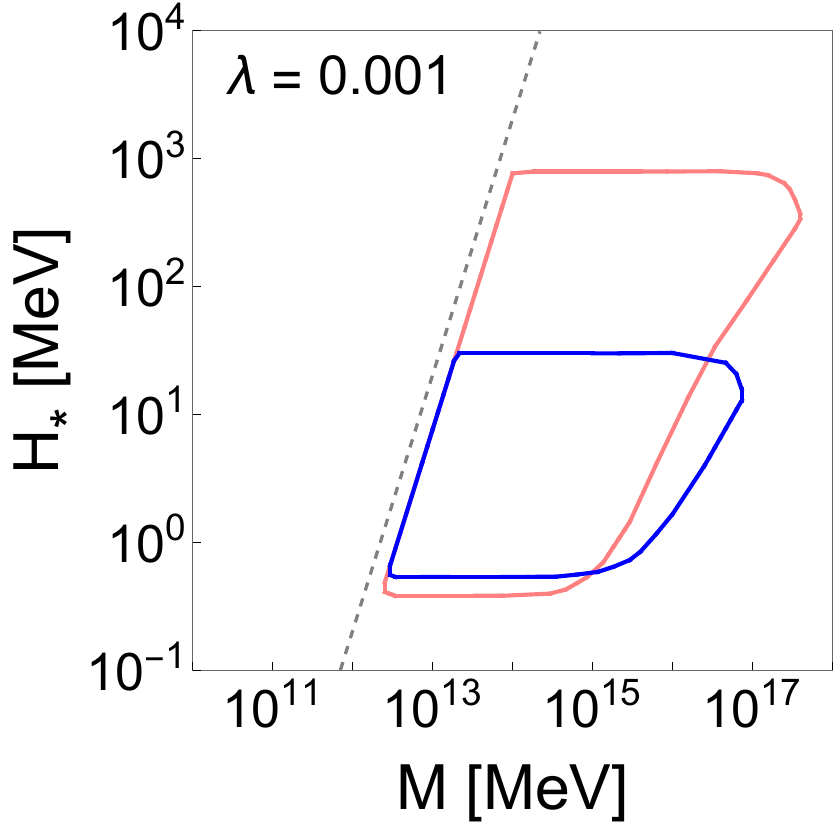}

\caption{Advanced LIGO: The contours on the $M-r$ plane (upper), the $M-\alpha$ plane (middle) and the $M-H_*$  plane (lower), inside which the parameters lead to GW spectra with the maximum inside the Advanced LIGO (aLIGO) band. The self-coupling is chosen as $\lambda=0.1,0.01,0.001$ (left to right). The different colors correspond to different values of $m_\chi/\mu=1.5,1.6,1.7,1.8$ (red, blue, black and brown respectively) for all panels. We see that the $m_\chi/\mu=0$ curve is absent, as expected (see discussion in the main text).
}
\label{fig:aLIGO}
\end{figure}

\subsubsection{Cosmic Explorer}

There is ongoing work on the design of next generation detectors that will reach sensitivities of $\Omega_{GW}={\cal O}(10^{-13})$ around the $10\,{\rm {Hz}}$ frequency band. Since the frequencies encompass those observed by aLIGO, we expect a similar (but broader) range of $M$ and since the sensitivity is better than ${\cal O}(10^{-12})$, we expect to be able to detect cases with $m_\chi/\mu=0$ as well as probe lower values of $\alpha$ and $r$ than aLIGO. 
In some sense, CE is expected to have similar sensitivity to LISA (the details depend on what assumptions are used for each of the two experiments, see e.g.~\cite{Evans:2021gyd}) and operate in LIGO frequencies. Therefore, we expect the CE-relevant parameter space to be (mostly) a ``shifted" version of the corresponding LISA-relevant one. The resulting parameter space is shown in Figure~\ref{fig:CE}. We first see that the scale accessible to CE, $M\gtrsim 10^9\,{\rm {MeV}}$, is shifted by around $5$ orders of magnitude compared to the corresponding values of LISA ($M\gtrsim 10^4\,{\rm {MeV}}$). This follows the simple scaling of Eq.~\eqref{eq:freqwithr}, where (if we neglect the logarithmic dependence) we get the overall scaling $f\propto M$.
Furthermore, we see that $1>r\gtrsim 0.01\lambda$ and the energy density ratio lies in the regime $10^{-3}\lesssim \alpha\lesssim {\cal O}(10)\lambda^{-2}$, similar to the values that arise in the case of LISA.
Finally, Eq.~\eqref{eq:LISAHubble} still defines the left edge of the allowed parameter space (approximated as a trapezoid) on the $M-H_*$ plane. Since in this equation $H_*\propto M^2$, the Hubble scale is expected to be $10$ orders of magnitude larger in CE than in LISA, which is verified by the results shown in Figures~\ref{fig:CE} and~\ref{fig:LISA2}.

We must also note that, since CE operates in a superset of the frequency range covered by aLIGO, we must incorporate the exclusion plots shown in Figure~\ref{fig:aLIGO}. 
The shaded contours in Figure~\ref{fig:CE} correspond to the parameter space that is excluded through the non-detection of stochastic GWs by aLIGO-VIRGO-KAGRA, shown in Figure~\ref{fig:aLIGO}. We see that a small  range of values in $m_\chi/\mu$ is excluded, creating a ``hole" in the four-dimensional parameter space; this hole
only exists for a few specific values of $m_\chi/\mu \sim 1.5-1.8$. 

\begin{figure}
\centering
\includegraphics[width=.32\textwidth]{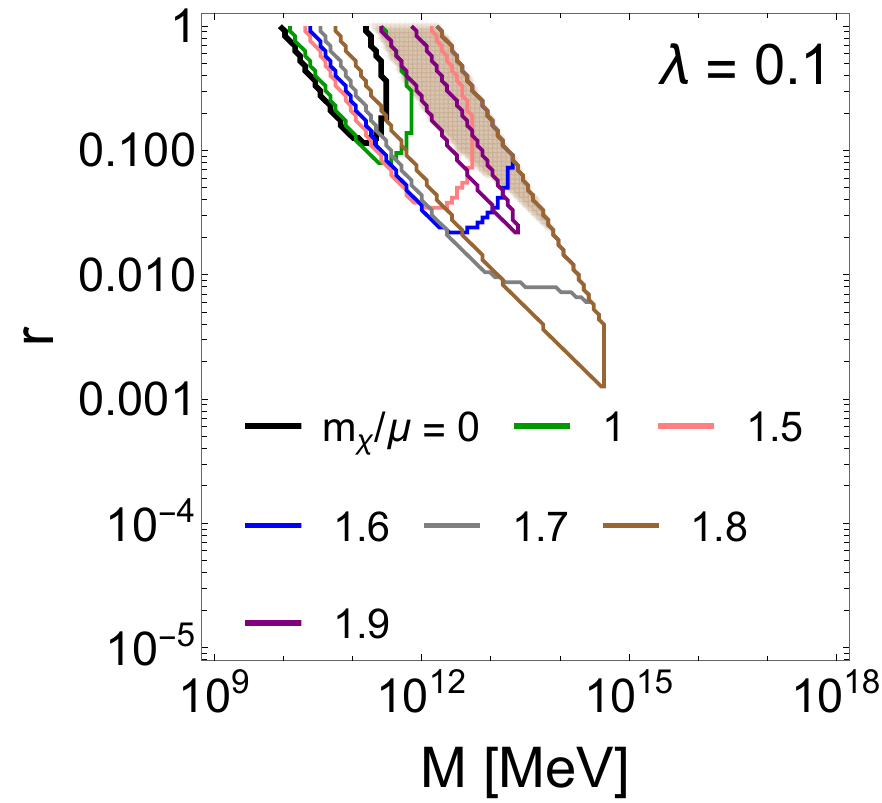}
\includegraphics[width=.32\textwidth]{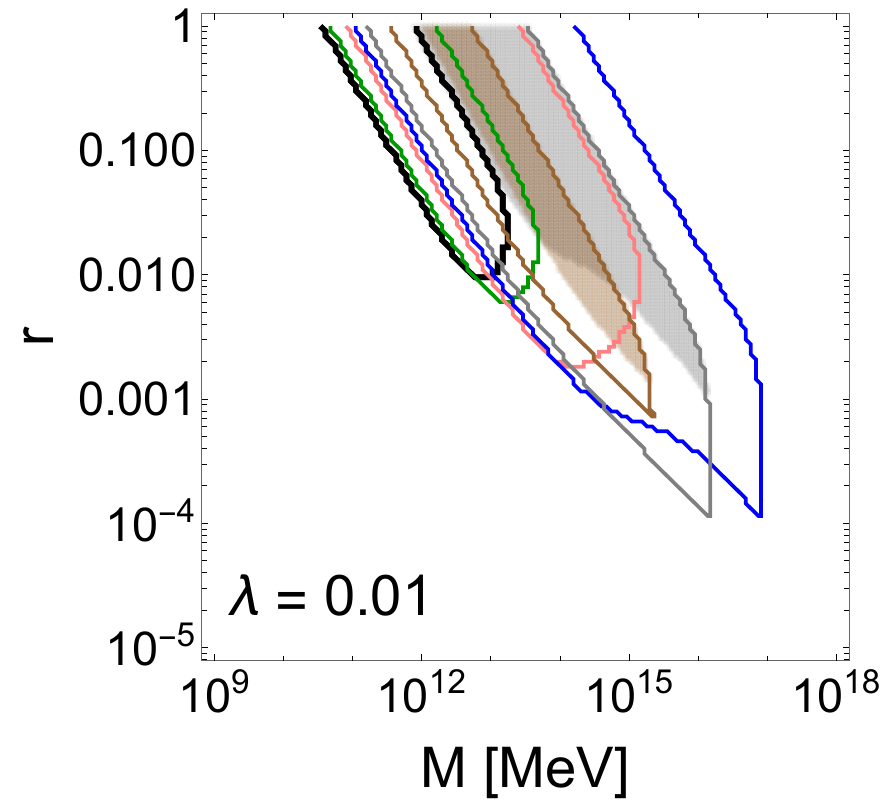} 
\includegraphics[width=.32\textwidth]{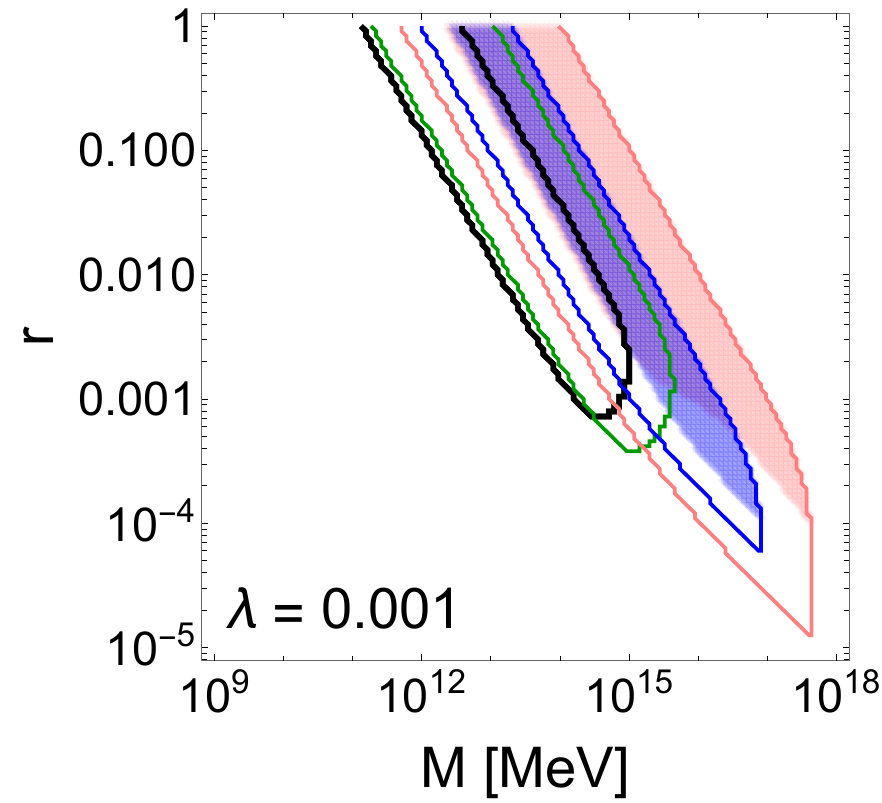}
    \\
\includegraphics[width=.32\textwidth]{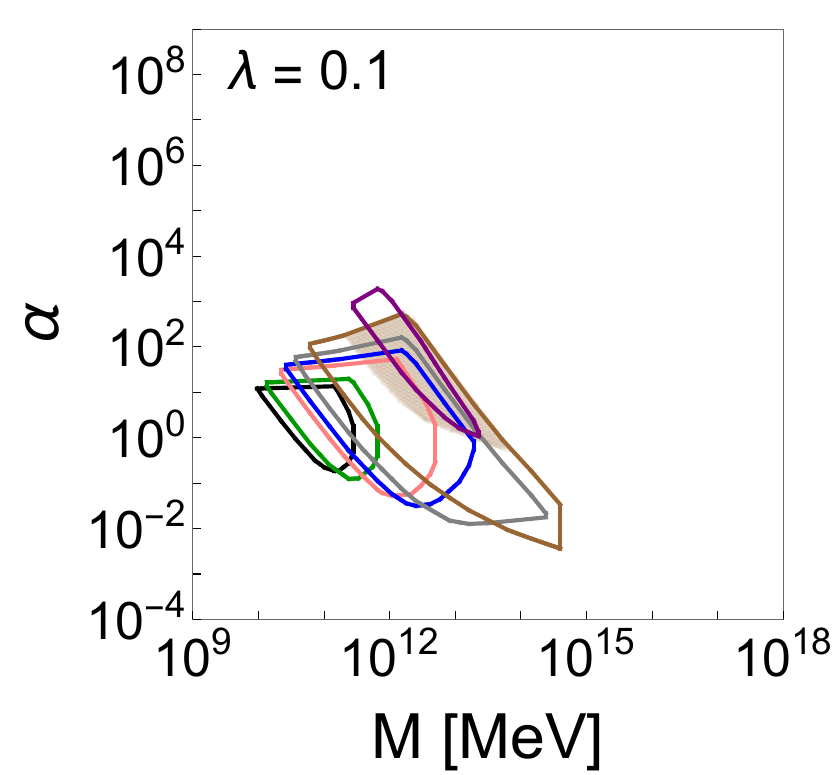}
\includegraphics[width=.32\textwidth]{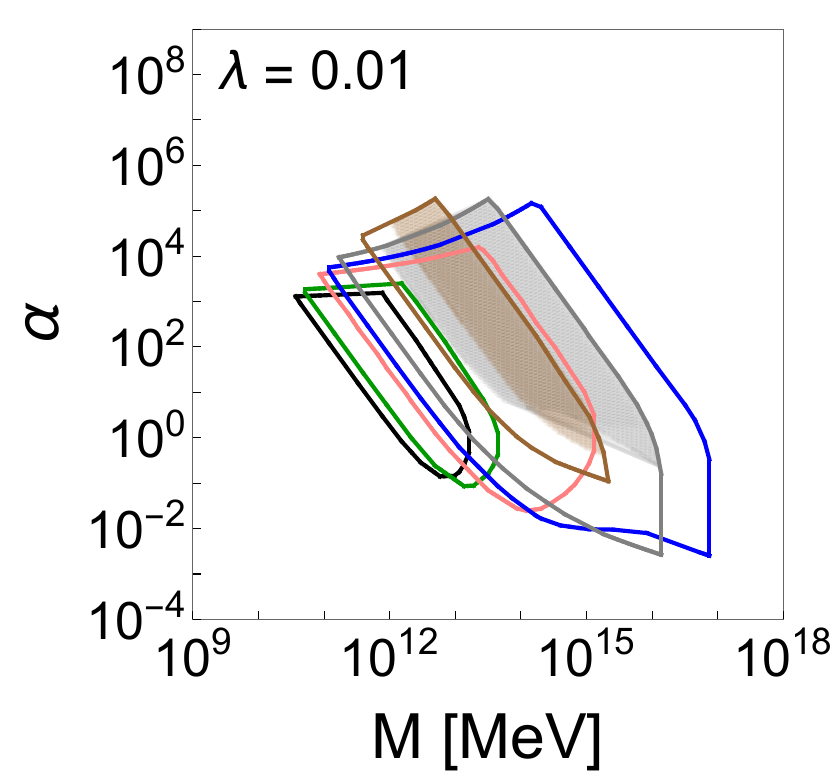} 
\includegraphics[width=.32\textwidth]{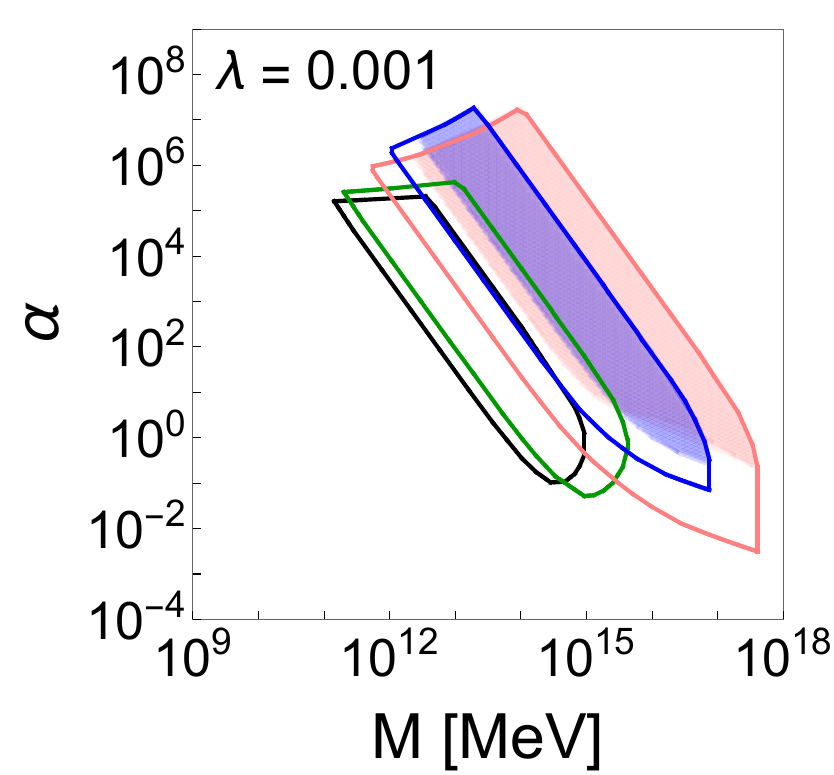}
    \\
\includegraphics[width=.32\textwidth]{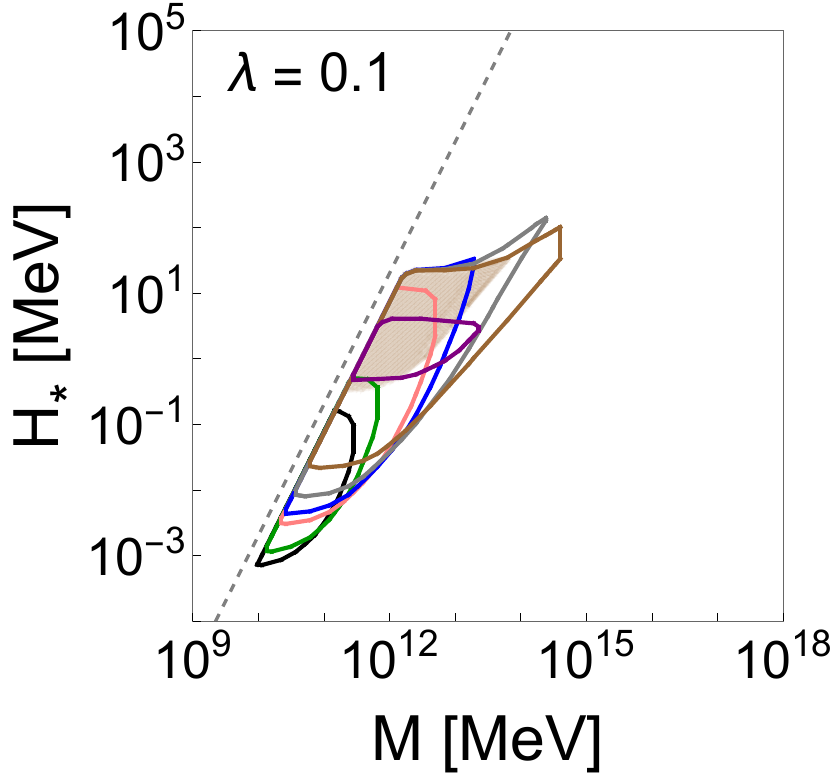} \includegraphics[width=.32\textwidth]{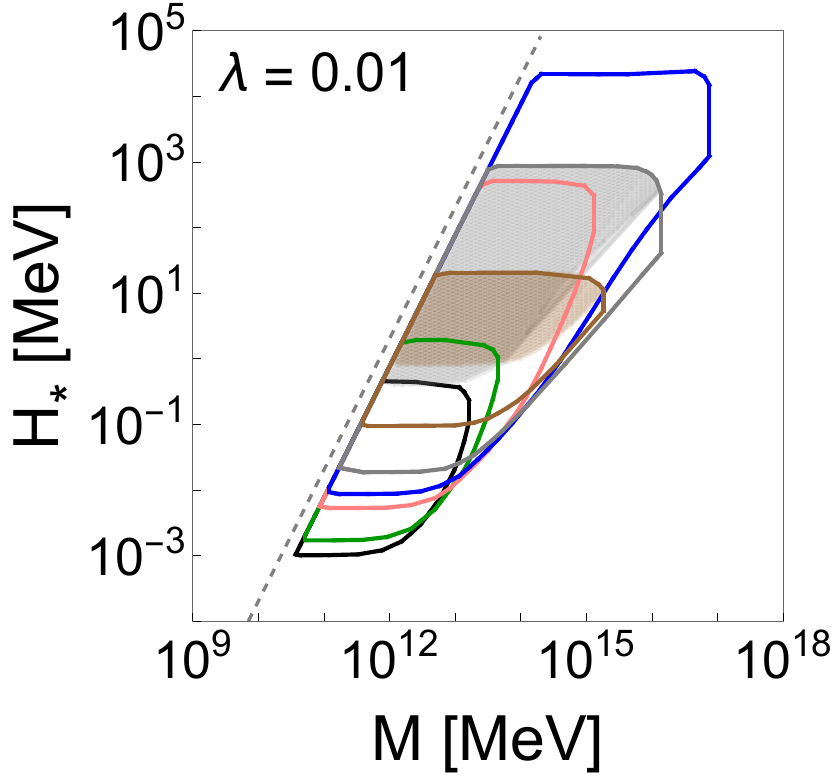}   \includegraphics[width=.32\textwidth]{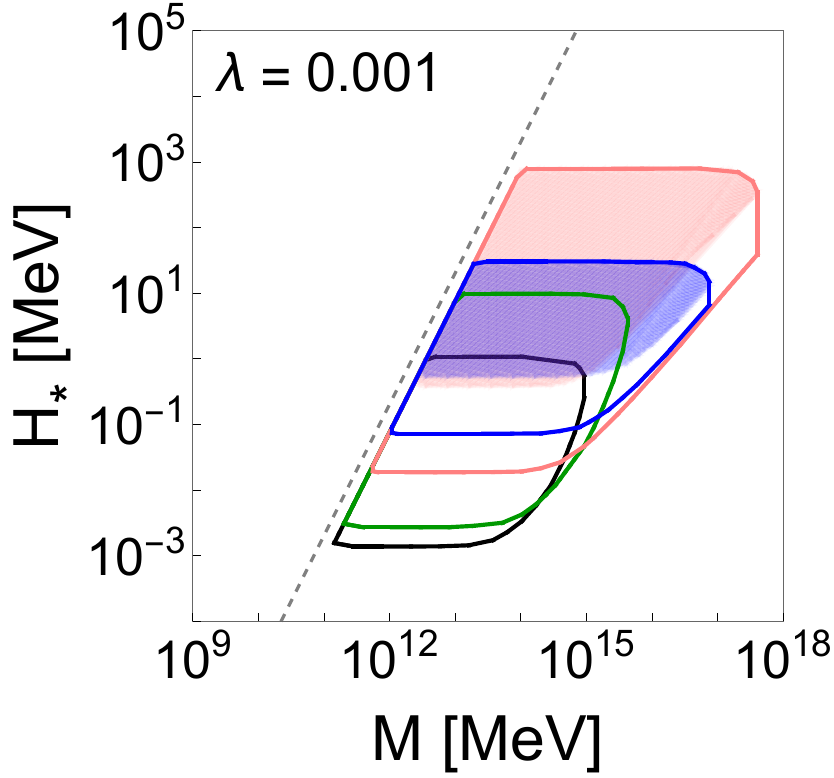}

\caption{CE: The contours on the $M-r$ plane (upper), the $M-\alpha$ plane (middle) and the $M-H_*$  plane (lower), inside which the parameters lead to GW spectra with the maximum inside the Cosmic Explorer (CE) band. The self-coupling is chosen as $\lambda=0.1,0.01,0.001$ (left to right). The different colors correspond to different values of $m_\chi/\mu $ as indicated in the legend.
{The shaded contours correspond to the parameter space that is excluded through the non-detection of stochastic GWs by aLIGO-VIRGO-KAGRA, shown in Figure~\ref{fig:aLIGO}.}}
\label{fig:CE}
\end{figure}

Summing up, CE aims for significantly improved sensitivity in the LIGO frequency band, reaching gravitational-wave energy densities of order $\Omega_{GW} \sim 10^{-13}$ for frequencies around $10\,\mathrm{Hz}$. Since CE operates in the same frequency range as aLIGO but with better sensitivity, it can probe a broader parameter range, including cases with $m_\chi/\mu = 0$, and access smaller values of the energy ratio  $\alpha$ and correspondingly of the parameter $r$. 
We find that  CE can probe FOPTs at Hubble scales $10^9\, {\rm {MeV}}\lesssim H_* \lesssim 10^{18}\, {\rm {MeV}} $ and mass parameters $10^{9} \,{\rm {MeV}}<M <10^{18} \,{\rm {MeV}} $ (for  $0.001\le \lambda\le 0.1$), as shown in  Figure~\ref{fig:CE}.

\subsubsection{Big Bang Observer (BBO)}

Before moving to the recent PTA results, we must discuss the final class of proposed GW experiments, frequently dubbed ``mid-band" detectors, since they are designed to probe deci-Hz frequencies, falling between those accessible to LISA and LIGO. The two proposals, DECIGO and BBO, share similar sensitivity curves, with BBO expected to reach lower values of $\Omega_{GW}={\cal O}(10^{-15})$, allowing us to probe even lower values of $r$ and $\alpha$ in our model. 
Figure~\ref{fig:BBO} shows the  parts of the parameter space that is accessible to BBO, based on the amplitude and frequency of the peak of the predicted GW spectrum. We can immediately see the qualitative similarities and quantitative differences with LISA (and also CE). Since BBO can detect GW signals with $\Omega_{GW}h^2 <10^{-12}$, it can detect model realizations with $m_\chi=0$, similar to LISA and CE. Because at $\Omega_{GW}h^2 \sim10^{-12}$ the frequency band covered by BBO is larger than other experiments, we expect the width of the observable parameter space in terms of $M$ to also be larger. This can be immediately seen by comparing the black contours on the $M-r$ plane between Figures~\ref{fig:BBO} and \ref{fig:LISA} for BBO and LISA respectively. In addition to the $m_\chi/\mu=0$ band being wider in BBO than in LISA, it can also cover smaller values of $r$, since the BBO sensitivity curve reaches GW amplitudes of $\Omega_{GW}h^2 \simeq 10^{-15}$. Remembering the relation between $r$ and the energy ratio $\alpha \approx 0.1 r^2/\lambda^2$ leads to the minimum value of $\alpha$ and $r$ for $m_\chi/\mu=0$ to be $\alpha\sim 0.02$ and $r \sim 0.5 \lambda$ respectively. These values agree with the numerical results of Figure~\ref{fig:BBO} in both the $M-r$ and $M-\alpha$ planes for all three values of $\lambda$ that we simulated. As before, the $m_\chi/\mu=1$ contour is almost identical to its $m_\chi/\mu=0$ counterpart. 
We can also compute the smallest accessible value of the energy ratio $\alpha_{min}$ by considering the simple relation $2\times 10^{-7} \alpha^2 \simeq 10^{-15}$, leading to $\alpha_{min}\sim 10^{-4}$. For $m_\chi/\mu>1$ this translates to a corresponding smallest accessible value of $r_{min} \simeq 0.003 \lambda$. Both these values are verified numerically in Figure~\ref{fig:BBO}.
Finally we see that the $M-H_*$ plots can be easily sketched using Eq.~\eqref{eq:LISAHubble} and the fact that the resulting (rough) trapezoid shifts by $M\propto 1/\sqrt{\lambda}$ and grows in width as $M\propto 1/\lambda$, as $\lambda$ gets smaller. These simple scalings, coupled with the frequency estimate given  in Eq.~\eqref{eq:freqwithr} allow us to quickly estimate the relevant parameter space for each GW detector.

To summarize  we note that 
BBO’s sensitivity down to $\Omega_{\rm GW} h^{2} \sim 10^{-15}$ allows it to access the largest 
and deepest region of parameter space among all detectors, 
extending the accessible range of both $r$ and $\alpha$ well beyond what is possible for LISA or CE. 
In particular, BBO can detect 
 model realizations with $m_\chi/\mu = 0$, where we find that BBO can reach values as small as $\alpha \sim 0.02$ and 
$r \sim 0.5\,\lambda$, while for $m_\chi/\mu > 1$ the minimum accessible values are 
$\alpha_{\min} \sim 10^{-4}$ and $r_{\min} \sim 0.003\,\lambda$. In addition, BBO’s broader 
frequency coverage at fixed GW amplitude implies a correspondingly wider range of $M$, a feature 
clearly visible when comparing the $M$--$r$ and $M$--$\alpha$ planes across Figures~\ref{fig:BBO} and~\ref{fig:LISA2}. 
In particular, for our range of $0.001\le \lambda\le 0.1$, BBO can access transitions occurring at  $10^{-12} \, {\rm {MeV}} < H_* \lesssim 10^{-2} \, {\rm {MeV}}$, corresponding to mass-scales $10^5 \, {\rm {MeV}} < M < 10^{14} \, {\rm {MeV}}$.
 
\begin{figure}
\centering
    \centering
    \includegraphics[width=.32\textwidth]{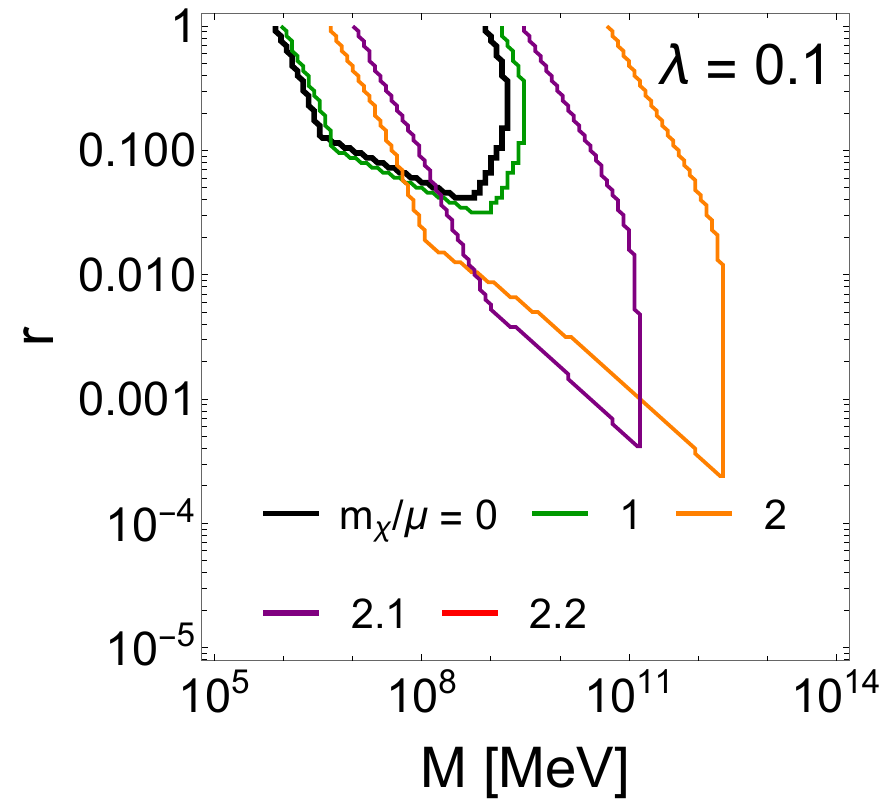}
\includegraphics[width=.32\textwidth]{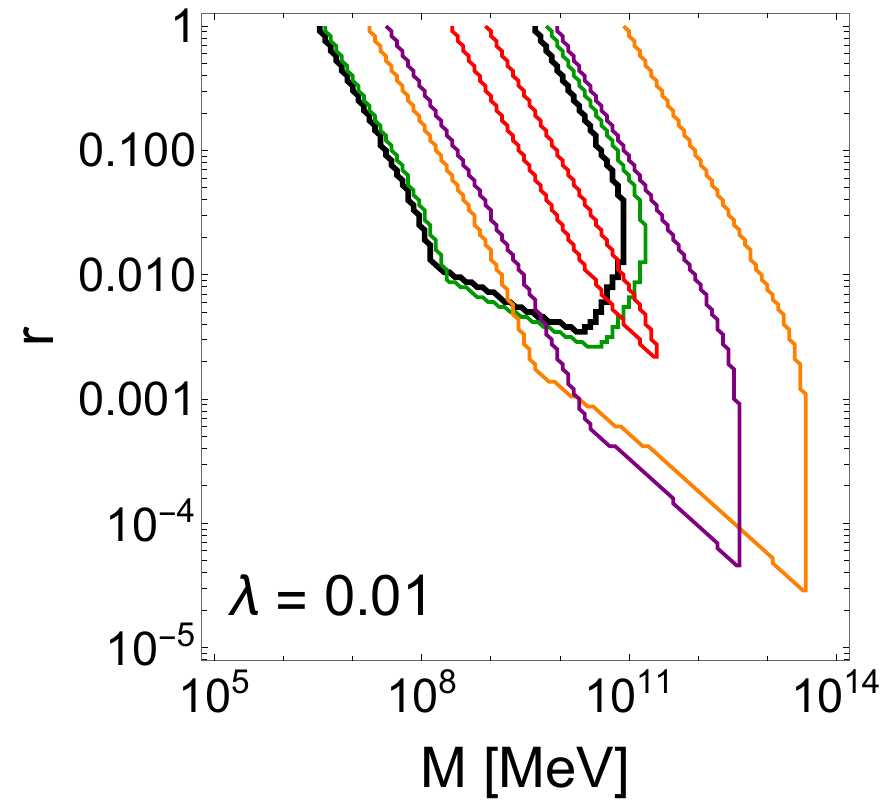} 
\includegraphics[width=.32\textwidth]{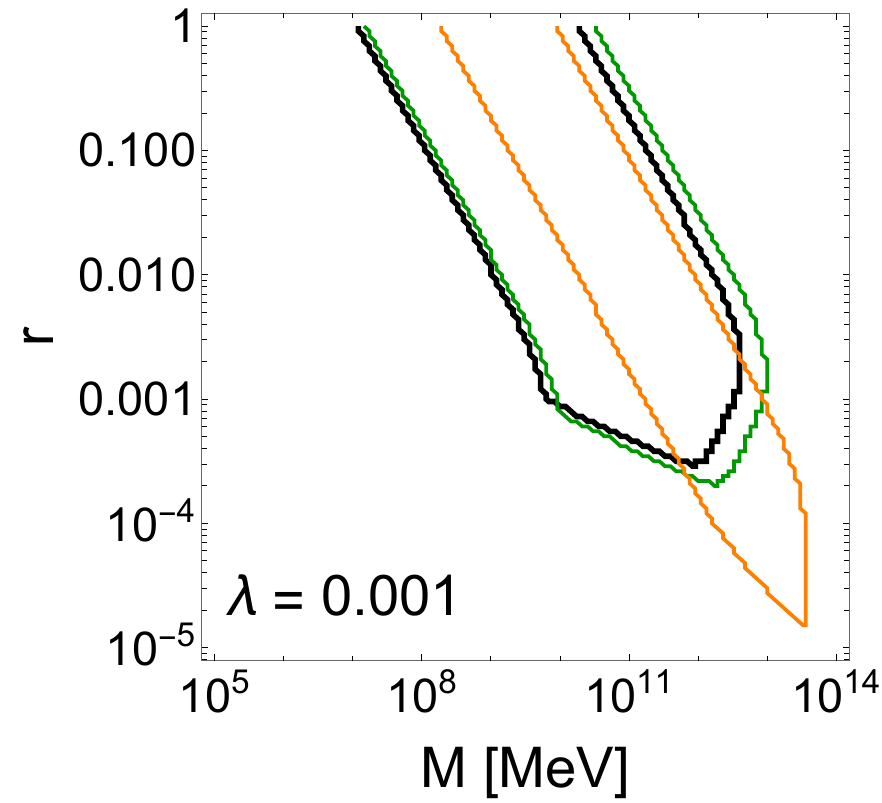}
    \\
\includegraphics[width=.32\textwidth]{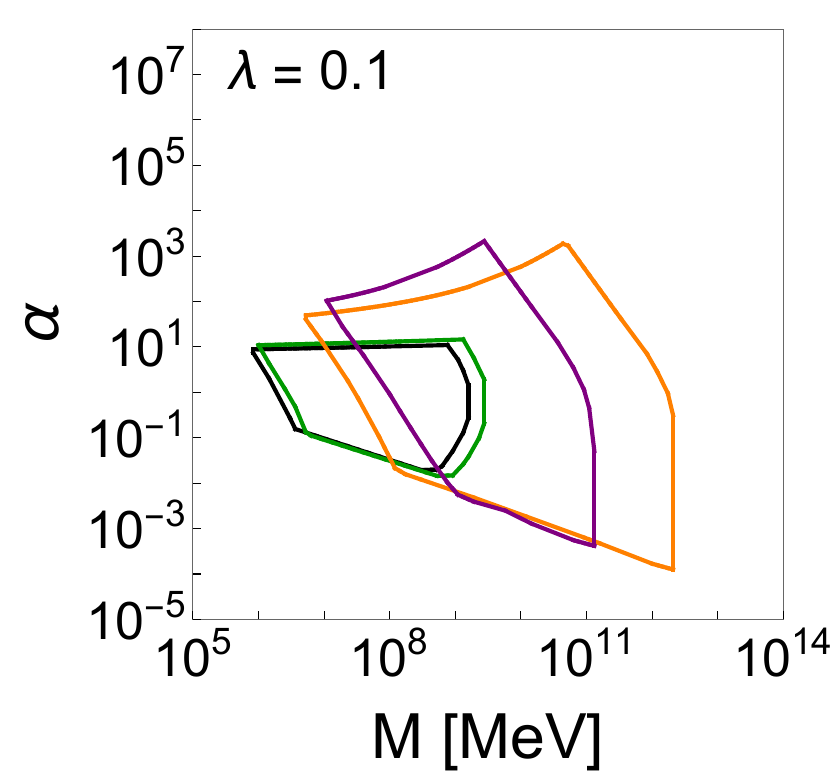}
\includegraphics[width=.32\textwidth]{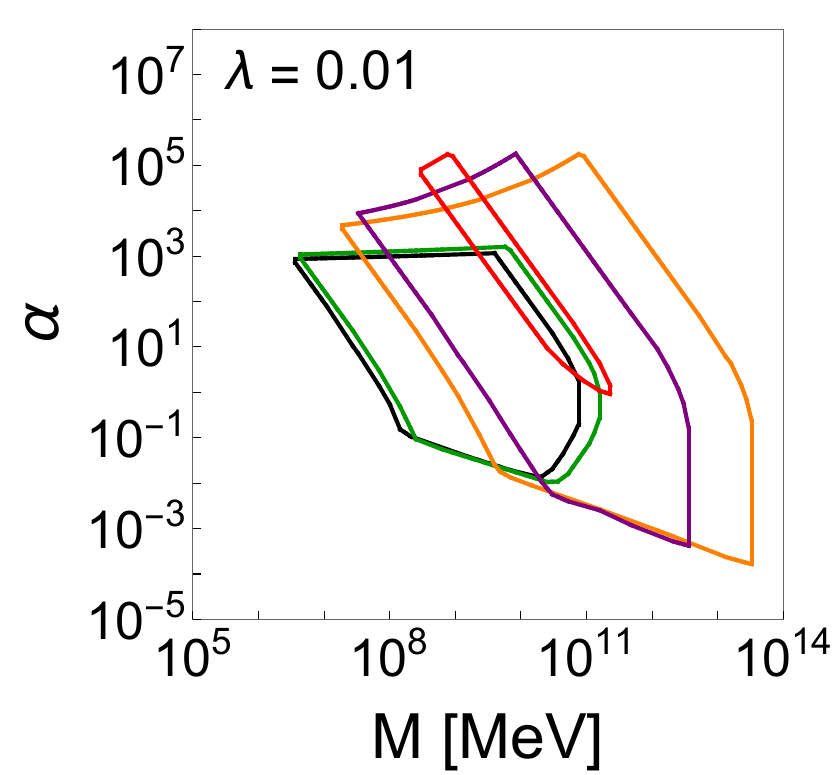} 
\includegraphics[width=.32\textwidth]{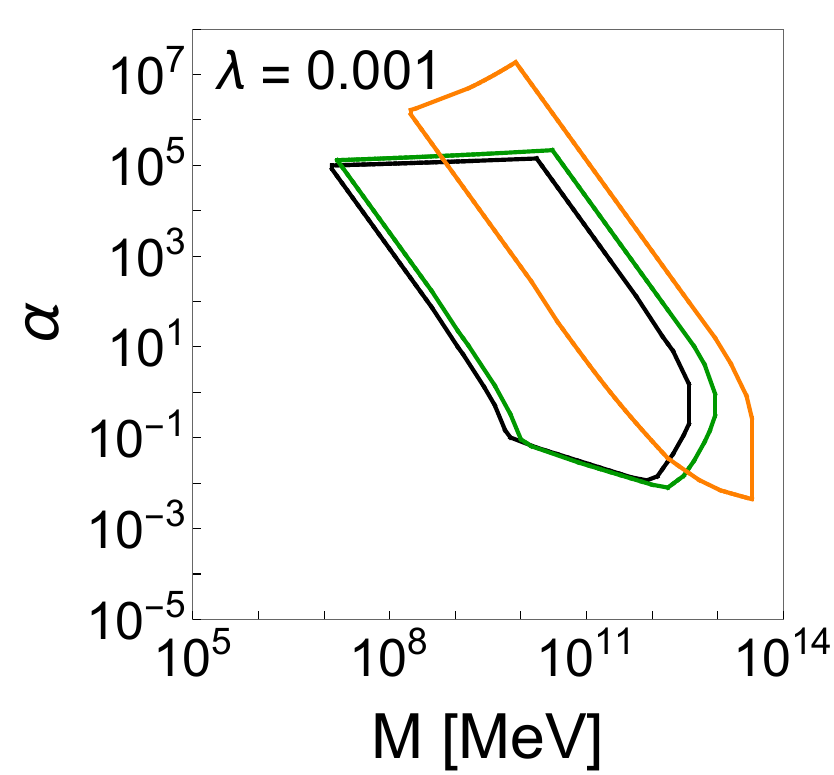}
    \\
\includegraphics[width=.32\textwidth]{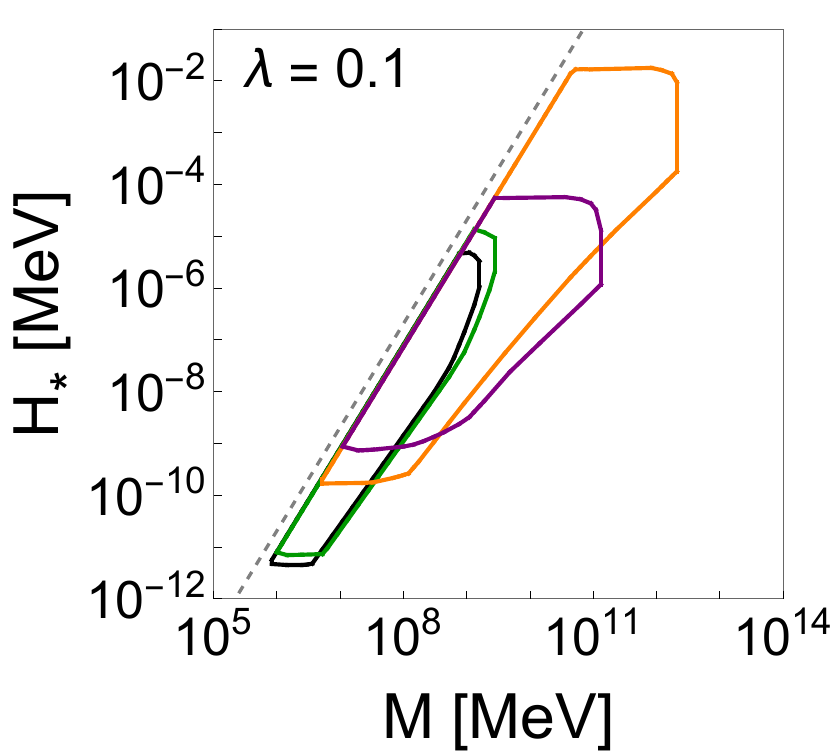} \includegraphics[width=.32\textwidth]{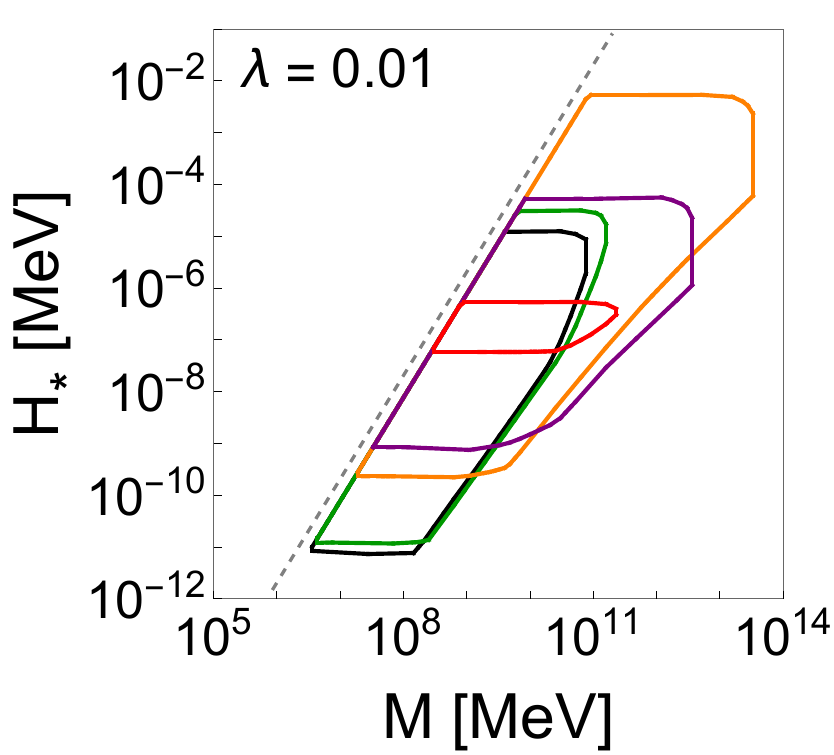}   \includegraphics[width=.32\textwidth]{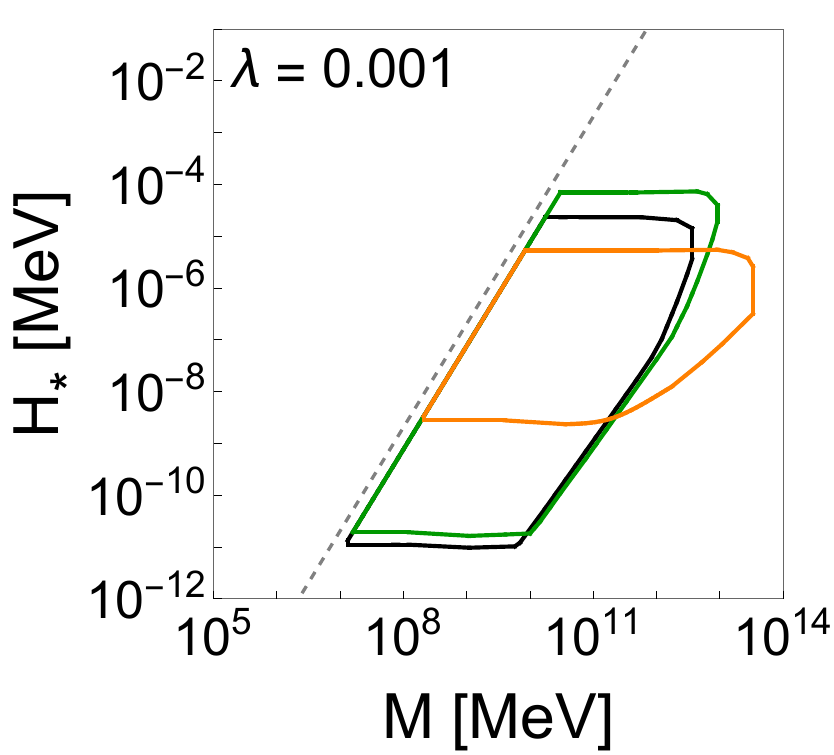}

\caption{BBO: The contours on the $M-r$ plane (upper), the $M-\alpha$ plane (middle) and the $M-H_*$  plane (lower), inside which the parameters lead to GW spectra with the maximum inside the Big Bang Observer (BBO) band. The self-coupling is chosen as $\lambda=0.1,0.01,0.001$ (left to right). The different colors correspond to different values of $m_\chi/\mu $ as indicated in the legend.
The dashed line in the lower panels corresponds to $H_*=0.5\lambda M^2/M_{\rm {Pl}}$, which is the limiting case of Eq.~\eqref{eq:LISAHubble} for $r=1$.
}
\label{fig:BBO}
\end{figure}

\subsection{Pulsar Timing Array results}
\label{sec:PTA}

Having quantified the parts of parameter space of our model that are discoverable with GW interferometers, current and future, we turn our attention to the GW signal reported by the pulsar timing array (PTA) experiments such as NANOGrav~\cite{NANOGrav:2023gor, NANOGrav:2023hvm}, EPTA+InPTA~\cite{EPTA:2023fyk, EPTA:2023xxk}, PPTA\cite{Reardon:2023gzh}, and CPTA\cite{Xu:2023wog}. 
{All PTA data point towards a  stochastic signal in the nanohertz gravitational-wave band ($\sim10^{-9}-10^{-7}$ Hz). 
The most commonly proposed  explanation for the observed low-frequency gravitational-wave (GW) background is the collective emission from a large population of coalescing supermassive black-hole (SMBH) binaries~\cite{Haehnelt:1994wt, Rajagopal:1994zj, Jaffe:2002rt, Wyithe:2002ep, Sesana:2008mz, Burke-Spolaor:2018bvk, Middleton:2020asl}. SMBHs, which reside in the centers of most galaxies, can form bound pairs following galaxy mergers. Once sufficiently close, these binaries efficiently lose orbital energy through GW emission, causing their separation to shrink until they eventually merge. For this to occur, the binary must first reach separations on the order of $0.1–0.001$ pc \cite{Begelman:1980vb}. However, no processes have been firmly established through which the BHs  can shed orbital energy when their relative distance is $\sim 0.1-1$ pc. 
 This challenge is commonly referred to as the ``final-parsec problem”~\cite{Milosavljevic:2002ht}.
Several ideas have been proposed to alleviate this bottleneck. These include the existence of more efficient stellar
relaxation mechanisms~\cite{Quinlan:1997qe, Yu:2001xp, Zhao:2001py, Milosavljevic:2002bn, Merritt:2003pf, Berczik:2006tz} or the formation of an accretion disc which can remove orbital energy from the BH binary when their distance falls below  $\sim 1$ kpc~\cite{Gould:1999ia, Begelman:1980vb}.  Such proposals require more detailed investigation and,  consequently, the expected amplitude of the GW background from SMBH mergers remains uncertain.
Given this uncertainty, a variety of alternative origins for the stochastic GW background have been considered. 
 Ref.~\cite{Winkler:2024olr} compared the fit to  the PTA data using a FOPT spectrum and the merging BH scenario and found a mild preference for the former.  
}

We will show that our model is capable of fitting the PTA results and provide the relevant parameter space. 
We take a different strategy than the one employed for LISA and aLIGO, since in the case of the PTA signal, the experiments have provided information about the strength of the GW signal in several frequency bins. Previously in this paper, we have focused on the peak amplitude and frequency of the GW signal.
In this subsection, we turn instead to the spectrum of GW predicted by FOPT and compare to the PTA data.  We consider only the NANOGrav for simplicity, but using any other experiment would give similar results (see e.g.~\cite{Winkler:2024olr}). We consider the amplitude of a GW spectrum $\Omega_{GW}$ as a function of the frequency $f$ arising from the generation and collision of thick-wall bubbles~\cite{Ellis:2024PTA, Winkler:2024olr} to have the triple power-law shape
\beq
\Omega_{GW}(f)  = \Omega_{GW}^{\rm{peak}} \times   \left(\frac{2.9(f/f_{\rm{peak}})^{0.7}}{2.2 + 0.7(f/f_{\rm{peak}})^{2.9}}\right)\times\left( 1 +\left(2\pi\frac{0.2\beta}{H_*}\frac{f}{f_{\rm{peak}}}\right)^{-2.3}\right)^{-1} \,
\label{eq:spectrum}
\eeq
where $\Omega_{GW}^{\rm{peak}}$ and $f_{\rm {peak}}$ are given in Eqs.~\eqref{eq:omegaGW} and \eqref{eq:frequency} respectively.
Note that at low frequencies $f\ll f_{\rm {peak}}$ the GW spectrum scales as $\Omega_{GW}\propto f^{3}$, as dictated by causality arguments.  At large frequency $f\gg f_{\rm {peak}}$ the spectrum drops as $f^{-2.2}$. The thick-wall bubble approximation introduces   a third ``knee" to the GW spectrum in frequency space at $f_{\rm{knee}} \sim f_{\rm {peak}}/(0.4 \pi  \beta/H_*)$. Between $f_{\rm{knee}}$ and $f_{\rm {peak}}$ the spectrum grows approximately as $f^{0.7}$. We must note that for $\beta/H_*\gtrsim 3$, the above formula introduces a correction to the peak amplitude a the few percent level, which is insignificant for all practical purposes.

We compare the predicted GW spectra for different model parameters against the NANOGrav data using two different criteria.
For the most conservative case, 
we only accept a specific GW spectrum if it goes through the $1\sigma$ bounds of all $15$ frequency bins of NANOgrav~\cite{NANOGrav:2023gor}. We then relax this requirement by considering the $2\sigma$ bounds. 
Figure~\ref{fig:NANOGrav_test} shows the difference in the allowed parameter space for fixed values of $\lambda$ and $m_\chi/\mu$ based on these two criteria, as well as different GW spectra arising in the two cases.  The shrinking of the allowed parameter space between our two criteria ($1\sigma$ and $2\sigma$)  demonstrates how a reduction of the experimental error-bars will help to extract model parameters with better accuracy.
Another  criterion for matching predictions to data would be the computation of the $\chi^2$ fit for each spectrum and the inclusion of only those parameter sets that exceed a predefined value. This increases computational cost without altering the basic picture of the results or building better physical intuition. It is the purpose of this paper to obtain predictions of our model for   a broad range of experiments, showing  the parameter space accessible to experiments in the different frequency bands.  A detailed fit of our model to the PTA data is outside  the scope of this work, but can be performed by combining the understanding developed here with the statistical analysis of Ref.~\cite{Winkler:2024olr}.
In what follows, we will require that the GW spectrum given by Eq.~\eqref{eq:spectrum} goes through all $2\sigma$ bins of NANOGrav.

\begin{figure}
\centering
\includegraphics[width=.36\textwidth]{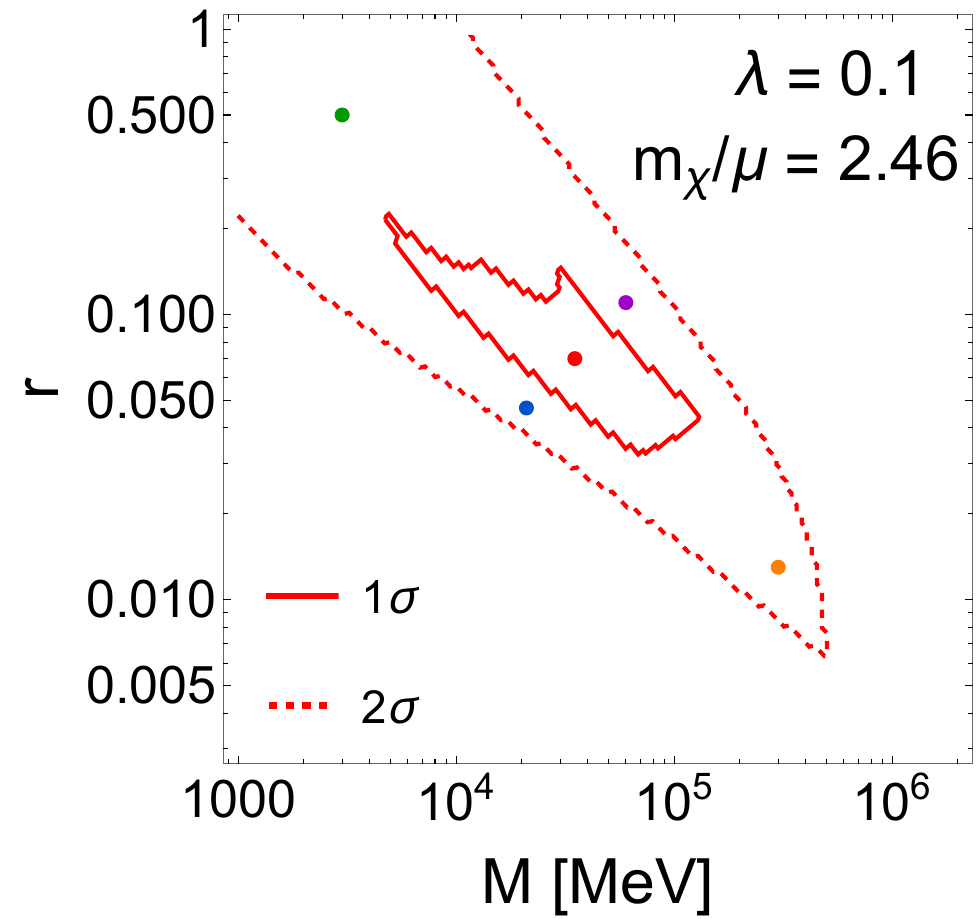}
  ~~  
\includegraphics[width=.53\textwidth]{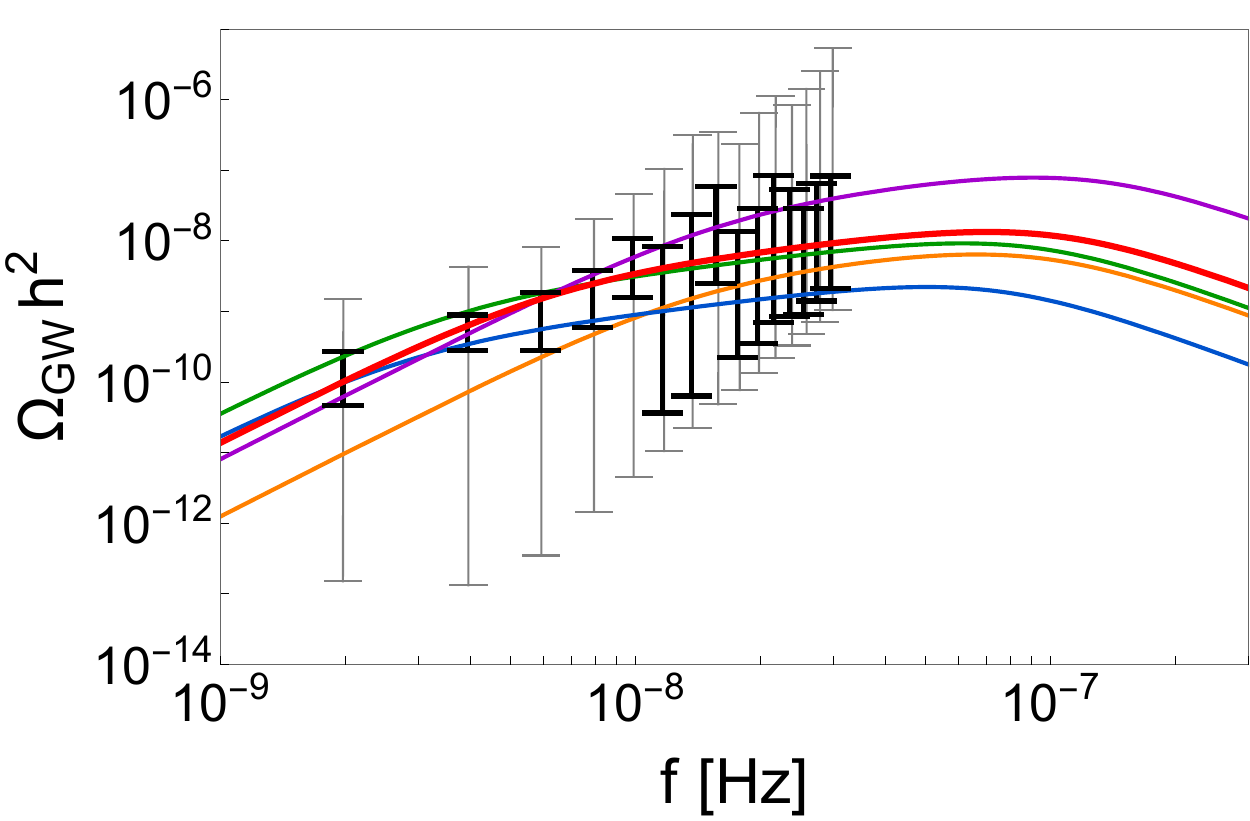} 
\caption{NANOGrav:
{\it Left:} The contours in the $M-r$ plane for $\lambda=0.1$ and $m_\chi/\mu=2.46$, inside which the parameters lead to GW spectra that pass through either all $1\sigma$ or $2\sigma$  bins of NANOGrav. In particular, points inside the solid (inner) contour lead to GW spectra that pass through all $1\sigma$ bins, whereas points inside the dashed (outer) contour lead to GW spectra that pass through all $2\sigma$ bins. 
The contours are computed by using a square grid of values $(M,r)$ and tracing the points that lead to viable GW spectra in light of PTA data. The discretization is evident in the shape of the contour, especially for the solid curve. The colored dots are representative points inside one or both contours.
{\it Right:} GW spectra corresponding to the colored dots of the left panel. We see that the red curve  passes through all $1\sigma$ bins, as it corresponds to the red dot, positioned inside the inner contour of the left panel. All  other curves pass either below or above some of the $1\sigma$ bins, while remaining inside the $2\sigma$ ones.
 }
\label{fig:NANOGrav_test}
\end{figure}

The full parameter scan is shown in  Figure~\ref{fig:NANOGrav}.
We choose the same scanning strategy as before, where we choose three values of $\lambda=0.1,0.01,0.001$ and for each one present the  parameter range that leads to NANOGrav-compatible spectra on the $M-r$, $M-\alpha$ and $M-H_*$ planes, for a discrete set of values of $m_\chi/\mu$.
The mass-scale $M$ that can account for this signal can be as low as  $1$ GeV (in order to safely allow for BBN to proceed after the FOPT) or approach $10^6$ GeV, depending on the choice of $\lambda \in [0.001,0.1]$.
We see that the large amplitude of the peak GW amplitude required to match the data leads to a narrow allowed range for $m_\chi/\mu \sim 2.5$. This is actually a feature of this model, where the detection of a large stochastic GW signal ($\Omega_{GW}>10^{-12}$) allows us to determine the mass ratio $m_\chi/\mu$ very accurately.

In addition, there are restrictions on the value of $\alpha$, due to the fact that the GW amplitude is suppressed as $\alpha^2$ for values of $\alpha<1$.  Too much suppression would push the GW spectrum below the detected values for several of the bins. The highest frequency bin requires $\Omega_{GW} h^2\gtrsim 10^{-9}$ and thus we expect  that $\alpha \gtrsim 0.1$. This is verified in the middle  row of panels of Figure~\ref{fig:NANOGrav}. The range of $\alpha$ leads to the corresponding allowed range  $r\gtrsim 0.1\lambda$. Finally, let us point out a qualitative difference between the bands shown in Figure~\ref{fig:NANOGrav} and their counterparts computed for future experiments examined in Section~\ref{sec:experiments}. For the latter, we required simply that the peak GW amplitude be above each experiment's sensitivity curve. 
Hence the value $r=1$ is allowed in all cases. However, this is no longer true in the case of the PTA data,
where the computed signal cannot be above the detected one.
 For example, for the largest value of $m_\chi/\mu$ shown in the blue curves in all panels of Figure~\ref{fig:NANOGrav}, the resultant GW signals 
with peak amplitudes  reaching $\Omega_{GW}h^2\sim 10^{-7}$ can only satisfy the NANOGrav bins if they have a suppression factor $\alpha<1$ and correspondingly $r<1$. This is shown by the blue curves in all panels of Figure~\ref{fig:NANOGrav}.

Summing up, a kinetically-induced FOPT can explain the recent NANOGrav data for a wide range of the mass parameter $10^3\, {\rm {MeV}}\lesssim M\lesssim  10^9\, {\rm {MeV}}$ (for $0.001\le \lambda\le 0.1$) and a highly restricted mass ratio around $m_\chi/\mu\sim 2.5$. This shows how the detection of a large signal ($\Omega^{\rm {peak}}_{GW}h^2 \sim 10^{-9}$ in the PTA case), allows for a sharp parameter extraction for $m_\chi/\mu$ in the context of this model.

\begin{figure}
\centering
    \includegraphics[width=.32\textwidth]{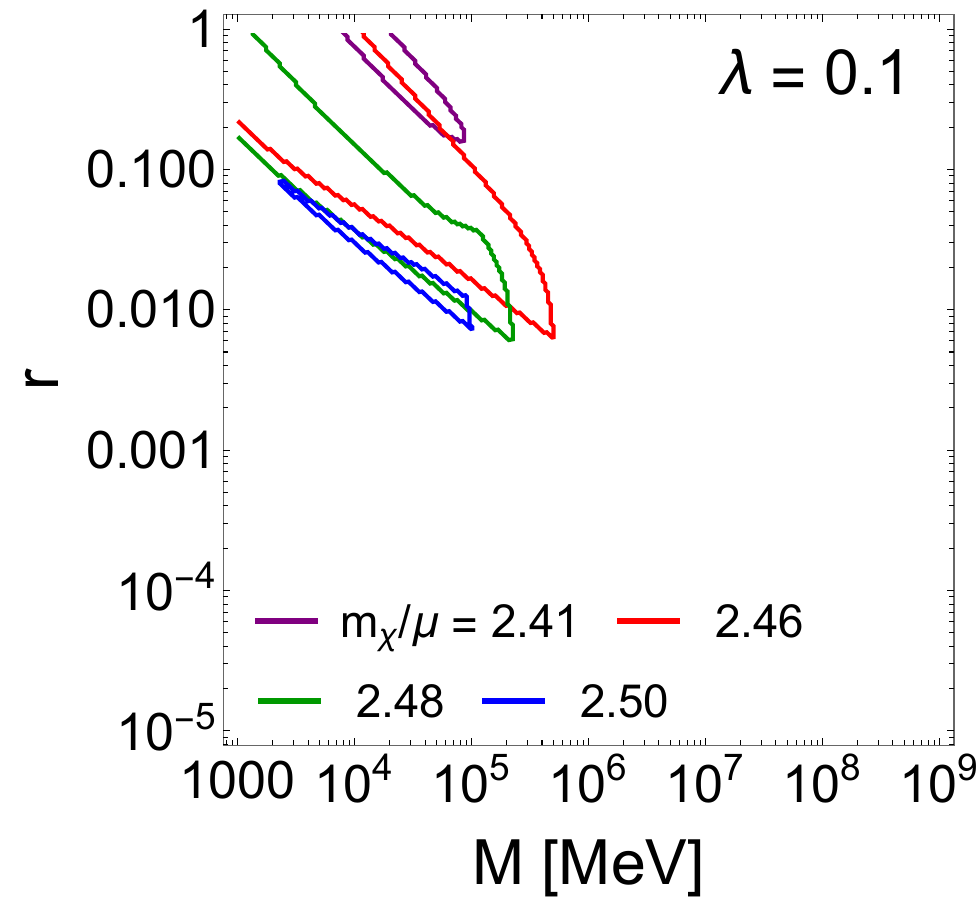}
    \includegraphics[width=.32\textwidth]{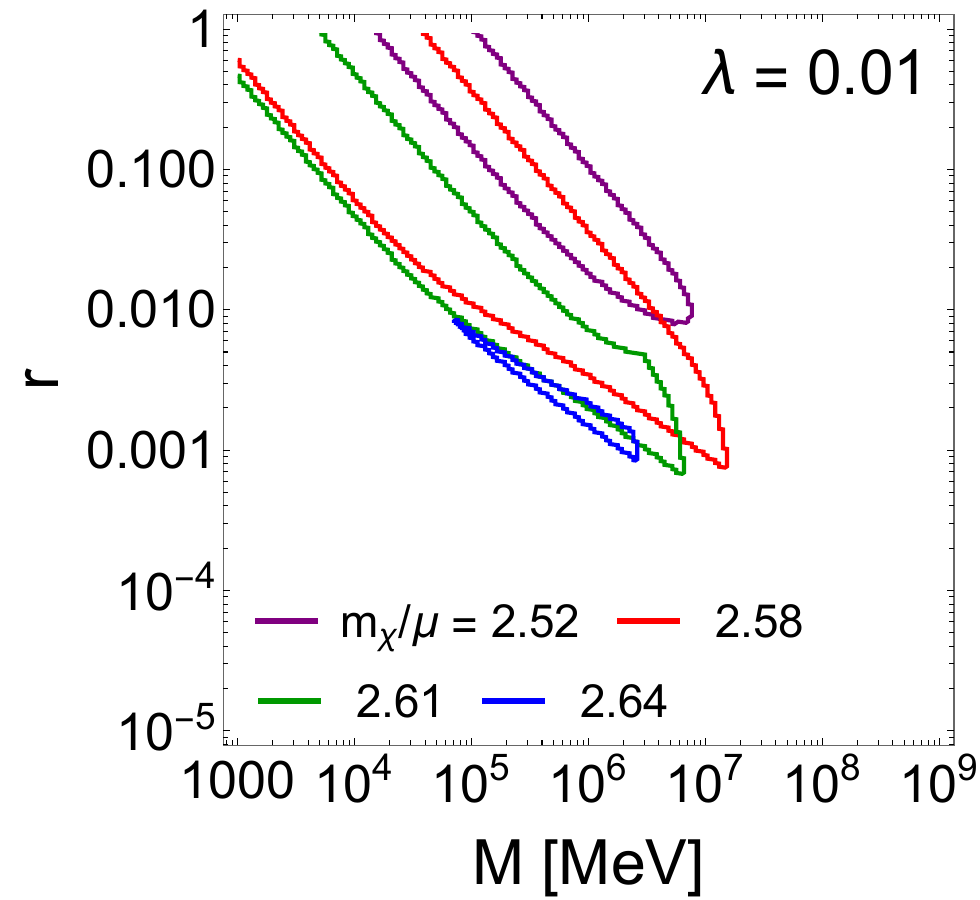}
    \includegraphics[width=.32\textwidth]{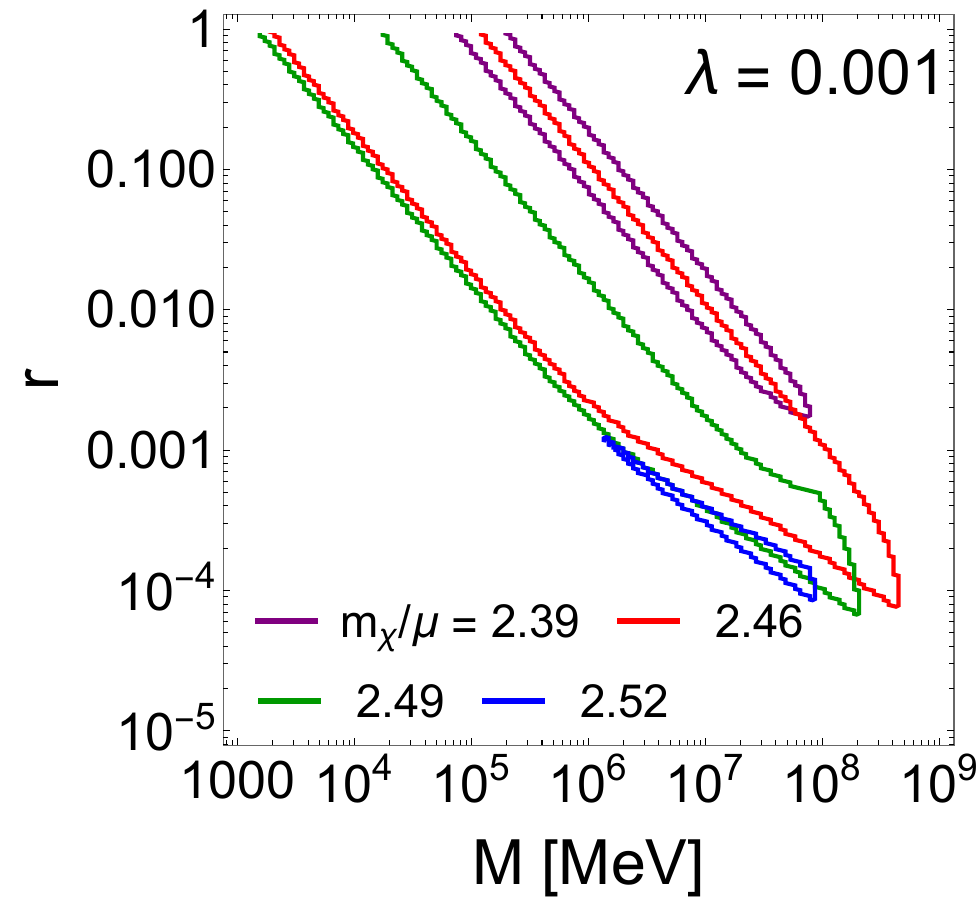}
    \\
    \includegraphics[width=.32\textwidth]{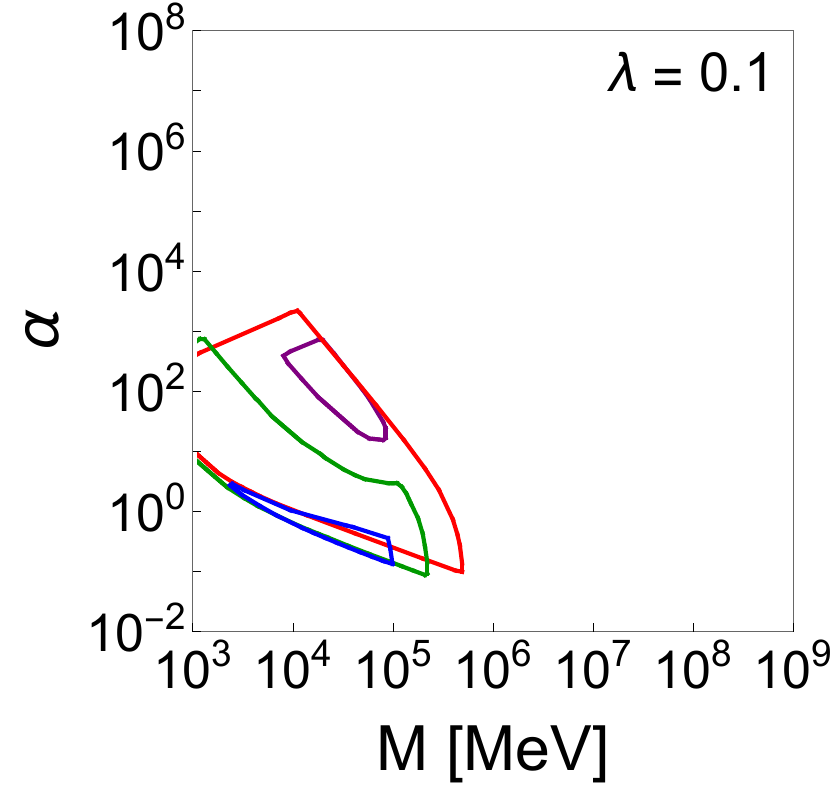}
    \includegraphics[width=.32\textwidth]{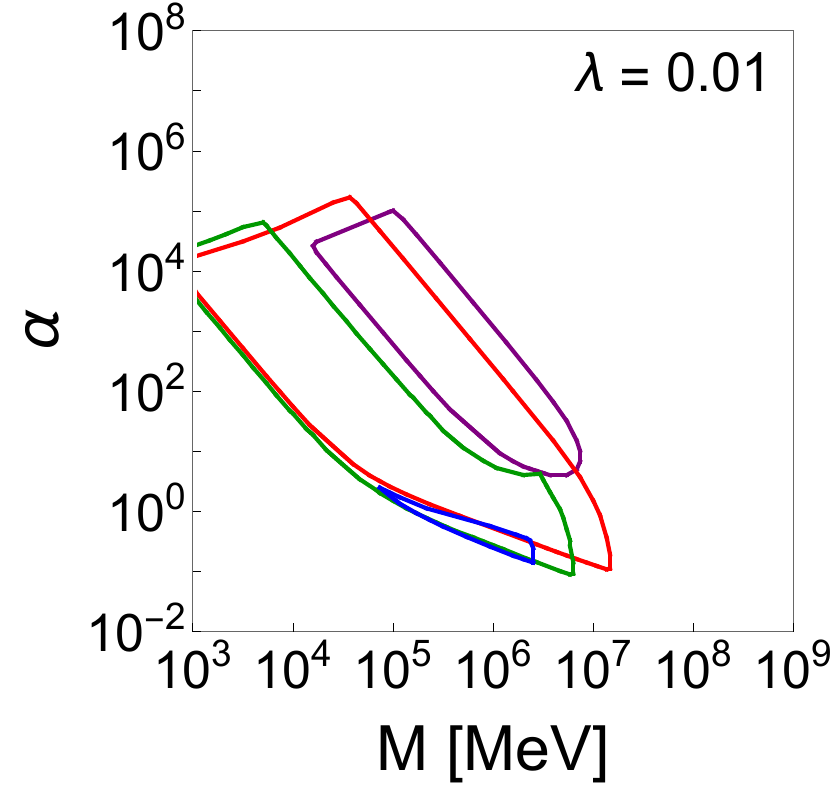}
    \includegraphics[width=.32\textwidth]{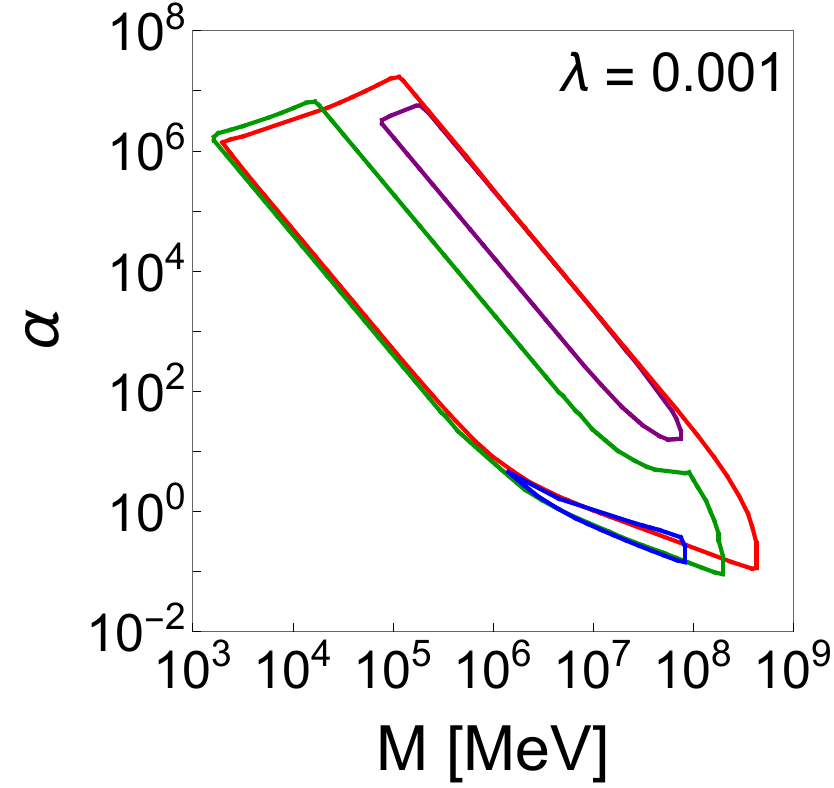}
    \\
    \includegraphics[width=.32\textwidth]{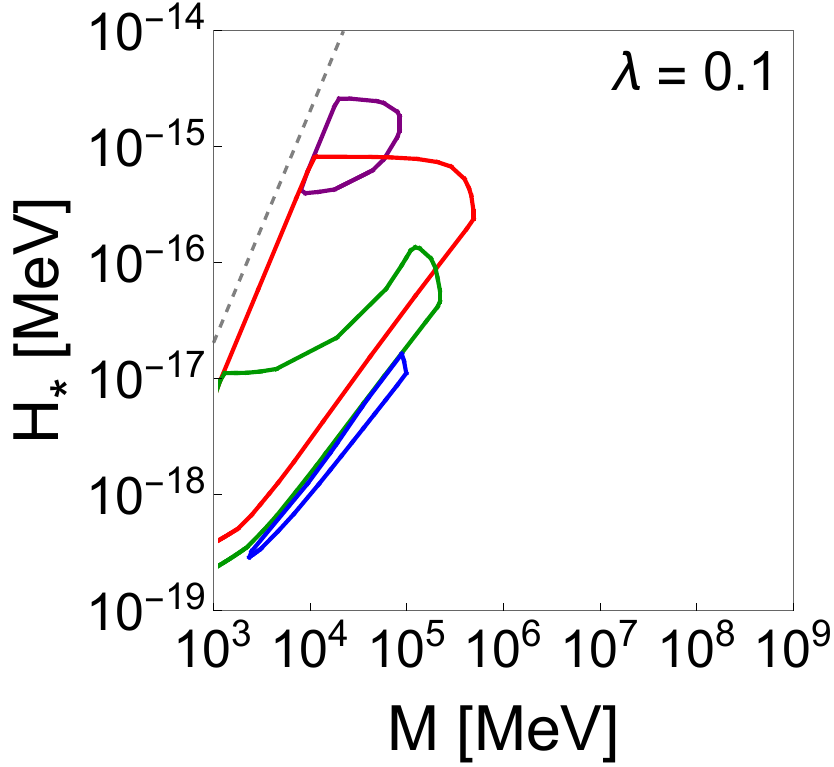}
    \includegraphics[width=.32\textwidth]{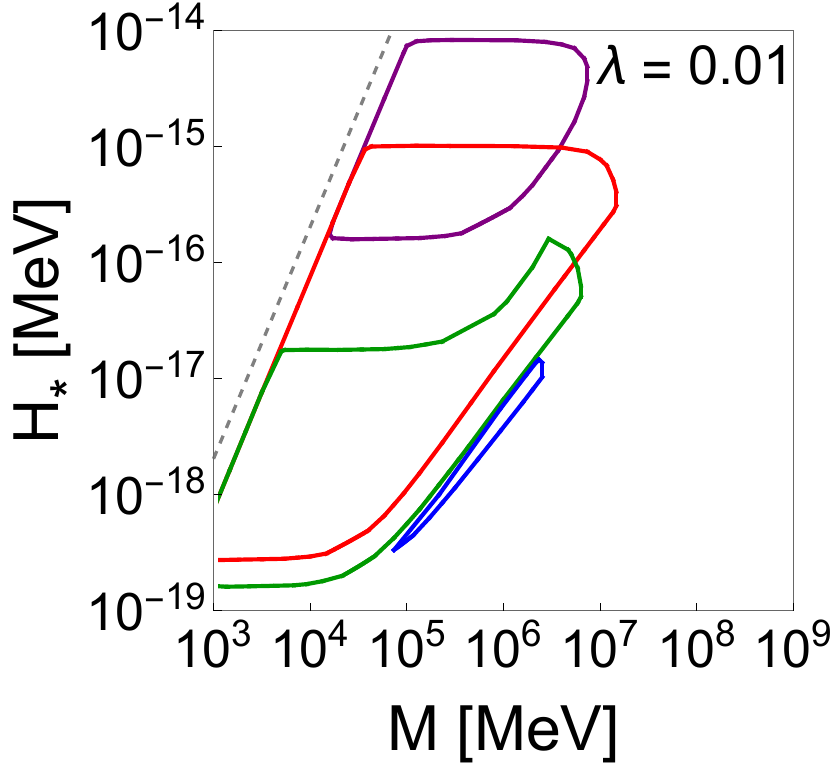}
    \includegraphics[width=.32\textwidth]{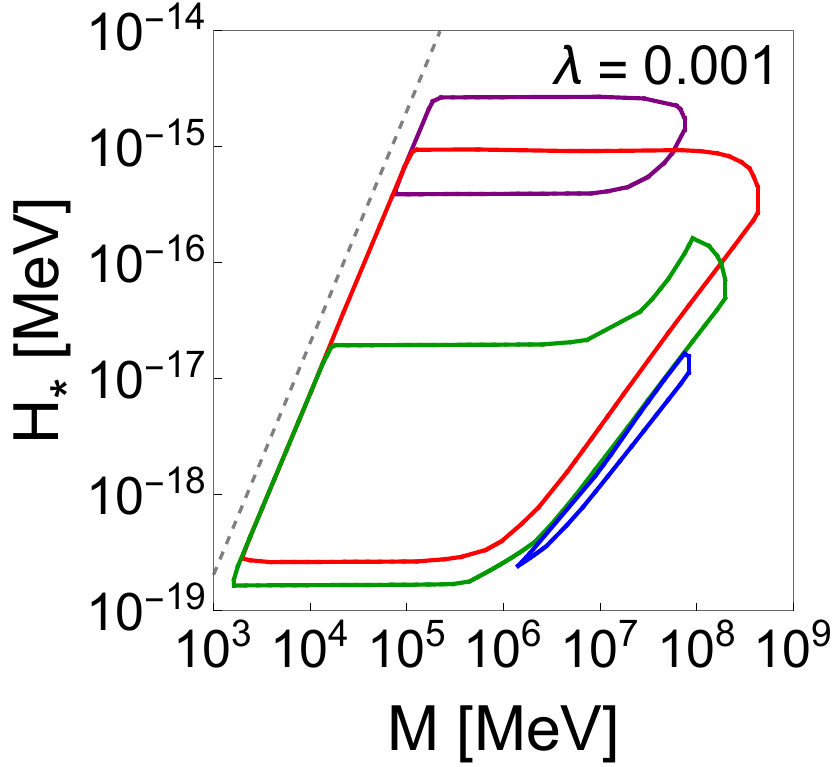}

\caption{NANOGrav:  The contours in the $M-r$, $M-\alpha$ and $M-H_*$ planes (from top to bottom), 
 inside which the parameters lead to GW spectra that pass through all $2\sigma$ bins of NANOGrav.
The self-coupling is chosen as $\lambda=0.1,0.01,0.001$ (from left to right.) The different contours correspond to different values of $m_\chi/\mu$. The chosen values are different for each value of $\lambda$ and span the entire $m_\chi/\mu$ range that fits the $2\sigma$ data, up to 0.01 precision. Specifically the purple, red, green and black curves correspond to $m_\chi/\mu=2.41, 2.46, 2.48, 2.5$ for $\lambda=0.1$,
$m_\chi/\mu=2.52, 2.58, 2.61, 2.64$ for $\lambda=0.01$
and 
$m_\chi/\mu=2.39, 2.46, 2.49, 2.52$ for $\lambda=0.001$, as shown in the top panel of each column. We restricted the values of $M\ge 1 \, {\rm {GeV}}$ to safely account for BBN after the transition.
}
\label{fig:NANOGrav}
\end{figure}

\subsection{Summary  of parameter space accessible to current and upcoming GW experimental searches}
\label{sec:experimentsall}

Having explored the parameter space of the model that is accessible by each class of experiments, from pulsar timing arrays to LISA, mid-band detectors (BBO/DECIGO) and finally kHz experiments (aLIGO and CE/ET), let us take a step back and use the intuition we built to construct an overview of the accessible parameter space of the model. We will only focus on the $M-H_*$ plane, which is a combination of the model parameter $M$ and the cosmological parameter $H_*$.
We consider these two parameters to be especially important when trying to embed a kinetically induced Big Bang into a complete cosmic history of our Universe. The Hubble scale $H_*$ defines the time after which the Universe became radiation dominated and allows us to compute the corresponding reheating temperature $T_* \sim \sqrt{H_* M_{\rm {Pl}}}$ and the redshift through $1+z_* \simeq 1.66\times 10^{20} \sqrt{H_*/{\rm {MeV}}}$. Further, as shown in Eq. (5.3) above, $H_*$ largely determines the peak frequency of the GW as
$f_{\rm peak} \propto \sqrt{H_*}$, within a factor of $\sim 300$.
Similarly, $M$ provides  important twofold microphysical information. On one hand, it can be viewed as the EFT limit, thereby suggesting the energy-scale until which our model is valid and above which new physics is required (for example heavy fields that are integrated out). But there is another important role that $M$ plays: it quantifies the strength of the coupling between the kination and  tunneling sectors, similarly to how the Fermi constant ($G_F\propto 1/m_W^2$) controls the strength of four-fermion interaction that was first used to explain $\beta$ decay. 

Figure~\ref{fig:experimentsall} 
presents the observable region in the $M-H_*$ plane for each experiment, overlaying the contours generated  for all values of $m_\chi/\mu$ and $\lambda \in [0.001,0.1]$. By doing so, we
essentially ``average over" the details of the tunneling potential (which are encoded in $m_\chi/\mu$, $\lambda$ and $r$) and present the results keeping $M$ as the only microphysical parameter. Therefore we single out the novel physical property of our model: the kinetic coupling between the two sectors (kination and tunneling). Simply put, Figure~\ref{fig:experimentsall} tells us, for a FOPT occurring at a specific time in cosmic history, what strength of the kinetic coupling between the two fields is in principle accessible by each experiment, regardless of the details of the tunneling potential. This  provides a simple guiding tool for quickly estimating the observability of a certain realization of this model within a complete framework for the early universe.

Figure~\ref{fig:experimentsall} shows our results for collective accessible parameter space for current and upcoming GW experiments in the $M-H_*$ plane.
The figure can be understood as follows. We first start by noting that all allowed regions of parameter space for any experiment can be roughly described as a trapezoid, bounded from the left by the line $H_* = 0.5 \lambda M^2/M_{\rm {Pl}}$, given in Eq.~\eqref{eq:LISAHubble} and having two edges almost parallel to the $M$-axis. The width scales as $M\sim 1/\lambda$, as we have seen. This is universally true, regardless of any specific GW experiment. 

Furthermore, we can estimate a  typical value of $M$ for each experiment, by identifying the central frequency of an experiment (the frequency where the sensitivity curve reaches the smallest value of $\Omega_{GW}h^2$) and using Eq.~\eqref{eq:freqwithr}. For the rest of this paragraph we take $r=1$ and $\lambda =0.1$ as an example.  For that case, by choosing the central frequency of LISA, BBO and CE as $f_{\rm {peak}}= 3\times 10^{-3}, ~0.1,~ 100$ Hz respectively, we can estimate the corresponding scale $M\sim 10^6,~ 5\times 10^7, ~3\times 10^{10}$ MeV.
We can also estimate the corresponding Hubble scale during the transition through Eq.~\eqref{eq:LISAHubble} as $H_* \sim 2\times 10^{-11}, ~5\times 10^{-8}, ~0.02$ MeV.
These are  shown by the  black dots on Figure~\ref{fig:experimentsall}, which reside on the thick black line, corresponding to Eq.~\eqref{eq:LISAHubble}. Overall, we see that the  dots (that we computed with our analytic estimations) fall within the numerically computed parameter range of their respective experiments, further cementing the usefulness of our analytic understanding for providing quick estimates for all relevant quantities in this model.  As expected, the observable parameter space of the model opens up significantly when we include mass ratios of $m_\chi/\mu>1$ compared to considering only $m_\chi/\mu\ll 1 $, shown by the light and dark regions of Figure~\ref{fig:experimentsall} respectively.

Finally, we can derive a lower  bound on the Hubble scale $H_*$ for a given value of $M$, as shown by the dashed black line in Figure~\ref{fig:experimentsall}.
The functional form of this lower bound  can be easily estimated, using the results of Section~\ref{sec:experiments}. 
In the case $\alpha<1$, which is close to the dashed line of Figure~\ref{fig:experimentsall}, the Hubble scale is dominated by the kinetic energy of the kination field, $H_*\simeq \dot\phi_c^2/3M_{\rm {Pl}}^2$.
To obtain an estimate of the location of the dashed line, we will utilize the results for the 
relevant parameter space for each experiment presented in the figures in Section~\ref{sec:experiments}.
For example, for the BBO experiment, we can look at the results presented in Figure~\ref{fig:BBO}.  
The lower panels show the $H_* - M$ plane; we seek to find the boundary of the allowed parameter space with the highest value of $M$ in those panels.
The upper panels can then give us the value of $r$ for various values of $\lambda$ that corresponds to the maximum $M$ in the lower panels.
We find that for the ``tip" of the BBO parameter space, $r\simeq 0.003\lambda$. Finally, we need to remember how the lowest $r$ or $\alpha$ value was derived: We considered a point with $\beta/H_*=3$ and asked what the maximum suppression of the GW amplitude can be that keeps the signal within the detector basin of detectability. Plugging $\beta/H_*=3$ into Eq.~\eqref{eq:betaeqn} we get
$\dot\phi^2_c \simeq 0.01 M^2\mu^2$.
We now simply use this value of the kination velocity, along with the definition of r in Eq.~\eqref{eq:definer} to derive the relation between the Hubble scale $H_*$ and the Lagrangian parameters
\beq
\label{eq:Hlowerbound}
H_* = {1\over \sqrt{6}M_{\rm {Pl}}}0.1 r \lambda^2 M^2
\simeq 10^{-4} \lambda^3 {M^2\over M_{\rm {Pl}} } \, ,
\eeq
where we used $r\simeq 0.003\lambda$ in the last step, taken from BBO. For $\lambda=0.001$, which is the smallest value we considered, the lower bound arises for the Hubble scale, namely
$H_*\gtrsim 10^{-13} M^2/M_{\rm {Pl}}$. This is the black-dashed line of Figure~\ref{fig:experimentsall}.
 We are thus able to provide a band on the $H_*-M$ plane, inside which this model can generate detectable GW signal for current and future experiments. This band is approximately given as 
\beq
\label{eq:Hbounds}
0.05 {M^2 \over M_{\rm {Pl}}} \gtrsim H_* \gtrsim 10^{-13}{M^2 \over  M_{\rm {Pl}}}
\eeq
for $0.1\ge\lambda\ge 0.001$.

Summing up, the $M-H_*$ plane is a useful tool for visualizing the parameter space of this model that is accessible by current and future GW experiments.  Our analytic arguments allow us to provide a band, given by Eq.~\eqref{eq:Hbounds}, inside which this accessible parameter space lies. 

\begin{figure}
\centering

\includegraphics[width=.8\textwidth]{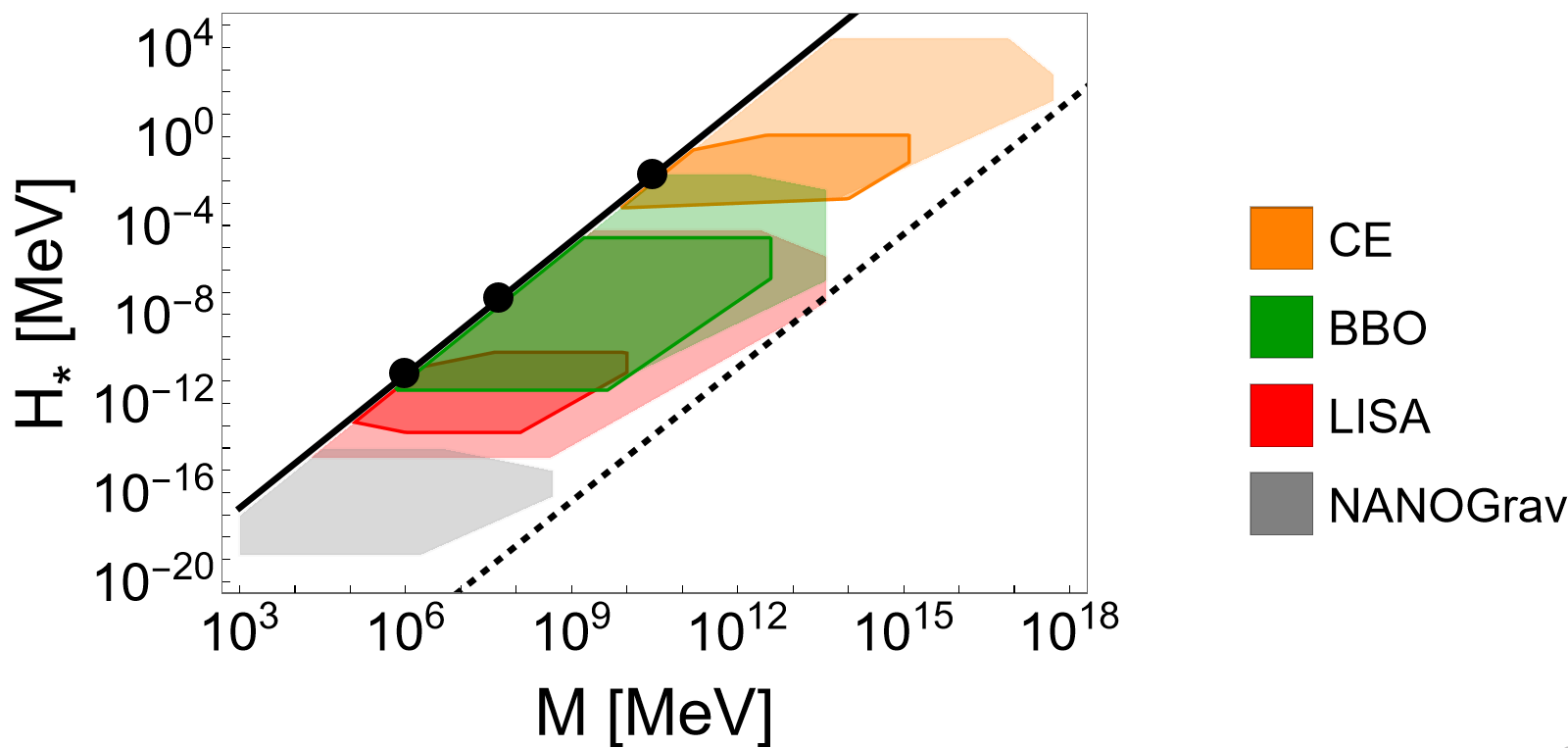}
\caption{
The collective accessible parameter space by several current and upcoming GW experiments: NanoGrav, LISA, BBO and CE.   The thick black line corresponds to the maximum Hubble scale for a given value of $M$, following  Eq.~\eqref{eq:LISAHubble} for $\lambda=0.1$. Smaller values of $\lambda$ parallel-transport the line to the right of the plot. The black dots correspond to the  scale $M$ and the Hubble scale $H_*$ estimated through Eqs.~\eqref{eq:freqwithr} and~\eqref{eq:LISAHubble} for $r=1$ and $\lambda=0.1$, using the central frequency of LISA, BBO and CE (left to right respectively). The lightly shaded regions for CE, BBO and LISA correspond to  overlaying the entire accessible parameter space for each experiment given in Figures~\ref{fig:LISA2}, \ref{fig:aLIGO}, \ref{fig:CE} and~\ref{fig:BBO} for all three values of $\lambda=0.001,0.01,0.1$. Choosing even smaller values of $\lambda<0.001$ would extend each of the colored regions horizontally to the right.
The dark shaded regions for these three experiments correspond only to the cases of $m_\chi/\mu \ll 1 $. In the case of NANOGrav, the  gray shaded region corresponds to the parameter space that leads to a spectrum that goes through all $2\sigma$ data bins of the experiment, as discussed in Section~\ref{sec:PTA}.
We see that the entirety of the parameter space is bounded above and below by the solid black line $H_* = 0.05 M^2/M_{\rm {Pl}}$ and the dashed black line  $H_* = 10^{-13} M^2/M_{\rm {Pl}}$. Both bounds are computed analytically; the former  in Eq.~\eqref{eq:LISAHubble} and the latter  in Eq.~\eqref{eq:Hlowerbound}. 
}
\label{fig:experimentsall}
\end{figure}

\subsection{Parameter Degeneracy}
\label{sec:degeneracy}

Before we conclude the discussion of the parameter dependence of the generated GW peak, we would like to perform in some sense the opposite exercise to Sections~\ref{sec:LISAetal} and \ref{sec:PTA}  and ask the following question:  ``If we measure the amplitude and frequency of the peak of the GW spectrum, what information do we get about the model?"
There are four
 parameters in the Lagrangian and we have access to only two observables: the peak frequency and amplitude of the stochastic GW spectrum from the FOPT.  Therefore, each pair of observables $(f^{peak},\Omega_{GW}^{peak})$ corresponds to a two-dimensional surface on the four-dimensional parameter space. We choose a specific point for concreteness $(f^{peak},\Omega_{GW}^{peak}) = (10^{-3}\, {\rm {Hz}}, 10^{-10})$, chosen to lie at the center of the LISA band. For reasons of better visualization, we again fix $\lambda =0.001,0.01, 0.1$, thus reducing the parameter space to three dimensions, instead of four. This does not hinder our understanding, since we have demonstrated that $\lambda$ 
 mostly affects the results through shifting the frequency (see Figure~\ref{fig:GWscandifflambda}) in a controlled and rather intuitive way. We can thus extrapolate the dynamics for any $\lambda$ by relying on these three values. 
Figure~\ref{fig:degeneracy} shows the combinations of parameters that lead to our chosen benchmark observation point. This provides a visual overview of the parameter range.  We see that for the chosen  point relevant for LISA and $\lambda=0.01$, the required mass ratio is $2\lesssim m_\chi/\mu\lesssim 2.2$, whereas the scale of the kinetic term is $10^7 \, {\rm{MeV}}\lesssim M\lesssim 10^{12} \, {\rm {MeV}}$, with $r$ varying similarly by around $4$ orders of magnitude. Therefore, a possible detection of a GW spectrum allows for a wide spread in model parameters. Because we chose a point with $\Omega_{GW}^{peak}>10^{-12}$, the range of masses will be $m_\chi/\mu>1$, making this the one parameter that is well constrained by such a measurement. This would not be true for $\Omega_{GW}^{peak}<10^{-12}$, which can be accounted for within our model by a potential with large $m_\chi/\mu$ and small $\alpha$, as well as by a potential with $m_\chi/\mu\ll 1 $ and less suppression through $\alpha$.

\begin{figure}
\centering

\includegraphics[width=.6\textwidth]{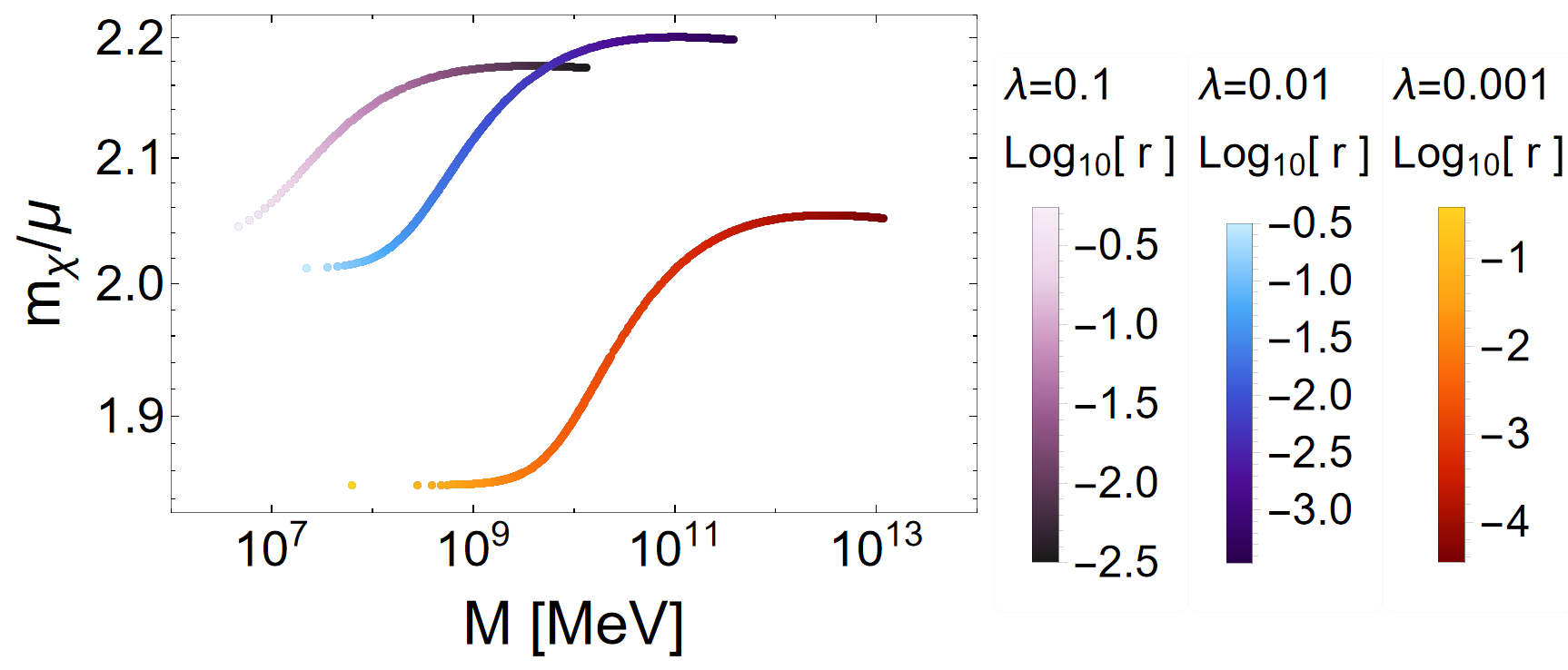}
\hspace{.05\textwidth}
\includegraphics[width=.3\textwidth]{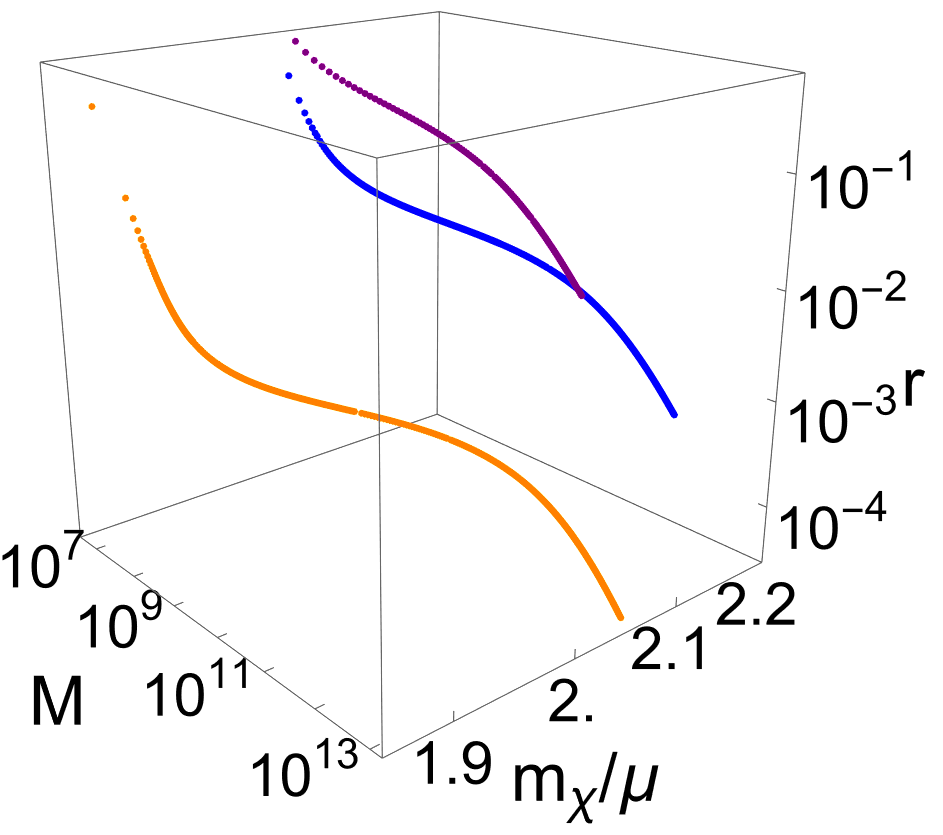}
\caption{
The degeneracy of the parameters for a hypothetical detection in  the LISA band. The left plot shows all the parameter combinations (up to specific choices of $\lambda$) that account for a GW signature of $\Omega_{\rm{GW}}^{peak}h^2=10^{-10}$ and $f_{peak}=10^{-3}$ Hz. 
 The three different lines correspond to $\lambda=0.001,0.01,0.1$ (colored in shades of orange, blue and purple respectively) and we chose to plot on two dimensions, $M$ and $m_\chi/\mu$ with the combination $r=\mu/(M\lambda^2)$ as the color. As a reminder, $0<r<1$  shows the proximity to the  EFT threshold,  derived by requiring positivity of the kinetic term of the $\phi$ field.  The right panel shows the same quantities, but on a three-dimensional space, where for each value of $\lambda$ (similarly in orange, blue and purple), the GW signal can be accounted for by points  along  a curve on the $(M,m_\chi/\mu,r)$ space.  
}
\label{fig:degeneracy}
\end{figure}

\section{Summary and Discussion}\label{sec:conclusions}

We examined the model of a Kination-Induced Big Bang, and more generally reheating after a kination phase in the early universe (in which
the kinetic energy of a rolling field $\phi$ dominates the energy density of the Universe).
Here a first order phase transition (FOPT) triggered by the rolling of the kination field successfully reheats the
universe.  The model can be realized through a derivative coupling of $\dot\phi$ to an auxiliary
scalar $\chi$ (the tunneling field).  Our specific focus in this paper was to calculate the resultant Gravitational Wave (GW) signal in light of current and upcoming GW detectors.

Quoting S. Coleman ``the decay of the false vacuum will lead to a sort of secondary big bang with interesting consequences"~\cite{coleman1977fate}.
The Lagrangian is chosen to correspond to the simplest scenario that can support such dynamics, one that is quite general in that typical
tunneling scenarios can get mapped onto the one we consider. 
The field that transitions from the false to the true vacuum is taken to obey a polynomial potential with quadratic, cubic and quartic terms. This is the simplest potential with two non-degenerate tunable minima. While this can be viewed as overly simplistic or constrictive, the rate of the vacuum tunneling can be always modeled using a quartic potential (see e.g. Ref.~\cite{Adams:1993zs}). In Appendix~\ref{appendix:fieldspace} we demonstrate the similarity between our quartic potential and one constructed using trigonometric functions, further cementing the generality of the chosen potential. Since we require the phase transition to follow a kination phase, the most straightforward operator is one that connects the effective mass of the tunneling field to the velocity of the kination field. 
We chose the lowest dimension operator ${\cal L}_{\rm{int}} = - (\partial \phi)^2\chi^2/M^2$, where the minus is essential in order to get the required behavior.
The effective mass of the tunneling field gets a contribution from this operator.
Since the kination field slows down as it rolls down the potential, the effective mass decreases, thereby destabilizing the tunneling field and causing a first order phase transition.

We demonstrated that a kination-induced first order phase transition (or kination-induced Big Bang) can lead to stochastic GW signals with peak amplitude up to  a maximum value of $ \Omega_{\rm GW}^{peak}h^2 \lesssim 2\times 10^{-7}$.  Here the upper limit is dictated by the requirement that the bubbles of true vacuum percolate successfully to fully put the whole observable universe in the true vacuum (see Eq. \eqref{eq:omegabound}).
A  ``benchmark" value for the GW amplitude is $\Omega_{GW}^{peak}h^2\sim 10^{-12}$. For the case where the energy density of the universe is dominated by the false vacuum during the transition rather than by the kinetic energy, we find that $\Omega_{GW}^{peak}h^2\gtrsim 10^{-12}$. 
We remind the reader that, even in the case where vacuum energy dominates at the time of the FOPT,  $\dot \phi$ nevertheless plays the dominant role in triggering the phase transition\footnote{The velocity $\dot\phi$ controls the tunneling rate  of the  $\chi$ field (see Eq.~\eqref{eq:simple}) due to its presence in $m_{\chi,\rm eff}^2$. 
}.

Alternatively, for the case where $\Omega_{GW}^{peak} h^2< 10^{-12}$, the rolling kination field must dominate (or contribute significantly to) the energy density of the universe at the time of the transition, i.e. $\alpha < 1.$ This can be translated to a very simple relation between the cubic term of the potential $\mu$, the quartic self-coupling $\lambda$ and the kinetic coupling $M$, specifically $\mu/M<3 \lambda^3$.
This means that detecting a GW signal with amplitude below $10^{-12}$ in the context of this model unveils a large hierarchy between the scales $\mu$ and $M$ or a quartic coupling $\lambda ={\cal O}(1)$,   very close to the perturbativity bound. This simple observation can strongly inform model building around this model. On the contrary, signals close to the upper cutoff of $ \Omega_{\rm GW}^{peak}h^2 \lesssim 2\times 10^{-7}$ must obey $\alpha > 1$ corresponding to $3\lambda^3\lesssim {\mu\over M}\lesssim \lambda^2$ (see Eq. \eqref{eq:alphabranches}.
For intermediate amplitudes $10^{-7}\gg \Omega_{GW}^{peak}h^2\gtrsim 10^{-12}$ the degeneracy of the parameters that produce the same signal is larger and no conclusion about the mass hierarchy can be made. 

Depending on the frequency (mostly determined by the pre-factor of the cubic term in the potential), different GW experiments can be used to probe the kination-induced big bang. Most experiments can only cover the ``upper" part of the observation space $\Omega_{GW} h^2\gtrsim 10^{-12}$. However DECIGO and BBO can cover about four orders of magnitude below that, getting deep inside the regime where the kination field dominates over the vacuum energy, pointing to a large hierarchy between the kinetic coupling mass-scale (equivalently the field space curvature) and the cubic term of the potential (related to the energy density of the false vacuum). 
The entire parameter space that is accessible by current and future GW experiments is shown in Figure~\ref{fig:experimentsall}, which shows our results  in the $M-H_*$ plane. This combines the most important macroscopic (cosmological) parameter, the Hubble scale at the time of the transition with the most important microphysical parameter, the strength of the interaction between the two fields, which is controlled by the mass-scale $M$.
Interestingly, all experiments on this plane fall into a well defined band ${\cal O}(0.1){M^2 / M_{\rm {Pl}}} \gtrsim H_* \gtrsim {\cal O}(10^{-13}){M^2 /  M_{\rm {Pl}}}$.

The work of this paper is particularly interesting in the context of quintessential inflation~\cite{Peebles:1998qn, Sahni:2001qp, Tashiro:2003qp, deHaro:2021swo}, which
proposes a unified explanation of inflation and dark energy. In quintessential inflation, there are two vacuum dominated epochs in the history of the Universe,
early inflation and late-time quintessence dark energy; these two periods are connected by a period of kination.  A single scalar field is responsible
for all three epochs. The potential for this field has a high-energy plateau where the rolling of the scalar field produces inflation; followed by a steeper part of the potential
where the potential energy is converted to the kinetic energy of a kination period; and then a low-energy plateau responsible for vacuum-dominated dark energy.
However, reheating has proven to be a difficult problem. In particular, gravitational particle production at the end of inflation \cite{Ford:1986sy}
 has been found to be too ineﬃcient to comply with BBN constraints (see e.g. \cite{Figueroa:2018twl}). {In an upcoming paper, we will  present the phenomenology of a kinetically induced first order phase transition as a reheating  mechanism for quintessential inflation.

Finally, the sign of the kinetic energy term in the Lagrangian that couples the two scalar fields in this paper leads to a potentially interesting connection to curved field-space manifolds in cosmology.
Specifically, the requirement for the kinetic coupling to induce a large effective mass for the tunneling field at early times, which decreases with time and eventually becomes small enough to allow for efficient tunneling, thus triggering the FOPT, specifies the sign of this term (for operators involving even powers of $\dot\phi)$. Viewing this coupling as part of the field-space metric, we can compute the relevant field-space Ricci scalar, which turns out to be positive. 
This observation provides an intriguing model-building opportunity in cosmology. Curved field-space manifolds have been extensively studied in the last several years mostly in the context of multi-field inflation and preheating, as well as for informing quintessence model-building (e.g.~\cite{Christodoulidis:2018qdw, Christodoulidis:2019mkj, Cicoli:2020noz, Carrasco:2015uma, Garcia-Saenz:2018ifx, Brown:2017osf, Iarygina:2018kee, Krajewski:2018moi}). The vast majority of these studies focus on negatively curved (hyperbolic) manifolds, which arise naturally in the context of string theory and super-gravity, or asymptotically flat manifolds with a localized curvature, which arises from non-minimal couplings to gravity (e.g.~\cite{DeCross:2015uza, DeCross:2016cbs, DeCross:2016fdz, Ema:2016dny}). Our current model points towards a positively curved manifold, which opens up a new direction for model-building in the context of inflation (if this ties to quintessential inflation) and  early universe scalar field dynamics in general.
We do not attempt to construct a concrete UV model and instead focus on the dynamics and GW production, using an EFT which contains all relevant parameters. Any concrete construction can easily translate our results to make them applicable to a particular potential or kinetic term.

\section*{Acknowledgments}
We thank 
Aaron Zimmerman for useful discussions regarding the non-detection of GWs by aLIGO. E.I.S. thanks R. Rosati for useful conversations regarding the LISA sensitivity curve.
K.F. holds the Jeff \& Gail Kodosky Endowed Chair at the University of Texas, Austin. We are grateful for support from this Chair.
We acknowledge support from the U.S. Department of Energy, Office
of Science, Office of High Energy Physics program under Award Number DE-SC-0022021. K.F. also 
acknowledges support from the Swedish Research Council (Contract No. 638-2013-8993).

\appendix

\section{Kination}
\label{appendix:kination}

To make our presentation self-contained, we summarize here the basic
dynamics of a kination phase, both in terms of a generic perfect fluid
and for the specific case of a canonical scalar field.
We work in a spatially flat Friedmann--Robertson--Walker (FRW) Universe,
with line element
\begin{equation}
ds^2 = -dt^2 + a(t)^2 d\vec{x}^{\,2} \, ,
\end{equation}
and Hubble parameter $H \equiv \dot{a}/a$. A homogeneous perfect fluid
with energy density $\rho$ and pressure $p$ is characterized by an
equation-of-state parameter
$w \equiv p/{\rho}$.
The continuity equation
$
\dot{\rho} + 3 H (\rho + p) = 0$ in the case of constant $w$ can be rewritten as
\begin{equation}
\dot{\rho} + 3H(1+w)\rho = 0 \, .
\end{equation}
Using $H = \dot{a}/a$ and changing variables from the cosmic time $t$ to the scale-factor $a$, one finds
\begin{equation}
\frac{d\rho}{da} = -\frac{3(1+w)}{a}\,\rho \, ,
\end{equation}
leading to the evolution of the energy density
\beq
\rho(a) \propto a^{-3(1+w)} \, .
\eeq
A kination phase is defined by an effective equation of state $w=1$,
where the pressure equals the energy density. In this case, the energy
density scales as
\begin{equation}
\rho_{\rm kin}(a) \propto a^{-6},
\end{equation}
which redshifts faster than both radiation ($\rho \propto a^{-4}$) and
pressureless matter ($\rho \propto a^{-3}$).
Assuming the kination component dominates the energy budget, the
Friedmann equation becomes
\begin{equation}
H^2 = \left(\frac{\dot{a}}{a}\right)^2
= \frac{\rho_{\rm kin}}{3 M_{\rm Pl}^2}
\propto a^{-6}\, .
\end{equation}
 This implies $H\equiv {\dot{a}} / {a} \propto a^{-3}$, which can be trivially integrated to give
$
a(t) \propto t^{1/3}$. 
Thus, during a kination-dominated era, the scale factor evolves as
$a(t) \propto t^{1/3}$ and the energy density dilutes as
$\rho_{\rm kin} \propto t^{-2}$.

A simple microscopic realization of kination is provided by a single
canonical scalar field $\phi(t)$ with Lagrangian density
\begin{equation}
\mathcal{L} = \frac{1}{2}\,\partial_\mu \phi \,\partial^\mu \phi - V(\phi) \, .
\end{equation}
At the background level, $\phi = \phi(t)$, the energy density
and pressure are
\begin{equation}
\rho_\phi = \frac{1}{2}\dot{\phi}^2 + V(\phi), 
\qquad
p_\phi = \frac{1}{2}\dot{\phi}^2 - V(\phi),
\end{equation}
so that the equation-of-state parameter can be written as
\begin{equation}
w_\phi = \frac{p_\phi}{\rho_\phi}
= \frac{\frac{1}{2}\dot{\phi}^2 - V(\phi)}{\frac{1}{2}\dot{\phi}^2 + V(\phi)}.
\end{equation}
A kination regime corresponds to the kinetic-energy domination limit
\begin{equation}
\frac{1}{2}\dot{\phi}^2 \gg V(\phi).
\end{equation}
In this limit, one has
\begin{equation}
\rho_\phi \simeq \frac{1}{2}\dot{\phi}^2,
\qquad
p_\phi \simeq \frac{1}{2}\dot{\phi}^2,
\end{equation}
and therefore
$
w_\phi \simeq 1
$.
The Friedmann equation for a  Universe dominated by a kination field becomes
\begin{equation}
H^2 = \frac{1}{3M_{\rm Pl}^2}\,\rho_\phi
\simeq \frac{\dot{\phi}^2}{6 M_{\rm Pl}^2}.
\end{equation}
At the same time, the Klein--Gordon equation simplifies because the
potential and its derivative are negligible,
\begin{equation}
\ddot{\phi} + 3H\dot{\phi} \simeq 0.
\end{equation}
Let us spend some time examining the approximation $dV/d\phi \ll 3H\dot \phi$. Kination requires $V\ll \dot\phi^2/2$, as we saw above. From the Friedman equation, we get that during kination
\beq
3H\dot{\phi} \simeq \frac{\dot{\phi}^2}{M_{\rm Pl}} \, .
\eeq
The ratio of the potential derivative (force) term to the friction term becomes
\begin{equation}
\frac{|V_{,\phi}|}{3H | \dot{\phi}|}
\simeq
\frac{|V_{,\phi}|\, M_{\rm Pl}}{\dot{\phi}^2}.
\end{equation}
During kination $\dot{\phi}^2 \gg V(\phi)$. For any potential that is not excessively steep, we can estimate the typical derivative as $dV/d\phi \sim V / \Delta\phi$ leading to
\beq
\frac{|V_{,\phi}|}{3H | \dot{\phi}|}
\ll { V\over \Delta\phi} {{M_{\rm{Pl}}} \over V } = {{M_{\rm {Pl}}}\over \Delta\phi}
\eeq
So unless the hierarchy between the Planck mass and the field excursion is larger than the hierarchy between the kinetic and potential energy, we are justified to neglect the potential term from the Klein-Gordon equation. That being said, since $\dot\phi \sim H M_{{\rm Pl}}$ we can estimate $\Delta\phi \sim \dot\phi \Delta t \sim (H M_{{\rm Pl}}) H^{-1}\sim M_{\rm {Pl}}$, thereby making our set of approximations self-consistent.

If we also want to compute the evolution of the field in terms of cosmic time $t$, we can insert our solution for the Hubble scale $H=\dot a/a=1/3t$ into the Klein-Gordon equation as
\begin{equation}
\ddot{\phi} + \frac{1}{t}\dot{\phi} = 0 \, .
\end{equation}
This can be integrated in two steps (first for $\dot\phi$ and subsequently for $\phi$), leading to

\begin{eqnarray}
    \dot{\phi}(t) &=& \pm \sqrt{\frac{2}{3}}\,\frac{M_{\rm Pl}}{t} \, ,
\\
\phi(t) &=& \phi_0 \pm \sqrt{\frac{2}{3}}\,M_{\rm Pl}
\ln\!\left(\frac{t}{t_0}\right) \, .
\end{eqnarray}
The corresponding energy density scales as
\begin{equation}
\rho_\phi \simeq \frac{1}{2}\dot{\phi}^2
= \frac{1}{3}\,\frac{M_{\rm Pl}^2}{t^2}
\propto t^{-2},
\end{equation}
which is consistent with the general scaling
$\rho_\phi \propto a^{-6}$ and $a(t)\propto t^{1/3}$.

In summary, a kination phase driven by a single canonical scalar field
is characterized by a stiff equation of state $w_\phi \simeq 1$, an
energy density $\rho_\phi \propto a^{-6}$, and scale-factor evolution
$a(t)\propto t^{1/3}$. The scalar field itself rolls
logarithmically in cosmic time, with its kinetic energy dominating over
the potential throughout the kination regime while becoming smaller over time.

\section{Potential structure}\label{appendix:potential}

Let us take a closer look at the potential of $\chi$,
\beq
V(\chi) = {m_\chi^2\over 2}\chi^2 -\mu \chi^3 +\lambda^2\chi^4 + V_0
\eeq
For visual clarity, we present a qualitative picture of the potential in Figure~\ref{fig:cartoonpotential}.
\begin{figure}
\centering
\includegraphics[width=.7\textwidth]{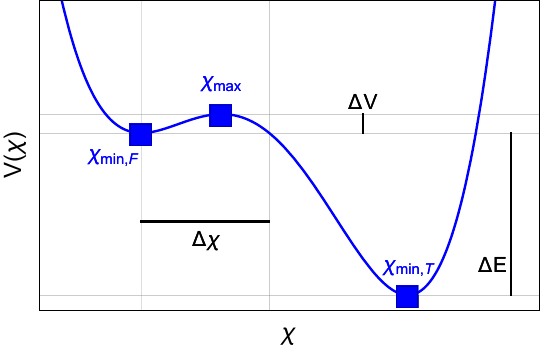}
\caption{The form of the tunneling potential $V(\chi)$, showcasing the two non-degenerate vacua separated by a potential barrier with height $\Delta V$ and width $\Delta\chi$.
The energy difference between vacua is $\Delta E$.
Since we take the energy of the true vacuum to be  $V(\chi_{\rm {min},T})=0$, we have $\Delta E = V_0$  in the potential in Eq. (A.1).
}
\label{fig:cartoonpotential}
\end{figure}
There are two minima at 
\begin{eqnarray}  
\chi_{\rm {min},F}&=&0 
\\\chi_{\rm {min},T} &=& \frac{3 \mu}{8 \lambda ^2}  \left(1 + \sqrt{1-\left (\frac{4 \lambda}{3} \frac{m_\chi}{ \mu } \right )^2}\right) 
 \simeq {3\mu \over 4\lambda^2}
- {m_\chi^2\over 3\mu} +{\cal O}(\lambda^2)
\label{eq:chiminT}
\end{eqnarray}
and one local maximum at
\beq
\chi_{\rm {max}} = \frac{3 \mu}{8 \lambda ^2}  \left(1 - \sqrt{1-\left (\frac{4 \lambda}{3} \frac{m_\chi}{ \mu } \right )^2}\right)
\simeq {m_\chi^2\over 3\mu} + {4m_\chi^4 \over 27\mu^3}\lambda^2 + {\cal O}(\lambda^4)
\, ,
\eeq
where the approximate expressions are valid for $\lambda\ll 1$. We thus see that the smallness of $\lambda$ creates a hierarchy between the position of the potential barrier and the true vacuum, 
$\chi_{\rm {\max}} / \chi_{\rm {min},T}
\simeq \lambda^2 (4m_\chi^2 / 9\mu^2)$. The smallness of $\lambda$ will be used to allow the derivation of simple (approximate) expressions for the dynamics of the phase transition and  the corresponding GW production. However, choosing $\lambda\ll 1$ is not fine-tuned, since $\lambda\sim 0.01-0.1$ is already enough for us to use the relevant approximations without sacrificing accuracy.

The height of the barrier (with respect to the metastable local minimum at $\chi=0$) is 
\beq
\Delta V \equiv V(\chi_{\rm {max}}) - V(\chi_{{\rm {min}},F}) \simeq \frac{m_\chi^6}{54 \mu ^2} +\lambda ^2  \frac{m_\chi^8}{81 \mu ^4}+{\cal O}\left (\lambda^4\right ) \, .
\eeq
The width of the barrier, the distance between $\chi_{{\rm {min}},F}=0$ and $V(\chi\ne 0)=0$, is
\beq
\Delta\chi = {\mu - \sqrt{\mu^2 - 2m_\chi^2 \lambda^2}\over 2\lambda^2} \simeq{m_\chi^2\over 2\mu} + {\cal O}(\lambda^2)
\eeq
When $\dot\phi\ne 0$, the position and height of the barrier is given by the simple substitution $m_\chi^2\to m_{\chi,{\rm {eff}}}^2\equiv m_\chi^2+2{\dot\phi^2/ M^2}$.
The constant $V_0$ is chosen such that the potential energy vanishes at the true vacuum $V(\chi_{\rm {min},T})=0$. Specifically
\beq
V(\chi_{\rm {min},T}) \simeq -\frac{27 \mu ^4}{256 \lambda ^6}
+\frac{9 \mu ^2 m_\chi^2}{32 \lambda ^4} -\frac{m_\chi^4}{8 \lambda ^2}+ V_0 + ...
\eeq
leading to
\beq
\label{eq:V0simplest}
V_0 \simeq \frac{27 \mu ^4}{256 \lambda ^6} 
\eeq
to lowest order in $\lambda\ll 1$.

Let us note that the above series expansion breaks if $m_\chi \gg \mu$, because in that case $\mu/m_\chi \ll 1$ is also a small parameter, whose ``smallness" must be compared to that of $\lambda \ll 1$. In the main text, the  calculations we perform are done in the opposite limit of $m_\chi/\mu\ll 1$ or at most $m_\chi/\mu = {\cal O} (1)$, making the small-$\lambda$ series expansion applicable.
Finally, the energy difference between the two vacua is
\beq
\Delta E=\frac{\mu^4  }{4 \lambda ^6}\left({9\over 16} - \lambda ^2 {m_\chi^2\over \mu^2}\right)^{3/2}
\eeq
where we recover the condition for degenerate minima, $\Delta E=0$ for $\lambda {m_\chi/ \mu}=3/4$.
Since we take $V(\chi_{\rm {min},T})=0$, we have $\Delta E = V_0$.

Analyzing the potential, we can also comment on the type of the bubbles of true vacuum that emerge during the phase transition. 
The requirement for the existence of a thin wall bubble is that the energy difference between the two vacua is much smaller than the height of the barrier. In the regime of $\lambda\ll 1$ this translates to
\beq
{V_0\over \Delta V} \simeq {1\over \lambda^6}{729\over 128} {\mu^6\over m_\chi^6} \, .
\eeq
We see that $\lambda\ll 1$ implies  ${V_0/ \Delta V}\gg 1$, putting us in the thick-wall regime for the generated bubbles of true vacuum. In fact, we must note that the mass of the $\chi$ field in these calculations is not the bare mass but rather the effective mass $m_{\chi,{\rm {eff}}}^2 = m_\chi^2 + 2\dot\phi^2/M^2$. Since  (generically) $\dot\phi^2/M^2 \sim \mu^2$ during the transition (as we showed in this work), we immediately see that $\mu/m_{\chi,{\rm{eff}}} = {\cal O}(1)$. In this case, it is impossible for the nucleation process to generate thin-wall bubbles. 
Let us make a brief comment here regarding the normalized barrier height $\Delta V/V_0$ which is generally small in our case,  meaning the barrier isn’t much taller than the false vacuum energy. 
In the thick-wall regime, the barrier can still be wide and the field may traverse a large range slowly, accumulating significant action. Thus, even with small 
$\Delta V/V_0$, the bounce solution can be broadly distributed in space, leading to a large Euclidean action and suppressed tunneling.

An alternative argument for the impossibility of thin-wall bubbles that does not rely on a series expansion, is based on the condition on the potential parameters for the two vacua to be degenerate. This translates to $\lambda = \mu /\sqrt{2} m_\chi$. Noting again  that in the relevant calculations we use the effective mass of the $\chi$ field,
$m_\chi\to m_{\chi,{\rm {eff}}} = {\cal O}(\mu)$, we see that having degenerate minima necessarily pushes $\lambda$ close to the perturbativity limit.

\section{EFT bound}
\label{sec:EFTlimit}

Let us comment on the value of $M$, which defines the quartic coupling ${\cal L}\subset -(\partial \phi)^2 \chi^2/M^2$.  $M$ is restricted by the requirement for a positive kinetic term for the $\phi$ field, leading to the constraint $\chi^2 / M^2 <1/2$. We require that this constraint holds for the entirety of the dynamics of the phase transition, including once the field $\chi$ has reached its true vacuum at $\chi^2_{\rm {min,T}}<M^2/2$. In the limit of $\lambda\ll 1$, we can approximate $\chi^2_{\rm {min,T}} \simeq 3\mu/4\lambda^2$
  leading to the constraint
  \beq
{\mu\over M\lambda^2 }< \sqrt{8\over 9}\simeq 0.94 \, .
  \eeq
The ratio on the left is defined in Eq.~\eqref{eq:definer} of the main text as $r$. For simplicity we round up the inequality as 
\beq
r = {\mu \over M\lambda^2} <1 \, . 
\eeq
We must note that the above inequality holds strictly for $\lambda m_\chi / \mu\ll 1$. We have seen in the main text that bubble percolation constraints limit the range of the $\chi$ field mass to be $m_\chi/\mu \lesssim 3$. With this in mind we can compute the correction that arises when using the full expression of Eq.~\eqref{eq:chiminT} for $\chi_{\rm {min,T}}$ versus the leading order contribution. For $\lambda=0.1$ the correction as the percent level or less and it quickly diminishes for smaller values of $\lambda$, becoming $\le {\cal O}(10^{-4})$ for $\lambda=0.01$. We will thus use the  simple expression $r <1$ throughout our work.

Before concluding this section, let us note that the scale $M$ is described as the limit of validity of the EFT presented in the Lagrangian of Eq.~\eqref{eq:lagrangian}, but it has also a ``tangible" particle physics interpretation. It controls the interaction strength between the $\phi$ and $\chi$ fields. Specifically, due to the coupling of the form $(\partial\phi)^2 \chi^2/M^2$ in the Lagrangian, the scattering cross section for $\phi\phi\to \chi\chi$ scales as $\sigma \propto s/M^2$, where $s$ is the center of mass energy and we neglected   $\langle \chi\rangle$, which produces extra diagrams. So we can  use a ``Fermi-like" coupling $g= 1/M^2$ instead of the scale $M$  as an equivalent description of the parameter space of the model.

\section{Field space metric}\label{appendix:fieldspace}

In Eq.~\eqref{eq:lagrangian}, we took an Effective Field Theory approach, using the dimension-$6$ operator ${\chi^2\over M^2} (\partial \phi)^2$, which could arise from integrating out some heavy fields at the scale $M$.
Considering the Lagrangian ${\cal L}\subset {1\over 2} {\cal G}_{IJ} \partial_\mu \varphi^I \partial_\mu \varphi^J$ with $\varphi^I =\{\phi,\chi\}$ the field-space metric is ${\cal G}_{II} =\{1-\chi^2/M^2 ,1\} $ with zero off-diagonal elements. The field-space Ricci curvature is 
\beq
{\cal R }=\frac{2 M^2}{\left(M^2-\chi ^2\right)^2}  = {2\over M^2} +{\cal O} (\chi^2/M^4)
\eeq
This is a positively curved field-space and it therefore introduces a positive contribution to the effective mass of the $\chi$ field (see e.g. Refs.~\cite{Kaiser:2012ak, Renaux-Petel:2015mga}). 

While negatively curved field-space manifolds typically arise in the context of string theory (through kinetic couplings of the form $e^\phi (\partial\chi)^2$) or supergravity  (in the form of $\alpha$-attractors), positively curved manifolds are found in the low energy EFT for light mesons in QCD.
Considering two flavors we can describe the meson fields $\pi^i$ using $SU(2)$ matrices \beq U(x^\mu) = e^{i {1\over f_\pi} \sigma^i \pi^i(x^\mu) } \, ,\eeq where $i=1,2,3$ and $\sigma^i$ are the Pauli matrices. 
The leading order chiral Lagrangian for pions is
\beq
\label{LagrU}
{\cal L} =  {f_\pi^2\over 4} {\rm Tr}\left[ \partial_\mu U^\dagger \partial^\mu U \right]  + {f_\pi^2 B \over 2} {\rm Tr} \left [
M(U+U^\dagger)
\right
]
\eeq
where $M= {\rm diag}(m_u, m_d)$ is the light quark mass matrix and $B = -\langle \bar q q \rangle / f_\pi^2$, where $\langle \bar q q \rangle $ is the chiral condensate (the vacuum expectation value  of the quark bilinear). We neglect the mass term, as we are interested in the geometrical structure of the kinetic term. 

We can parametrize the pion fields as
\beq
\vec \pi(x^\mu) = 
f_\pi \rho(x^\mu) \left (
\sin\theta(x^\mu)\cos\phi(x^\mu),
\sin\theta(x^\mu)\sin\phi(x^\mu),
\cos\theta\phi(x^\mu)
\right )
\eeq
leading to
\beq
{\cal L}\supset
{f_\pi^2 \over 2} \left [
(\partial \rho)^2 + \sin^2\rho (\partial\theta)^2 + \sin^2\rho \sin^2\theta (\partial \phi)^2
\right
]
\eeq
with the spacetime Ricci curvature being ${\cal R}=6/f_\pi^2$. If we consider only two fields, thus truncate the kinetic term to only include $\rho,\theta$, the Ricci scalar becomes ${\cal R}=2/f_\pi^2$.  If we make the trivial field redefinition $\rho \to \rho \pm \pi/2$ and absorb $f_\pi$ in the definition of the fields $\rho$ and $\theta$, the Lagrangian becomes
\beq
{\cal L}\supset {1 \over 2} (\partial \tilde \rho)^2
+ {1 \over 2} \cos^2(\tilde \rho /f_\pi^2) (\partial\tilde \theta)^2
\simeq {1 \over 2} (\partial \tilde \rho)^2
+ {1 \over 2} \left (1 - {\tilde\rho^2\over f_\pi^2}
\right )(\partial\tilde \theta)^2
+{\cal O}\left ({\tilde\rho^4\over f_\pi^4}\right )(\partial \tilde\theta)^2
\eeq
We thus see that we recover the kinetic term of the  Lagrangian given in Eq.~\eqref{eq:lagrangian}.

If we use a pion-like Lagrangian, the potential must also respect the geometrical structure of the three-sphere, meaning that the potential must be periodic. Using we trigonometric functions, it is very easy to construct a potential with a false and true minimum, for example $V(\chi) = 1-\cos^2\chi - \sin^3\chi - 54\sin^6(\chi/4)$ is very similar to the potential $V(\chi) = \chi^2 - \chi^3 + 0.1\chi^4$. In particular, the energy barrier (which determines the dynamics of the first order phase transition) is practically identical, as shown in Figure~\ref{fig:toymodels}. Of course, the existence of two identical true vacua can cause issues like domain walls. This can be alleviated by the introduction of the third field, connecting the two vacua similarly to the SM Higgs potential. This will leave behind cosmic strings instead of domain walls. While such constructions can provide more observable predictions, these become highly model dependent. More importantly, the dynamics of the phase transition, and therefore the GW signal, is insensitive to such details and is thus a ``universal" prediction of a kinetically induced first order phase transition, which is described by the effective Lagrangian of Eq.~\eqref{eq:lagrangian}.

\begin{figure}
\centering
\includegraphics[width=.45\textwidth]{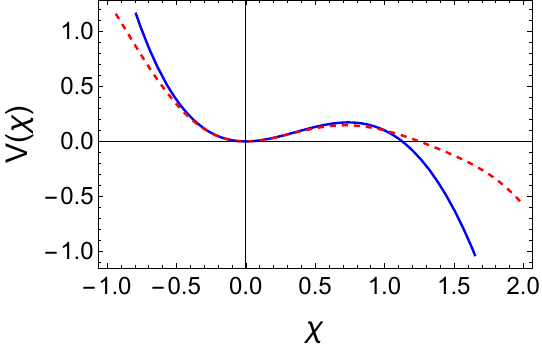}
\includegraphics[width=.45\textwidth]{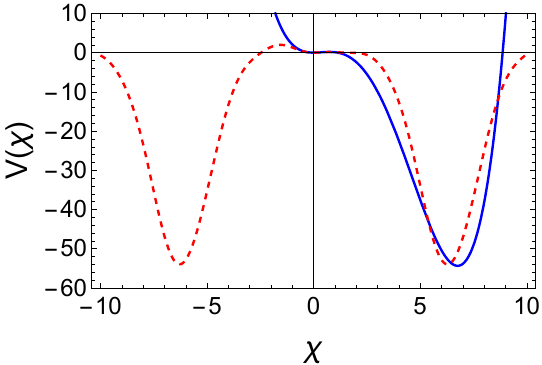}
\caption{Polynomial (blue solid) and trigonometric (red-dashed) potential for the $\chi$ field, demonstrating that the energy barrier between true and false vacua can be realized in a periodic potential, such that the kinetically-induced phase transition can be realized in a Lagrangian with  a pion-like kinetic term.}
\label{fig:toymodels}
\end{figure}

\bibliographystyle{JHEP}
\bibliography{GW_kination.bib}

\end{document}